\documentclass[reprint,superscriptaddress,notitlepage,amsmath,amssymb,aps,onecolumn,floatfix,tightenlines,longbibliography,11pt]{revtex4-1}
\usepackage{graphicx}
\usepackage{amssymb,amsmath}
\usepackage{natbib}
\usepackage{subfigure}
\usepackage{epstopdf}
\usepackage{color}
\usepackage{tikz}

\begin{document}

\title{Instability and disintegration of vortex rings during head-on collisions and wall interactions}
\author{Aakash Mishra}
\affiliation{Department of Mechanical Engineering, University of Houston, Houston, TX 77040, USA}
\author{Alain Pumir}
\affiliation{Universit\'e de Lyon, ENS de Lyon, Universit\'e Claude Bernard, CNRS, Laboratoire de Physique, 69342 Lyon, France}
\affiliation{Max Planck Institute for Dynamics and Self-Organization, 37077 G\"ottingen, Germany}
\author{Rodolfo Ostilla-M\'onico}
\affiliation{Department of Mechanical Engineering, University of Houston, Houston, TX 77040, USA}
\date{\today}

\begin{abstract}
The head-on collision of two vortex rings can produce diverse phenomena: a tiara of secondary rings, vortex sheets which flatten and interact iteratively, or the violent disintegration of the rings into a turbulent cloud. The outcome of the interaction is determined by the nature of the instability affecting two impinging vortex rings. Here, we carry out a systematic study to determine the dominant instability as a function of the parameters of the problem.  To this end, we numerically simulate the head on collision of vortex rings with circulation Reynolds numbers between $1000$ and $3500$ and varying slenderness ratios $\Lambda=a/R$ ranging from $\Lambda=0.1$ to $0.35$, with $a$ the core radius and $R$ the ring radius. By studying the temporal evolution of the energy and viscous dissipation, we elucidate the role azimuthal instabilities play in determining what the outcomes of the collision are. We then compare these collisions to the head-on impact of a vortex ring on a free-slip and a no-slip wall. The free-slip wall imposes a mirror symmetry, which impedes certain instabilities and at sufficiently large Reynolds numbers leads to the formation of a half-tiara of vortices. Impact against a no-slip wall results in the process where a secondary vortex ring is formed after the ejection of the resulting boundary layer. When the Reynolds number is above a certain threshold, which increases with $\Lambda$, the vortices disintegrate through azimuthal instabilities, resulting in a turbulent cloud.
\end{abstract}

\maketitle

\section{Introduction}

In a fluid stirred at large scale, turbulence sets in when the Reynolds number is large enough. The motion is then 
characterized by a hierarchy of eddies, from the forcing scale down to very small scales. The mechanism describing the formation of these small eddies can be pictorially described by the notion of Richardson cascade~\cite{ric20}. Taylor and Green made the essential remark that generation of small scales in a fluid can be traced back to 
the nonlinear interaction in the Navier-Stokes equations, which in fact could conceivably lead to the formation of singularities in the Navier-Stokes equations~\cite{tay37}. A considerable amount of work has been devoted to a description of the cascade in statistical terms~\cite{bat53,pop00}, and more recently, into providing a mechanistic description of it \cite{got12,keo20}.

Understanding quantitatively the formation of higher-harmonics, or equivalently of large velocity gradients, is a multi-faceted problem. By solving an instantaneous optimization problem, Lu and Doering \cite{lu08} proposed that head-on ring collisions are maximally enstrophy producing at high Reynolds numbers. This result was later extended to finite time optimization by Kang \emph{et al.}~\cite{kan20}. Thus, it is not surprising that historically vortex tube or vortex ring interactions have been used as a framework for studying the formation of large velocity gradients~\cite{pum87,ker93s,hou06,bre16,moff19a,Yao20} and the turbulent cascade \cite{sha92}. In particular, the classical work of Lim \& Nickels \cite{lim92} showed that at sufficiently high Reynolds numbers, the head-on collision of two vortex rings led to a rapid disintegration of the coherent rings to form a turbulent cloud, and this has inspired further experimental and numerical studies of this flow \cite{keo18,keo20}.

How such a turbulent cloud forms crucially depends on which azimuthal instability is predominant \cite{keo20}. Below the threshold Reynolds number for cloud formation, Lim and Nickels \cite{lim92} observed the formation  of a tiara-like structure of secondary rings. The interaction process is started by azimuthal instabilities consistent with the mechanism originally discovered by Crow to explain the growth of perturbation along antiparallel vortices~\cite{cro70}, and is later mediated through local reconnection events that result in the formation of the secondary vortices. We stress that the Crow instability acts, in certain circumstances, as a precursor to local reconnection, but it cannot be identified with reconnection as it can sometimes lead to disintegration of significant regions of the vortices, leaving behind a turbulent cloud \cite{ost21}.

In other circumstances \cite{keo20}, the formation of turbulent clouds after vortex collision and  disintegration was associated to the Tsai-Widnall-Moore-Saffman (TWMS) instability \cite{moo75,tsai76}, also known as the elliptical instability. McKeown \emph{et al.}~\cite{keo18,keo20} further investigated the system in Ref.~\cite{lim92} using high-speed imaging and simulations, observing diverse mechanisms of interaction, ranging from the iterative formation and breaking of vortex sheet to a cascade of elliptical instabilities, finding that at higher Reynolds number the elliptical instability tends to dominate \cite{keo20}. However, the competition between elliptical and Crow-like instabilities depends not just on Reynolds number, but also on ring slenderness and the vorticity distribution \cite{chu95,keo20}. For example, for vanishingly slender filaments, the Crow instability, and not the elliptical instability dominates at high Reynolds numbers, as seen for example in an aeroplane's wake \cite{lew16}. Thus, the regions of phase space corresponding to various dynamical regimes are delimited by more than a single parameter.

The interaction of a single vortex ring with a wall is related to the head-on collision between two rings. In particular, a ring approaching a free-slip wall can be thought of as interacting with its mirror image. It is therefore interesting to consider the instabilities of the ring during the interaction with both stress-free and no-slip walls. In the former case, the presence of a mirror image imposes a strong symmetry, which favors the development of the Crow instability and favors a certain type of reconnection. Existing studies of this geometry have distinguished the three instabilities happening during the interaction: the ring's elliptical instability stemming from self-interaction, a long-wavelength Crow-type instability between the image and the ring, and the a short wavelength elliptical-like instability between image and ring which is not present for two tubes with imposed symmetry \cite{arc10}. Using this, Archer \emph{et al.} \cite{arc10} proposed that the interaction between the initial TWMS seeding and the Crow instability controls the transition between secondary ring pinch off seen by Lim and Nickels \cite{lim92} and complete disintegration, further complicating the parameter space. A similar phenomena was observed by Laporte \& Corjon \cite{lap00} in the interaction between two vortex tubes.

On the other hand, head-on collision of a vortex against a no-slip wall forms boundary layers close to the surface, which significantly affect the flow by forming opposite signed vorticity structures which later lift-off to produce secondary and tertiary rings, and cause the vortex ring to ``rebound'' off the wall \cite{wal87,orl93,swearingen1995dynamics,che10}. While this appears to be remarkably distinct from the two vortex collision, at sufficiently high Reynolds numbers the interaction between primary and secondary vortex rings leads to complete disintegration of the ring after substantial deformations of the secondary vortex \cite{wal87}. The large strains arising from the primary (original) vortex are responsible for destabilizing the secondary ring \cite{swearingen1995dynamics}, so this process has been related to the elliptical instability \cite{lut97}, which is responsible for the disintegration of vortex ring pairs during head-on collision. However, more recent work has called this strict identification into question, as the calculations in Ref.~\cite{lut97} do not account for the angular rotation effects \cite{har12}. In addition, Ref.~\cite{thompson2007sphere} showed that centrifugal instabilities can arise even when the Reynolds number is not large enough to curl the lifted boundary layer up into a secondary vortex due to the changing vorticity sign. Regardless of the exact instability type, at high Reynolds numbers the impact of a ring on a no-slip wall results in flow dynamics are that share many similarities with the head-on collision and this merits that this case is included alongside the other two as the parameter space is mapped.

This brief review makes it clear that a range of diverse instabilities and their interactions control the outcome of the head-on collision of two vortex rings and of vortex rings impacting a flat surface. Understanding when the rings disintegrate to form a turbulent cloud is crucial for understanding the transfer of energy across scales in a fluid flow, and to relate it to the question of maximal ensthropy growth \cite{lu08,kan20}. In this manuscript, we conduct a series of direct numerical simulations of vortex ring head-on collision, as well as vortex ring impact against a flat wall to explore the parameter space and classify the different outcomes of the process, noting which regions of parameter space produce the fastest transfer of energy across scales.

The manuscript is laid out as follows. Section 2 describes the code used, the simulation parameters and the geometry used, as well as supporting the choices by conducting a series of additional simulations of a single ring evolving in time. Section 3 describes the head-on collision between two thin rings, and the collision between a thin ring and a stress-free wall. Section 4 describes the effect of ring thickness on these two cases. Section 5 describes the interaction between a no-slip wall and a ring. Finally, Section 6 presents the summary and conclusions of the study. 

\section{Simulation setup}

\subsection{Code details}

Both configurations studied in this paper (head-on collision and ring-wall impact) are simulated using the same code. This code solves the incompressible Navier-Stokes equations using an energy-conserving second-order centered finite difference scheme in cylindrical coordinates. We emphasize that the choice of cylindrical coordinates is crucial to avoid artifacts coming from a Cartesian discretization of a toroidal fluid structure \cite{ver96}. We will return to this point later, see Sec.~\ref{subsec:ring_thick}.

Fractional time-stepping is implemented such that a third order Runge-Kutta scheme is used for the non-linear terms and a second order Adams-Bashforth scheme is used for the viscous terms \cite{ver96}.  The solver uses $q_r=r v_r$ as a primitive variable to avoid singularities near the axis. Spatial discretization is taken as uniform in the azimuthal direction, while points are clustered in the radial and axial directions around the regions of interest (the collision region or the wall). The time-step was dynamically chosen so that the maximum Courant-Friedrich-Lewy (CFL) condition number was $1.2$, with the stability limit of the code being $\approx\sqrt{3}$ \cite{ver96}.

\begin{figure}
\centering
\begin{tikzpicture}
\usetikzlibrary{arrows.meta}
\draw (2,7) circle(1cm);
\draw (2,6) -- (8,6);
\draw (2,8) -- (8,8);
\draw (8,7) circle(1cm);

\draw [-latex, thick] (7.3,7) arc(180:380:0.7cm); 
\node at (8,7) {$\Gamma$};

\draw [latex-latex, thick] (2.5,7.866) -- (1.5,6.134);
\node [above left] at (2,7) {$2a$};
\draw[fill] (2,7) circle [radius=0.05];

\draw [latex-latex, thick] (5,8.5) -- (2,8.5);
\node [above] at (3.5,8.5) {$R$};
\draw [dotted] (2,7)--(2,8.5);
\draw [dashed] (5,9) -- (5,5.5);
\draw [-latex,thick] (5,6) -- (5,5.5);
\node [right] at (5,5.5) {$V_a$};

\end{tikzpicture}
\caption{Schematic of vortex ring. The figure represents a cut of the ring by a plane containing the axis of symmetry, shown as a dashed line. The ring slenderness $\Lambda$ is defined as $\Lambda=a/R$. The ring self-advects downwards with a velocity $V_a$. } 
\label{fig:schemaring}
\end{figure}
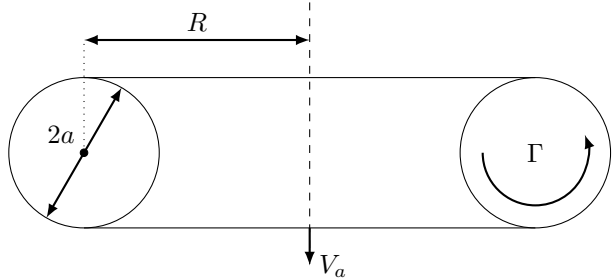

Vortex rings are implemented as initial conditions. We use vortex rings, starting from the standard
Gaussian (Lamb-Oseen) vorticity profile for a tube:

\begin{equation}
 \omega_\theta(\rho) = \displaystyle\frac{\Gamma}{\pi a^2}\exp\left( -\frac{\rho^2}{a^2} \right)
\end{equation}

\noindent with $\rho$ the distance to the vortex center. By using this vorticity distribution and extending it to a ring, we obtain solutions with a circulation $\Gamma$ and a core radius $a$, with the core radius defined as the second moment of the vorticity around the vortex center \cite{lew16}. This geometry results in the ring self-advecting with a velocity $V_a\sim\log(a/R)$ \cite{saf79}. We can characterize the rings using two non-dimensional control parameters: the circulation Reynolds number $Re_\Gamma=\Gamma/\nu$, and the slenderness ratio $\Lambda=a/R$, where $R$ is the ring, or outer radius. A schematic of how these parameters are defined can be seen in Fig.~\ref{fig:schemaring}.

The above definition of vorticity results in the following rotational velocity profile when the radius $R$ is very large (a straight vortex tube), i.e.~when $\Lambda \rightarrow 0$: 

\begin{equation}
 V_\theta(\rho) = \displaystyle\frac{\Gamma}{2\pi\rho} \left ( 1-\exp \left (-\displaystyle\frac{\rho^2}{a^2} \right) \right ).
 \label{eq:lamboseen}
\end{equation}

\begin{figure}
 \centering
 \includegraphics[width=0.49\textwidth]{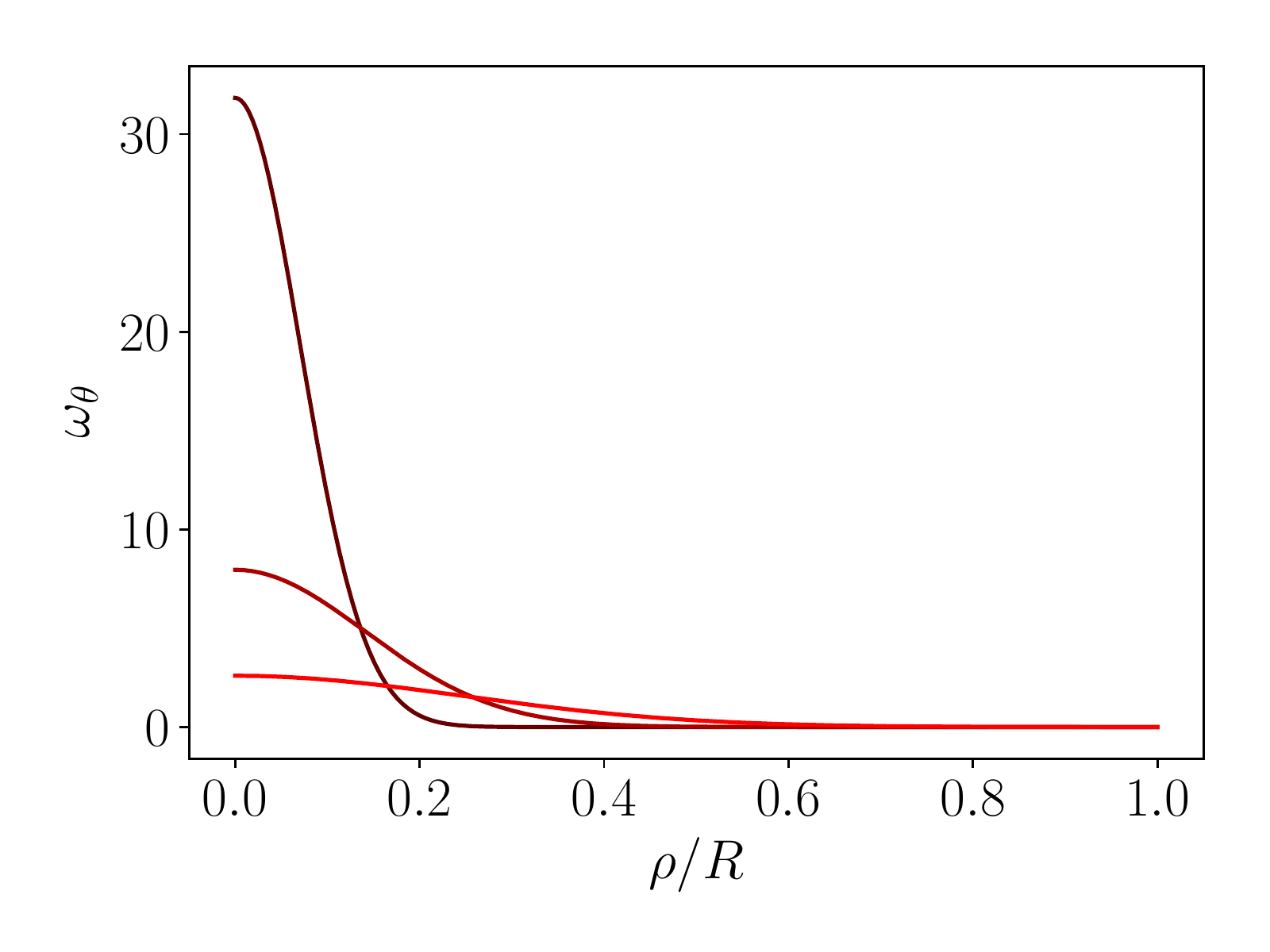}
 \includegraphics[width=0.49\textwidth]{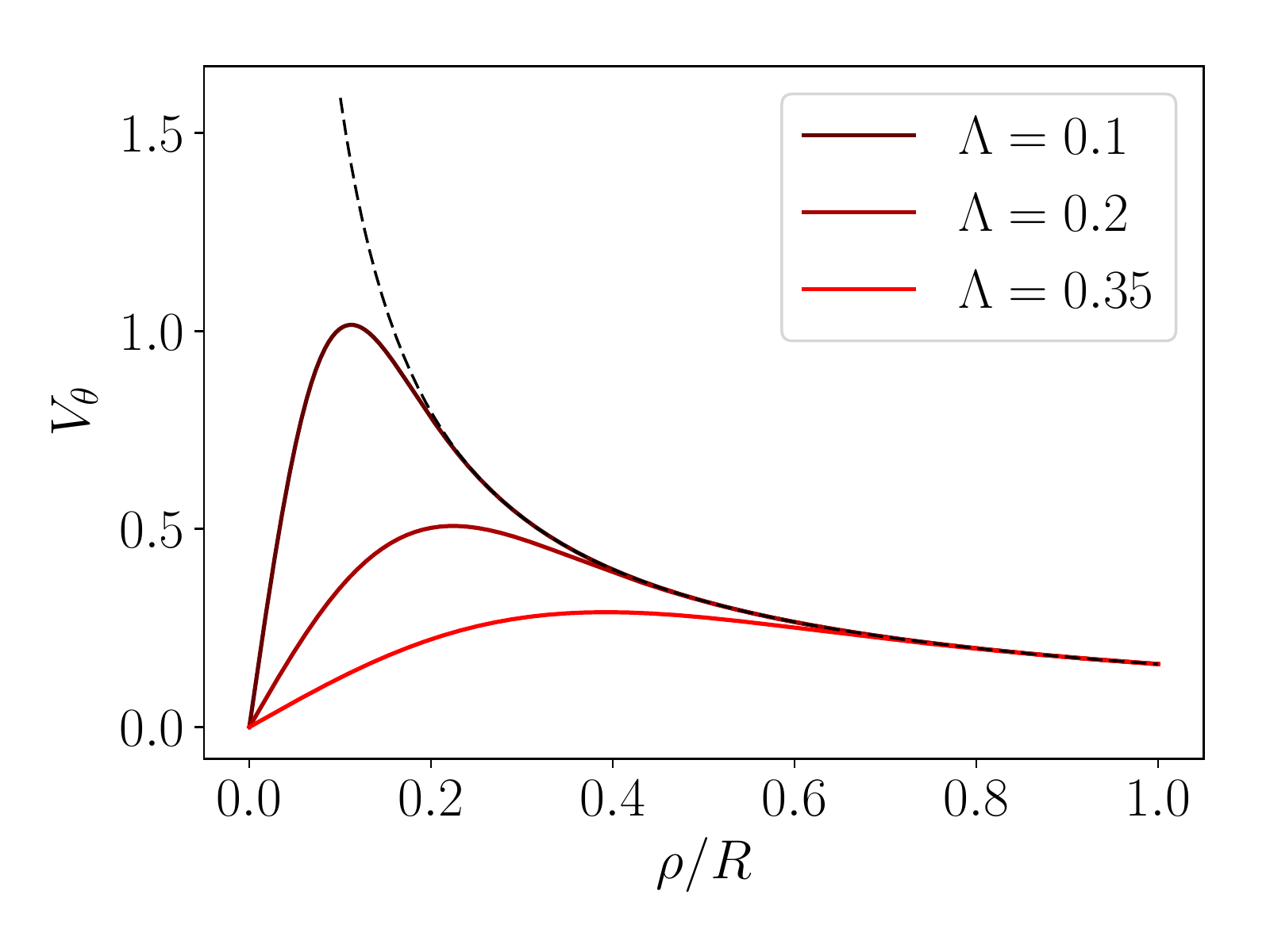}
 \caption{Vorticity (left) and rotation velocity (right) profiles for rings of different slenderness. The dashed black line on the right panel denotes the point free-vortex asymptote $V_\theta=\Gamma/[2\pi\rho]$.}
 \label{fi:vortprof}
\end{figure}

\noindent We show the initial profiles for several values of $\Lambda$ in Fig. \ref{fi:vortprof}. In a ring with a finite radius, the profiles of $\omega_\theta$ and $V_\theta$ will be modified after a brief transient, as the Lamb-Oseen solution is only an exact solution for the case of tubes. The ``inner'' side of the ring interacts with other parts of the vortex, the more so as the slenderness ratio, $\Lambda$, is larger. With time, the rings relax to a vorticity distribution which is different from the original, while still being dependent on $\Lambda$. How this relaxation occurs and the resulting vorticity profiles is discussed in more detail in the following section. 

We can see that even if the circulation is constant inside each ring (by definition), the vorticity is much more concentrated for smaller values of $\Lambda$, and the maximum velocity in each ring also increases with decreasing $\Lambda$. This will be later reflected in certain instabilities having a lower onset value of $Re_\Gamma$ for slender rings than for thicker rings. 

From this point onwards, space and time variables will be non-dimensionalized with the ring circulation $\Gamma$ and the initial ring radius $R_0$.
We also highlight that the choice of $Re_\Gamma$ and $\Lambda$ as control parameters is one usually done in numerics where the initial velocity profile is easily controlled. In experiments, a vortex ring is usually generated using a piston, of diameter $D_p$, which moves a stroke $L_p$ at velocity $U_p$. This leads to two dimensionless control parameters: a Reynolds number defined using piston scales $Re_p=U_pD_p/\nu$, and a stroke ratio $L_p/D_p$. While both Reynolds numbers are close to each other, with $Re_p$ usually being slightly higher, the relationship between the ring thickness and the two experimental control parameters is much more complicated \cite{gha98,keo20}. As such, numerical simulations have a much more direct control over the slenderness of a ring.

\subsection{Simulation geometry and white and coloured noises}

\begin{figure}
\centering
\begin{tikzpicture}[scale=1.4]
\usetikzlibrary{arrows.meta}
\draw (1.5,1.5) circle(0.5cm);
\draw (1,1.5) -- (1,3.5);
\draw (2,1.5) -- (2,3.5);
\draw (1.5,3.5) circle(0.5cm);
\draw[fill] (1.5,3.5) circle [radius=0.02];

\draw (5.5,1.5) circle(0.5cm);
\draw (5,1.5) -- (5,3.5);
\draw (6,1.5) -- (6,3.5);
\draw (5.5,3.5) circle(0.5cm);
\draw[fill] (5.5,3.5) circle [radius=0.02];

\draw [-latex,thick] (2,2.5) -- (2.4,2.5);
\draw [-latex,thick] (5,2.5) -- (4.6,2.5);
\node [above] at (2.2,2.5) {$V_a$};
\node [above] at (4.8,2.5) {$V_a$};

\draw [stealth-] (1.2,1.5) arc(180:360:0.3cm); 
\node at (1.5,1.5) {$\Gamma$};

\draw [stealth-] (5.8,1.5) arc(360:180:0.3cm); 
\node at (5.5,1.5) {$\Gamma$};

\draw [dashed] (0.5,2.5) -- (6.5,2.5);

\draw [latex-latex] (0.5,2.5) -- (0.5,3.5);
\node [left] at (0.5,3) {$R(\theta)$};
\draw [dotted] (1.5,3.5)--(0.5,3.5);

\draw [dotted] (1.5,3.5)--(1.5,4.3);
\draw [dotted] (5.5,3.5)--(5.5,4.3);
\draw [latex-latex] (5.5,4.3) -- (1.5,4.3);
\node [above] at (3.5,4.3) {$L_z$};

\end{tikzpicture}
\caption{Schematic of the head-on collision. The figure represents a cut of the two vortex rings by a plane containing the axis of symmetry, shown as a dashed line. The two vortices are moving towards each other with equal velocities.} 
\label{fig:schemaringcol}
\end{figure}
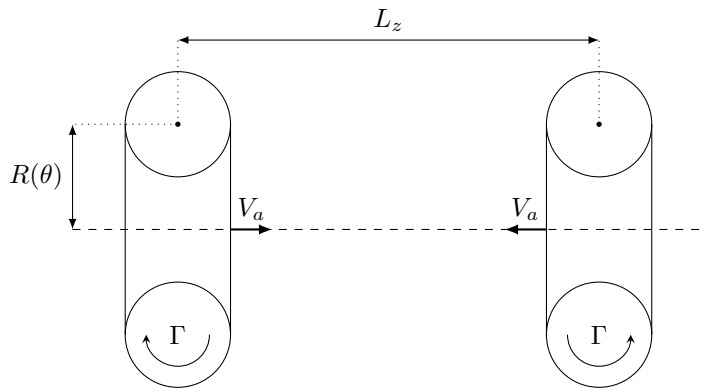

For head-on vortex ring collision, two aligned vortex rings of identical but opposite vorticity are placed a distance $L_z$ away, as shown in Fig.~\ref{fig:schemaringcol}. A rotational symmetry ($n_{sym}$) of order five was forced on the simulation to reduce computational costs. This value was chosen following Ref.~\cite{sha94}, which postulate that modes with an azimuthal wavenumber $m$ of five are most unstable modes for the instabilities associated to a ring interacting with itself. We further justify this choice later.

The initial position of the cores are perturbed using a low-band noise: 

\begin{equation}
 R(\theta,t=0)=1+\sum_{k=1}^{20} \epsilon_k \sin(k n_{sym} (\theta+\phi_k)),
 \label{eq:pert}
\end{equation}

\noindent 

\noindent with $\epsilon_k$ taken from a Gaussian distribution of zero mean and $\phi_k$ taken from a uniform distribution between $0$ and $2\pi$. 

Different random values were used for the two rings, and two types of perturbation were used, for reasons we explain in the next paragraph. For the first, which we refer to as  white noise, we take $3\times10^{-3}$ as the variance of the Gaussian distribution that gives $\epsilon_k$ for all the values of $k$ ($1 \le k \le 20$). This is done for all runs discussed in Sections~\ref{sec:IIIa}, \ref{sec:IIIc}, \ref{sec:IVb} and \ref{sec:V}, as well as the first half of \ref{sec:IVa}. For the second, which we denote colored noise, the amplitudes of the two largest wavelengths in the system ($\epsilon_1$ and $\epsilon_2$) are amplified ten times after being obtained from a random distribution. This is done for the runs discussed in Sections~\ref{sec:IIIb}, and in the second part of \ref{sec:IVa}. The noise is calculated once per value of $\Lambda$ and per geometry, but is kept the same as $Re_\Gamma$ is increased. We note that with the form of the perturbation used, Eq.~\ref{eq:pert}, the azimuthal wavenumbers in the system $k$ are a subset of the total azimuthal wavenumbers $m$, and are related by $m=k n_{sym}$. So the smallest wavenumber ($k=1$) we can seed if $n_{sym}=5$ is $m=5$, the second smallest is $m=10$, and so on.

The vortices were started at a distance of $L_z=2.5$. This value is sufficient to relax the ring from the initial conditions, as most of the dynamics we are analyzing in this paper happens away from the axis (cf.~Section \ref{sec:relax} for a discussion on relaxation). Larger values of $L_z$ would also provide stronger seeding for the Crow-type instabilities due to the ring's self-instability acting for a longer time \cite{arc10}. However, increasing $L_z$ comes at the cost of computational resources, so as a proxy for increased seeding of the long wavelength cases due to large values of $L_z$ we simulated the colored noise cases. Furthermore, a discussion of how long wavelength modes grow with increasing $L_z$ is provided in the Section \ref{sec:lzstudy}.

The range of Reynolds numbers based on the circulation $Re_\Gamma$ considered was between $Re_\Gamma=1000$ and $Re_\Gamma=3500$. Three values for ring slenderness $\Lambda$ were considered: $0.1$, $0.2$ and $0.35$.  The cylindrical computational domain was bounded by stress-free lateral walls at a distance $R_{ext}$, sufficiently far from the rings to not affect them significantly. This was chosen as one torus radius below and above the rings, and between five to six torus radius from the ring axis in the radial direction ($R_{ext}$), depending on $Re_\Gamma$ and $\Lambda$. Resolution adequacy was checked by monitoring the viscous dissipation and the energy balance. For a full list of the numerical parameters and resolutions used see Table \ref{tbl:numdetring}. 

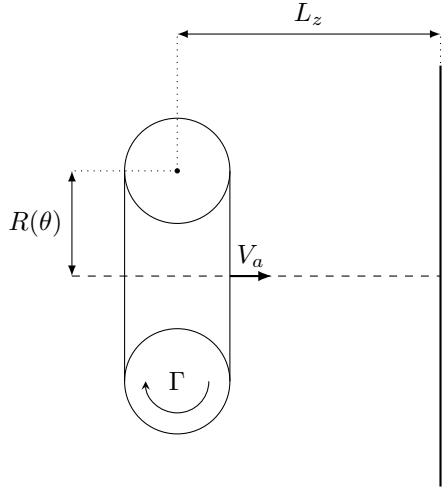
\begin{figure}
\centering
\begin{tikzpicture}[scale=1.4]
\usetikzlibrary{arrows.meta}
\draw (1.5,1.5) circle(0.5cm);
\draw (1,1.5) -- (1,3.5);
\draw (2,1.5) -- (2,3.5);
\draw (1.5,3.5) circle(0.5cm);
\draw[fill] (1.5,3.5) circle [radius=0.02];

\draw [stealth-] (1.2,1.5) arc(180:360:0.3cm); 
\node at (1.5,1.5) {$\Gamma$};

\draw [dashed] (0.5,2.5) -- (4,2.5);

\draw [latex-latex] (0.5,2.5) -- (0.5,3.5);
\node [left] at (0.5,3) {$R(\theta)$};
\draw [dotted] (1.5,3.5)--(0.5,3.5);

\draw [dotted] (1.5,3.5)--(1.5,4.8);
\draw [dotted] (4,3.5)--(4,4.8);
\draw [latex-latex] (4,4.8) -- (1.5,4.8);
\node [above] at (2.75,4.8) {$L_z$};
\draw [thick] (4,0.5) -- (4,4.5);

\draw [-latex,thick] (2,2.5) -- (2.4,2.5);
\node [above] at (2.2,2.5) {$V_a$};

\end{tikzpicture}
\caption{Schematic of the head-on wall impact. The figure represents a cut of the ring by a plane containing the axis of symmetry, shown as a dashed line. The vortex is moving towards the wall, shown as a thick line, at a distance $L_z$ from the vortex center.} 
\label{fig:schemaringwall}
\end{figure}

For the vortex ring-wall impact, a ring is released a distance $L_z=4$ away from the wall, undergoing a head-on impact as shown in Figure \ref{fig:schemaringwall}. The initial position of the ring is perturbed in a similar manner to the rings in the head-on collision, and similarly only a fraction of the azimuthal domain is considered. The rest of the cylindrical computational domain is bounded by stress-free walls at a sufficient distance from the rings to not affect them significantly. For this case, it is one torus radius below and above the rings, and four torus radii for no-slip walls and between eight to ten torus radii for free-slip walls from the ring axis in the radial direction, depending on $Re_\Gamma$ and $\Lambda$. Resolution adequacy was again checked by monitoring the viscous dissipation and the energy balance. These are provided in Table \ref{tbl:numdetring}.

\begin{table}
    \centering
\begin{tabular}{|c|ccc|ccc|ccc|}
\hline
$Re_\Gamma$ & $\Lambda$ & $R_{ext}$ & $N_\theta\times N_r\times N_z$ & $\Lambda$ & $R_{ext}$ & $N_\theta\times N_r\times N_z$  & $\Lambda$ & $R_{ext}$ & $N_\theta\times N_r\times N_z$ \\
\hline
 \multicolumn{10}{|c|}{Head on collision with white noise} \\
1000 & 0.1 & 5 & $384\times512\times264$ & 0.2 & 5 & $384\times512\times264$ & 0.35 & 6 & $384\times512\times264$ \\ 
2000 & 0.1 & 5 & $384\times512\times264$ & 0.2 & 5 & $384\times512\times264$ & 0.35 & 6 & $384\times512\times264$ \\ 
3500 & 0.1 & 5 & $384\times512\times264$ & 0.2 & 5 & $384\times512\times264$ & 0.35 & 6 & $384\times512\times264$ \\
\hline
\multicolumn{10}{|c|}{Head on collision with coloured noise} \\
1000 & 0.1 & 5 & $384\times512\times264$ & 0.2 & 5 & $384\times512\times264$ & 0.35 & 5 & $384\times512\times264$ \\ 
2000 & 0.1 & 5 & $384\times512\times264$ & 0.2 & 5 & $384\times512\times264$ & 0.35 & 5 & $384\times512\times264$ \\ 
3500 & 0.1 & 5 & $384\times512\times264$ & 0.2 & 5 & $384\times512\times264$ & 0.35 & 5 & $384\times512\times264$ \\
\hline 
 \multicolumn{10}{|c|}{Stress-free wall impact} \\
1000 & 0.1 & 8 & $192\times256\times256$ & 0.2 & 10 & $192\times192\times256$ & 0.35 & 10 & $192\times192\times256$ \\ 
2000 & 0.1 & 8 & $192\times256\times256$ & 0.2 & 10 & $192\times192\times256$ & 0.35 & 10 & $192\times192\times256$
\\ 
3500 & 0.1 & 8 & $256\times512\times256$ & 0.2 & 10 & $256\times256\times256$ & 0.35 & 10 & $192\times192\times256$\\ 
\hline 
 \multicolumn{10}{|c|}{No-slip wall impact} \\
1000 & 0.1 & 4 & $128\times192\times256$ & 0.2 & 4 & $128\times192\times256$ & 0.35 & 4 & $128\times192\times256$\\ 
2000 & 0.1 & 4 & $128\times192\times256$ & 0.2 & 4 & $128\times192\times256$ & 0.35 & 4 & $128\times192\times256$\\ 
3500 & 0.1 & 4 & $128\times192\times256$ & 0.2 & 4 & $128\times192\times256$ & 0.35 & 4 & $128\times192\times256$\\ 
5000 & ~ & ~ & ~ & ~ & ~ & ~ & 0.35 & 4 & $192\times192\times256$\\ 
\hline 
\end{tabular}
    \caption{Summary of resolutions used for all simulations discussed in Secs.~\ref{sec:III}, \ref{sec:IV} and \ref{sec:V}.}
    \label{tbl:numdetring}
\end{table}

\subsection{Single Vortex relaxation}
\label{sec:relax}

To analyze the initial relaxation of the vorticity profile, we consider a series of single-ring simulations where we impose complete axisymmetry (i.e.~$n_{sym}\to\infty$), remove the azimuthal noise, and vary $Re_\Gamma$ and $\Lambda$. The ring is allowed to self-advect and relax in cylinder for which the collision plane is very far away from the initial position ($L_z=40$). This makes the geometry for our purposes effectively infinite ($L_z>>1$) as the ring never reaches the collision plane. A resolution of $N_r\times N_z=192\times256$ is used for all cases. We note that as the ring travels, more and more vorticity will diffuse from the core due to viscosity, and one of the effects this has is to increase the effect thickness of the ring, and to slow down the axial translation due to self-induced velocity. So to distinguish between relaxation and viscous diffusion, it is necessary to compare not only various values of $\Lambda$ but also various values of $Re_\Gamma$. 

\begin{figure}
 \centering
 \includegraphics[width=0.32\textwidth]{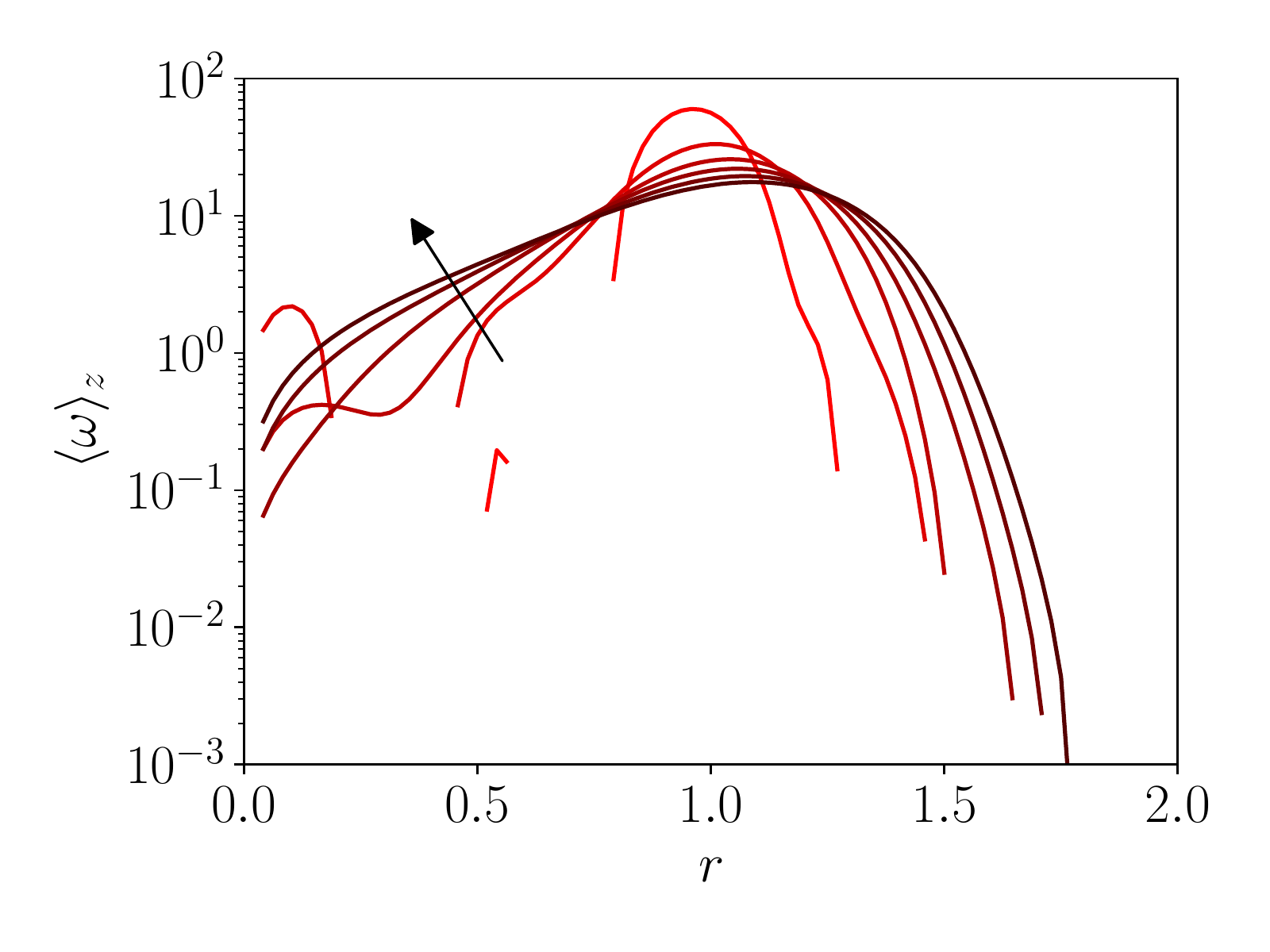}
 \includegraphics[width=0.32\textwidth]{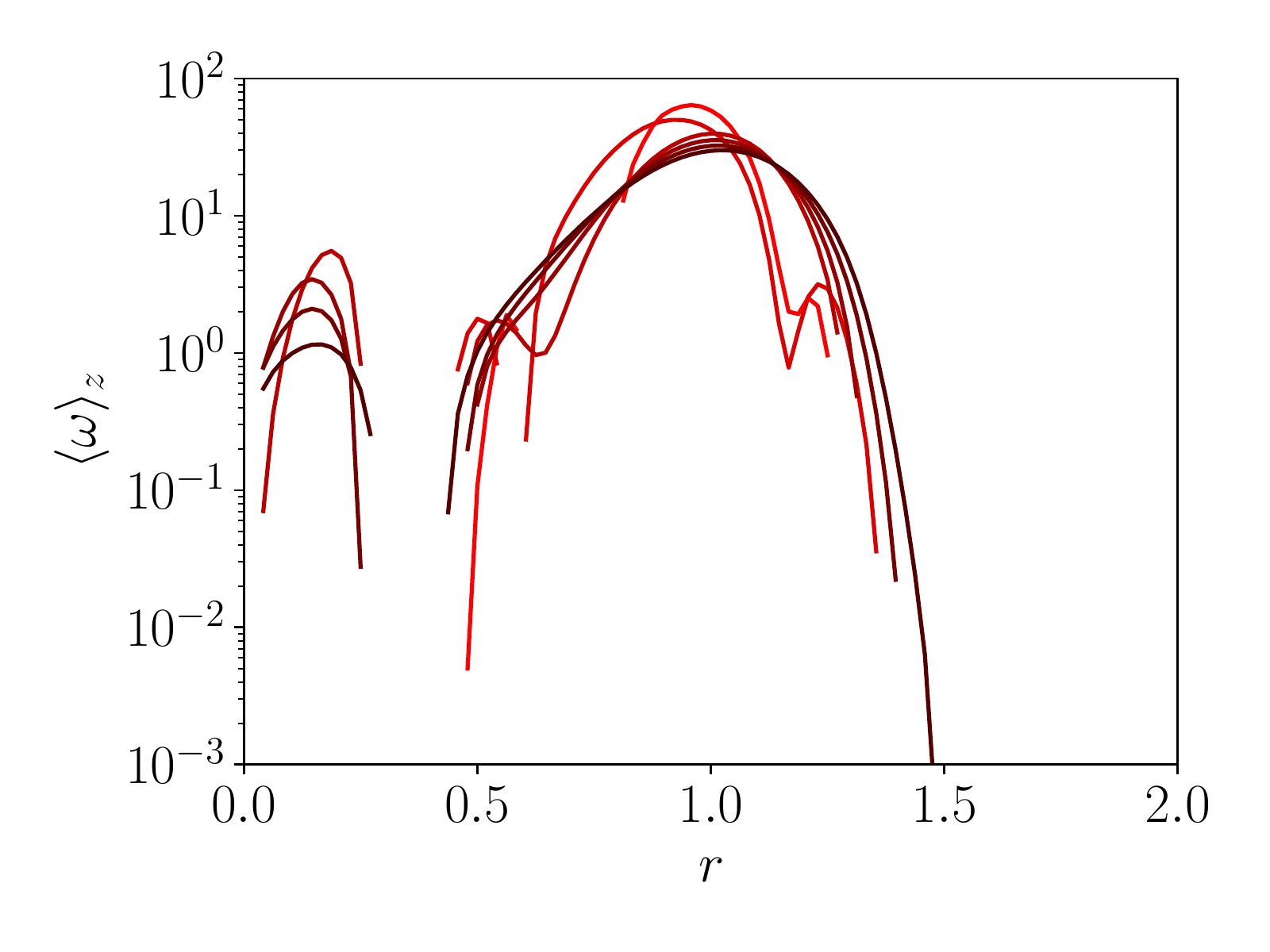}
 \includegraphics[width=0.32\textwidth]{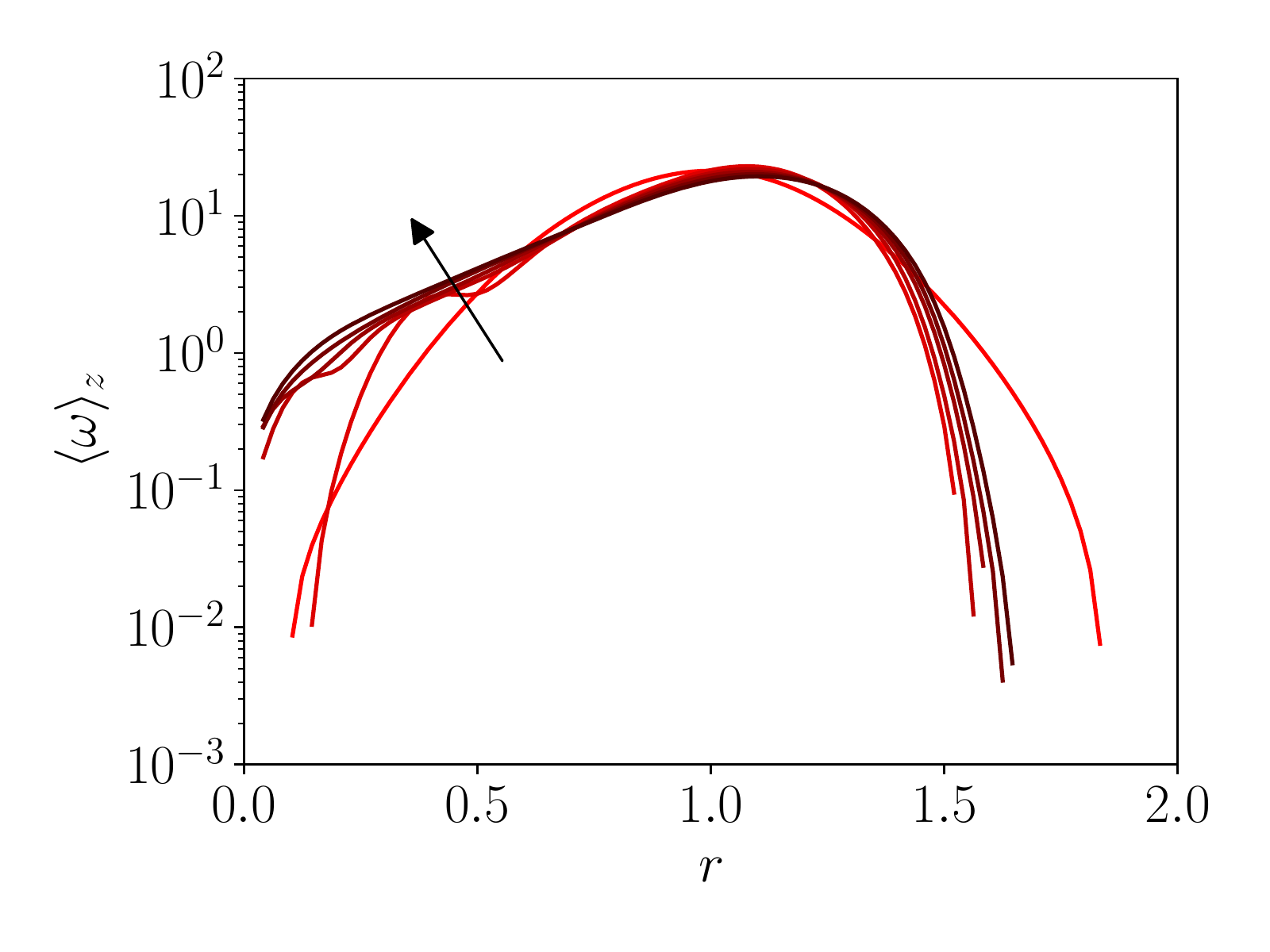}
 \caption{Temporal evolution of the axially ($z$) integrated vorticity $\langle \omega \rangle_z$. Lines are plotted every $10$ time units from $t=1$ to $51$, increasing in darkness as time progresses. Left: $\Lambda=0.1$, $Re_\Gamma=1000$. Center: $\Lambda=0.1$, $Re_\Gamma=3500$. Right: $\Lambda=0.35$, $Re_\Gamma=3500$. Arrows also indicate time progression.}
 \label{fi:vortrelax}
\end{figure}

In Figure \ref{fi:vortrelax}, we show the temporal evolution of the vorticity modulus integrated along the cylinder axis ($z$-coordinate) for these simulations. Just after initialization, some remnants of vorticity appear near the axis. This is especially prominent for slender rings, and is a consequence of initializing vorticity and velocity with Eq.~\ref{eq:lamboseen} which are inexact solutions for rings. As the ring evolves, the vorticity distribution flattens from a Gaussian (a parabola in semi-log scales), to a broader distribution, extending towards the axis of symmetry. The ring also widens out, as vorticity is diffused from the core outwards. This is especially visible in the left panel at $Re_\Gamma=1000$ and $\Lambda=0.1$, where the final distribution of vorticity resembles that at the latest time of the right panel ($Re_\Gamma=3500$ and $\Lambda=0.35$) much more than they resemble the middle panel which has the same slenderness ratio $\Lambda=0.1$ but a larger $Re_\Gamma$ equal to $Re_\Gamma=3500$. We also note that the vorticity close to the axis in the middle panel are the remnants of vorticity that appear in the region close to where the vortex ring is placed initially, and are left behind as the vortex advects itself. By taking $L_z\geq 2.5$, we ensure that this trail is sufficiently far away from the collision plane that it does not interfere with the dynamics. 

In summary, Fig.~\ref{fi:vortrelax} shows that the Lamb-Oseen solution is far from the relaxed solution, especially for thick vortices and in the direction towards the direction of the axis of symmetry. As the rings expand rapidly when interacting, we do not expect the differences in the inner side to be very relevant. We note that whatever our choice of $L_z$ results from a compromise: the rings thicken as $L_z$ increases in a manner that is dependent on $Re_\Gamma$. Therefore, larger values of $L_z$ would not only come at increased computational costs, but also complicate comparisons across $Re_\Gamma$, as rings which are started up with the same value of $\Lambda$ would arrive in the collision plane with effectively different thicknesses. 

\subsection{Single Vortex self-instability}
\label{sec:lzstudy}

We also analyzed the self-elliptical instability of a single ring, i.e.~that originating due to the interaction of various points of the ring with other parts of the ring, as a function of $\Lambda$. These simulations were performed to check how the self-instability would enter into play in our simulations if we were to increase $L_z$ beyond $L_z=2.5$. For this, we ran the same cases as before: a ring self-advecting and relaxing in a very long cylinder ($L_z=40>>1$), but removed any rotational symmetries and simulated the full azimuthal extent of the domain ($n_{sym}=1$) which also means we can capture all azimuthal modes ($m=k$). To ensure that we do not miss out any mode, we use a different seeding procedure for these cases and add white noise of equal magnitude (i.e.~not randomly sampled) to the first twenty azimuthal modes ($\epsilon_k =7\times10^{-4}$). A resolution of $N_\theta\times N_r\times N_z=192\times192\times256$ is used for all cases. 

\begin{figure}
 \centering
 \includegraphics[width=0.32\textwidth]{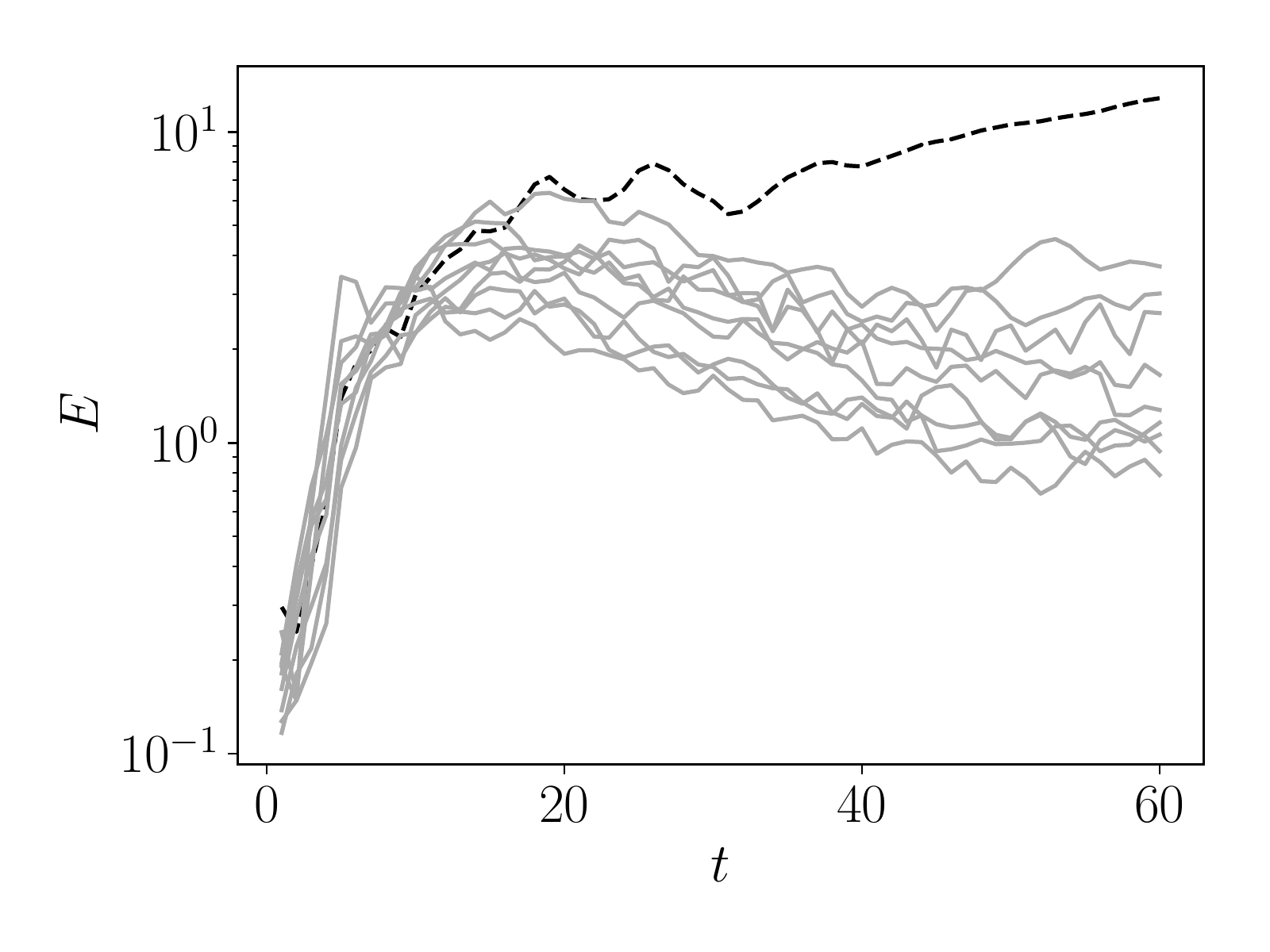}
 \includegraphics[width=0.32\textwidth]{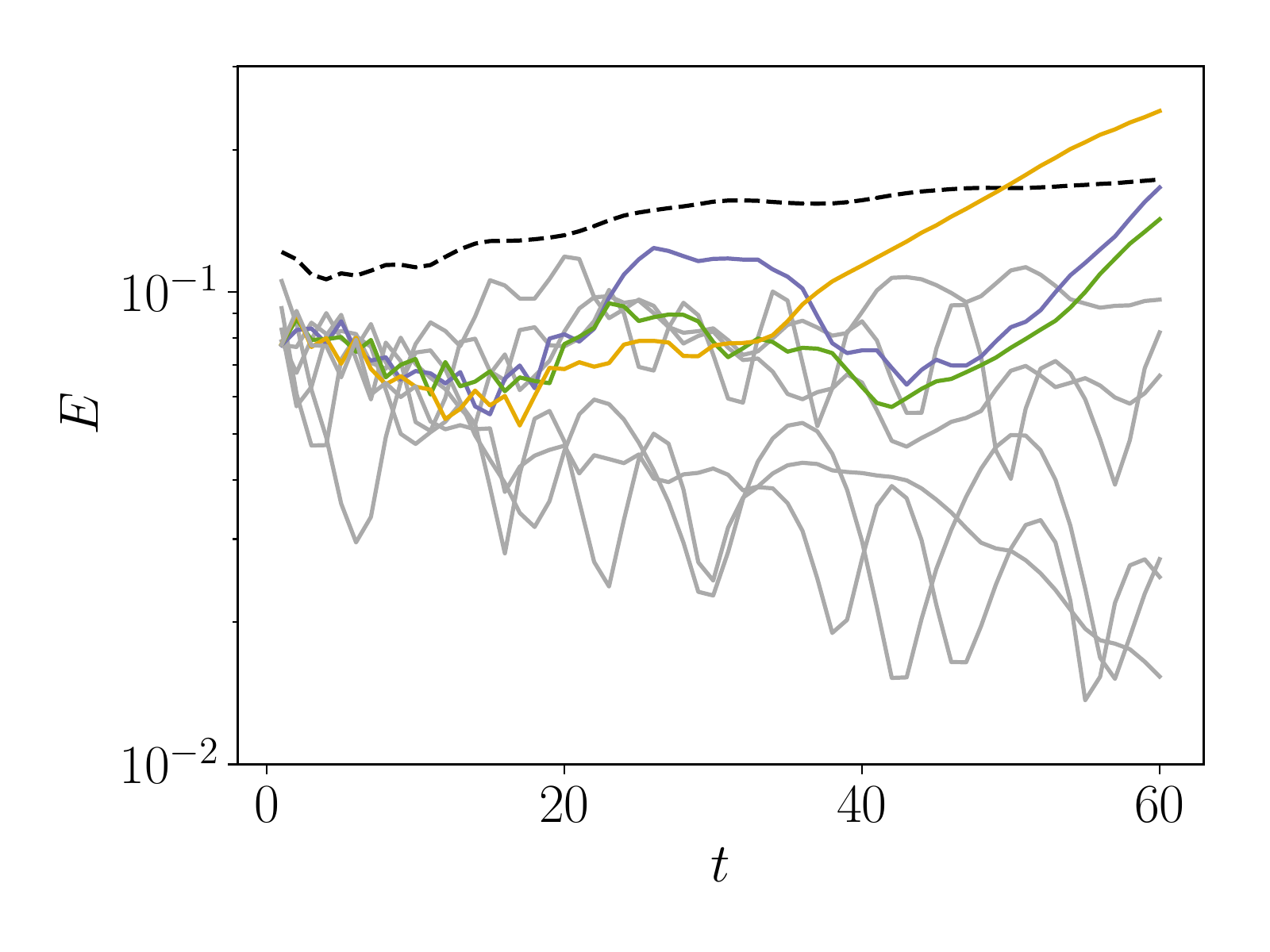}
  \includegraphics[width=0.32\textwidth]{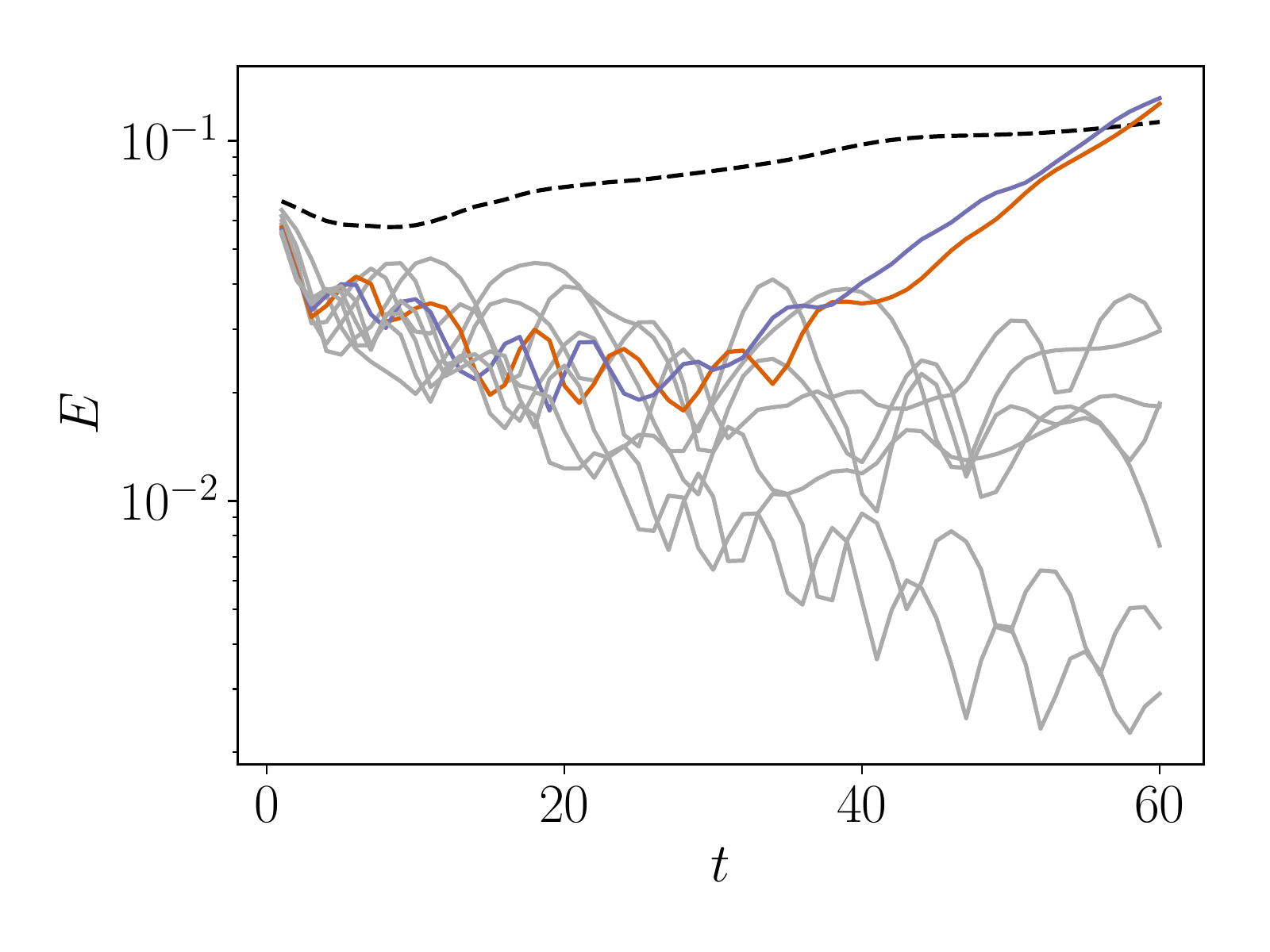}
   \caption{Temporal evolution of the energy in the first 10 azimuthal modes ($k=m=1$ to $k=m=10$), for the full azimuthal extent, $Re_\Gamma=3500$ and $\Lambda=0.1$ (left), $\Lambda=0.2$ (center) and $\Lambda=0.35$ (right). The translation mode $m=1$ is shown in a black dashed line. All other modes are shown in grey except when highlighted due to their growth: $k=m=5$ (brown), $k=m=6$ (purple), $k=m=7$ (green), $k=m=8$ (dark yellow).}
 \label{fi:vortexms}
\end{figure}

We then measure the energy contained in the azimuthal modes which evolves over time for the first 60 time units. This can be used to estimate how larger values of $L_z$ could bias the initial noise. The results of this are shown in Figure \ref{fi:vortexms}. The first thing to note is the prominence of the $m=1$ mode, which we have marked with a dashed line. This mode does not decay or grow much, and it can be understood as representing a displacement of the ring from the axis as a whole. As such, it does not decay much due to viscosity. In the simulations below, we take $n_{sym}>1$ and eliminate this mode. This means we do consider the effects of possible ring misalignment in this study.

The other result we obtain is that it takes a long time for certain modes to begin to grow, if at all. For $\Lambda=0.1$, all modes seem to behave similarly, with a fast growth and a slow decay. For $\Lambda=0.2$, at later times the energy in a bandwidth of $m$ modes $m\in[6,9]$ shows some exponential growth, with the strongest growth for $m=9$. However, the growth is very slow (or the growth rate is very small). A similar amplification is observed
for the $m=5$ and $m=6$ modes for $\Lambda=0.35$, showing some growth at later times. The self-instability modes roughly correspond to smaller values of $m$ with increasing $\Lambda$, something that makes sense from an analysis of the TWMS mode: the characteristic wavelength is the core radii. Thicker rings 
have a larger wavelength, and thus a smaller value for the most unstable $m$. 

This analysis convinced us that
the energy of the modes fluctuate substantially before much growth is seen, and the observed amplification, in all cases, remains small. It is not significant before $t=30$, which roughly corresponds to 8-9 ring radii from the launch point for $\Lambda=0.2$ and 7-8 ring radii for $\Lambda=0.35$, and would take another 40-50 more time units to increase the energy in these modes by an order of magnitude. This means that the coloured noise simulations are effectively a proxy which captures a noise growth comparable to $L_z \approx 50$, without the problems associated to ring thickness growth mentioned in the section above. We thus feel confident to proceed with our study, and to use the coloured noise model as a proxy for long $L_z$. 

\section{Instability and disintegration for $\Lambda=0.1$}
\label{sec:III}

\subsection{Head-on collision with white noise}
\label{sec:IIIa}

\begin{figure}
 \centering
 \includegraphics[trim={3cm 0 3cm 0},clip,width=0.32\textwidth]{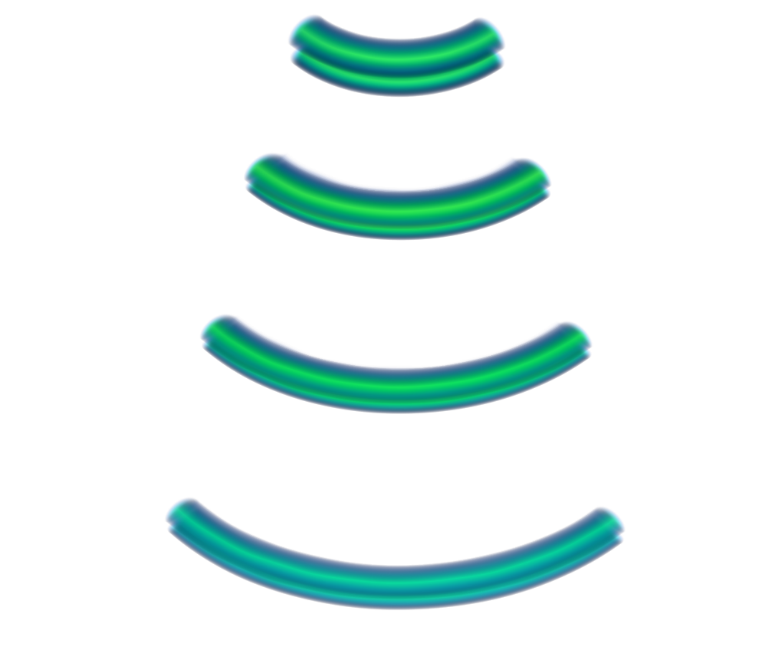}
 \includegraphics[trim={3cm 0 3cm 0},clip,width=0.32\textwidth]{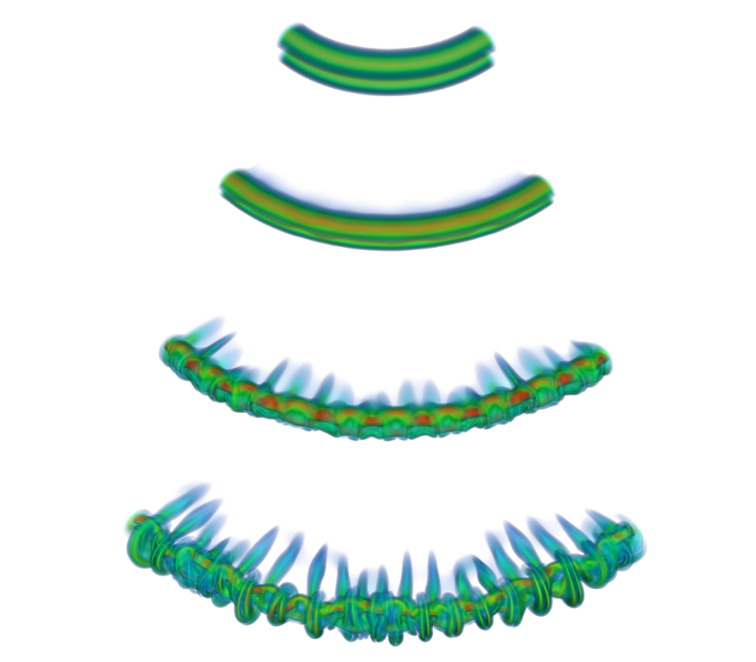}
 \includegraphics[trim={3cm 0 3cm 0},clip,width=0.32\textwidth]{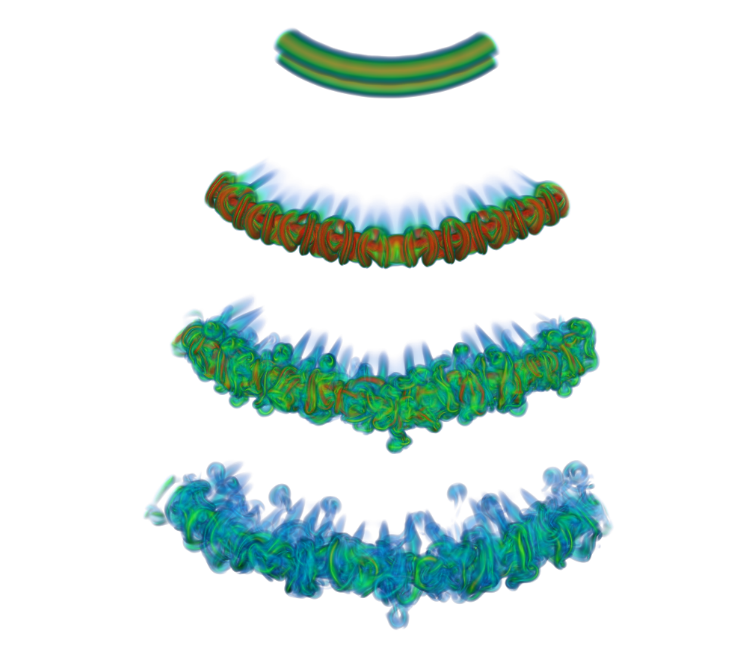}
 \caption{Vorticity volume visualization at several instances of time for $\Lambda=0.1$ and white noise perturbations. Reynolds number increases from left to right ($Re_\Gamma=1000$, $Re_\Gamma=2000$ and $Re_\Gamma=3500$), while time increases from top to bottom ($t=6$, $8$, $10$ and $12$). Red denotes regions of particularly high vorticity, while blue denotes regions of low vorticity. The colors are the same for all panels. }
 \label{fi:vortcp1}
\end{figure}

We first focus on the cases of a slender ring with $\Lambda=0.1$ and white noise ($\langle \epsilon_k^2 \rangle = 3\times10^{-3}$), and study the effect of the Reynolds number, $Re_\Gamma$. Figure \ref{fi:vortcp1} shows the vorticity magnitude for the three Reynolds numbers studied. For $Re_\Gamma=1000$, left column, the two vortices stretch each other, expanding out while remaining relatively axisymmetric. Due to conservation of circulation, we can observe that the total vorcitity magnitude increases as the vortices are stretched and their core becomes smaller. There is no disintegration or significant instabilities arising and the rings will eventually decay due to viscosity. 

As the Reynolds number is increased to $Re_\Gamma=2000$, middle column, the dynamics changes. We can observe the onset of an azimuthal instability marked by the deformation of the vortex core. This deformation arises from the elliptical instability, whose signature is shown in the small instability wavelength as well as its general anti-symmetric character during the initial, or  ``cooperative'' phase of the instability \cite{lew98}. As the rings twist, small vortex filaments which are perpendicular to the primary rings are formed. The rings slow down as they lose circulation to the newly forming secondary perpendicular filaments. However, the rings do not completely disintegrate into a turbulent cloud. 

Only by further increasing the Reynolds number to $Re_\Gamma=3500$ does the interaction become sufficiently strong to result in a turbulent cloud, see the right column. This happens through a similar process that starts with the formation of secondary perpendicular filaments, but in this case they contain more circulation which appears as a stronger red color in the visualization at $t=8$. The ring eventually stops and disintegrates into fine turbulent structures. This is already in the asymptotic regime for vortex ring collision, where fine scales are generated through an iterative process of cascading instabilities seen in Ref.~\cite{keo20}. We also note that the vorticity visualizations for $Re_\Gamma=3500$ show strong qualitative similarities to the photographs of the experiment in Ref.~\cite{lew98} of two anti-parallel vortex tubes undergoing the elliptical instability, in both the early, or ``cooperative'' stage, and the late stage. However, there are some differences, which we will elaborate on in the next section. 

\begin{figure}
 \centering
  \includegraphics[width=0.49\textwidth]{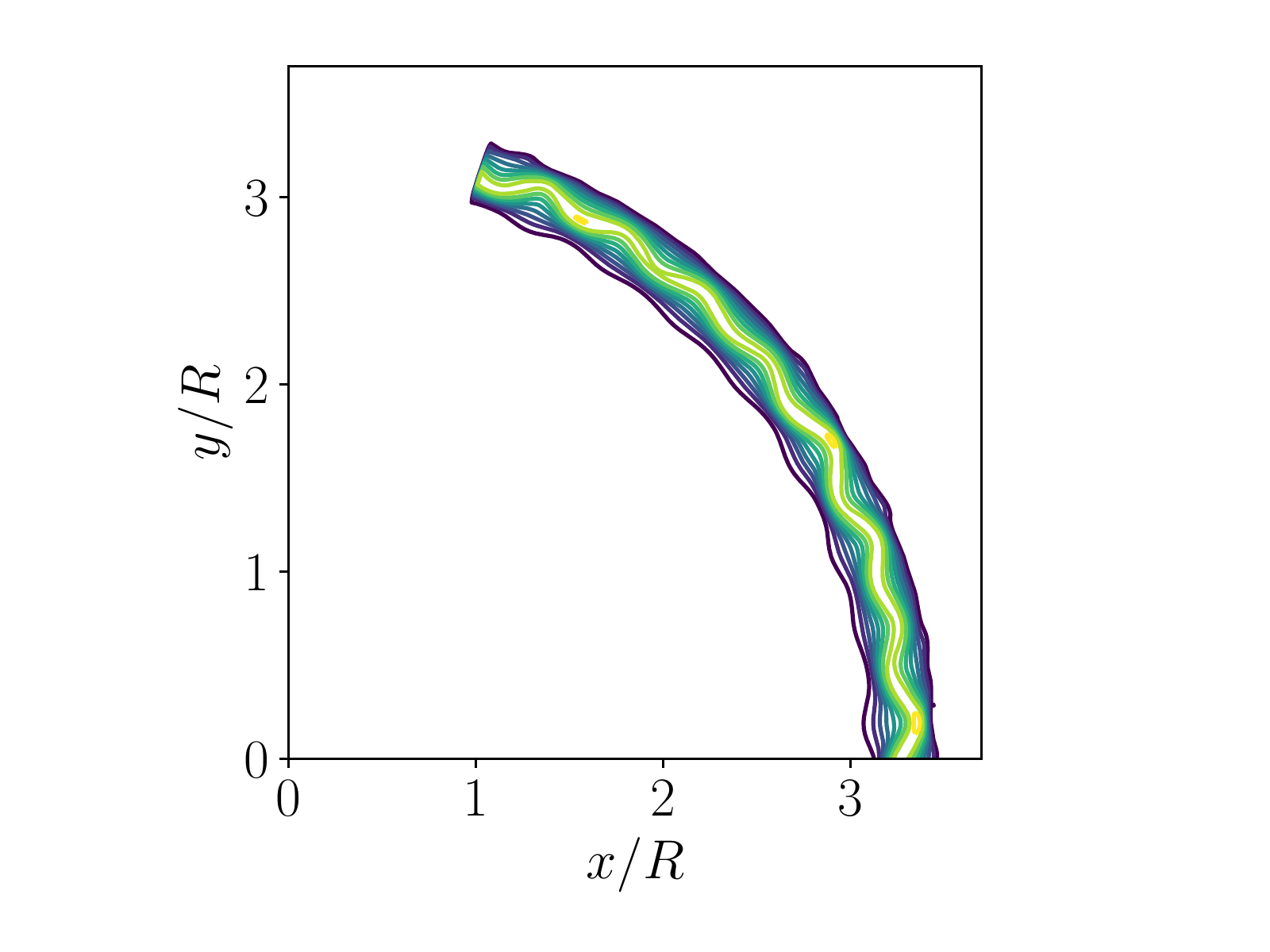}
  \includegraphics[width=0.49\textwidth]{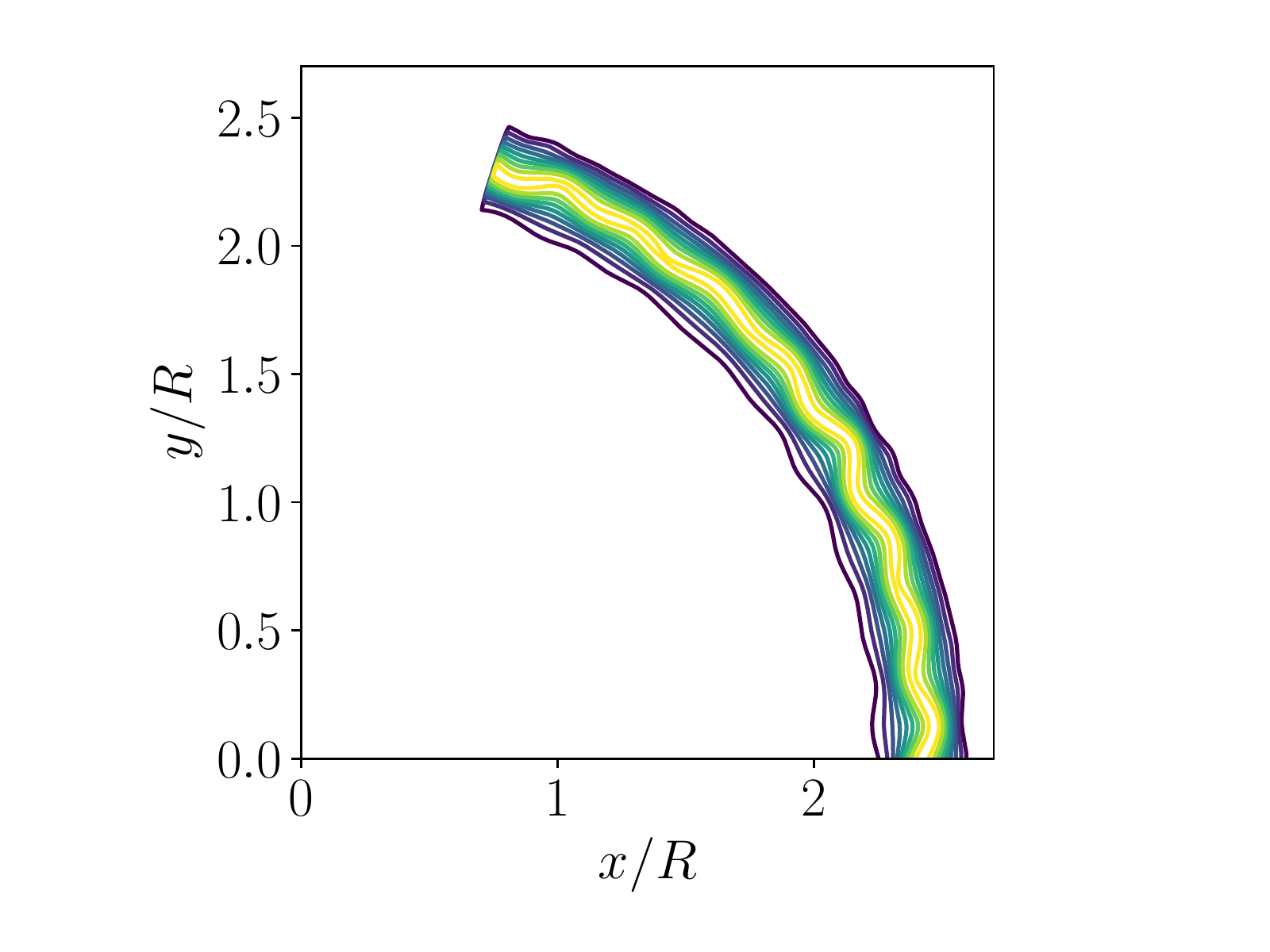}
 \caption{Contour plots of vorticity modulus at constant $z$ for the top vortex at $\Lambda=0.1$ and $Re_\Gamma=2000$ at $t=9$ (left) and $Re_\Gamma=3500$ at $t=7$ (right).}
 \label{fig:contwwcrp1}
\end{figure}

To confirm that the asymptotic regime is reached once the elliptical instability is fully developed, we first isolate the instability, confirming that it is indeed the elliptical instability. Evidence for this can be seen by considering a series of iso-contours of vorticity magnitude in the top vortex of the simulation, see Figure \ref{fig:contwwcrp1}. We note that the vortex core and the periphery appear to deform in opposite directions. This coincides with the deformation pattern predicted for an ``invariant streamtube'' in Ref.~\cite{lew98}, and suggests a possible mechanism for the formation of the secondary perpendicular vortices, from the peripheral regions which deform away from the core. Further evidence for the elliptical instability is provided by determining, in the following paragraphs, the growth rates of the instability.

\begin{figure}
 \centering
 \includegraphics[width=0.49\textwidth]{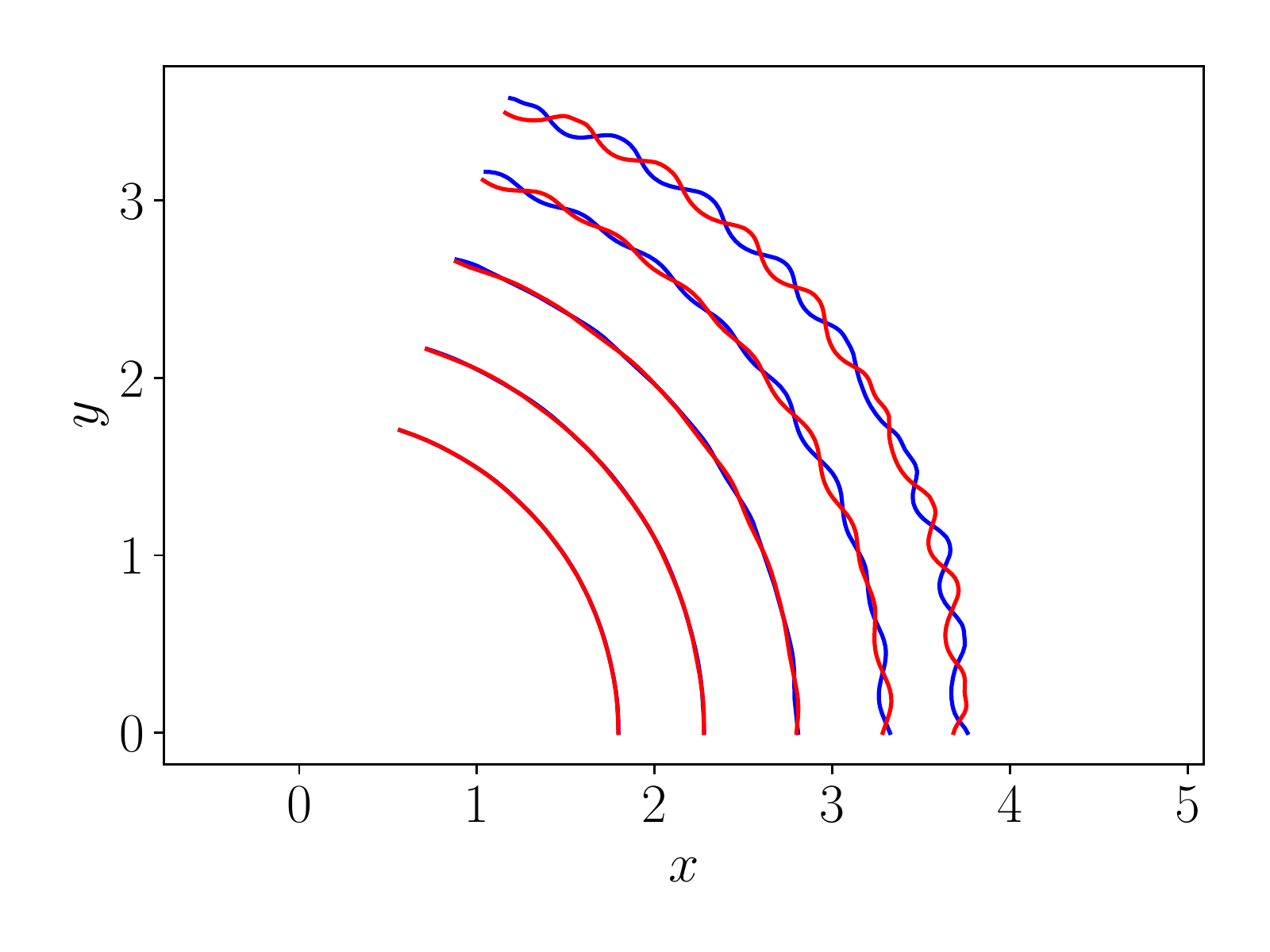}
 \includegraphics[width=0.49\textwidth]{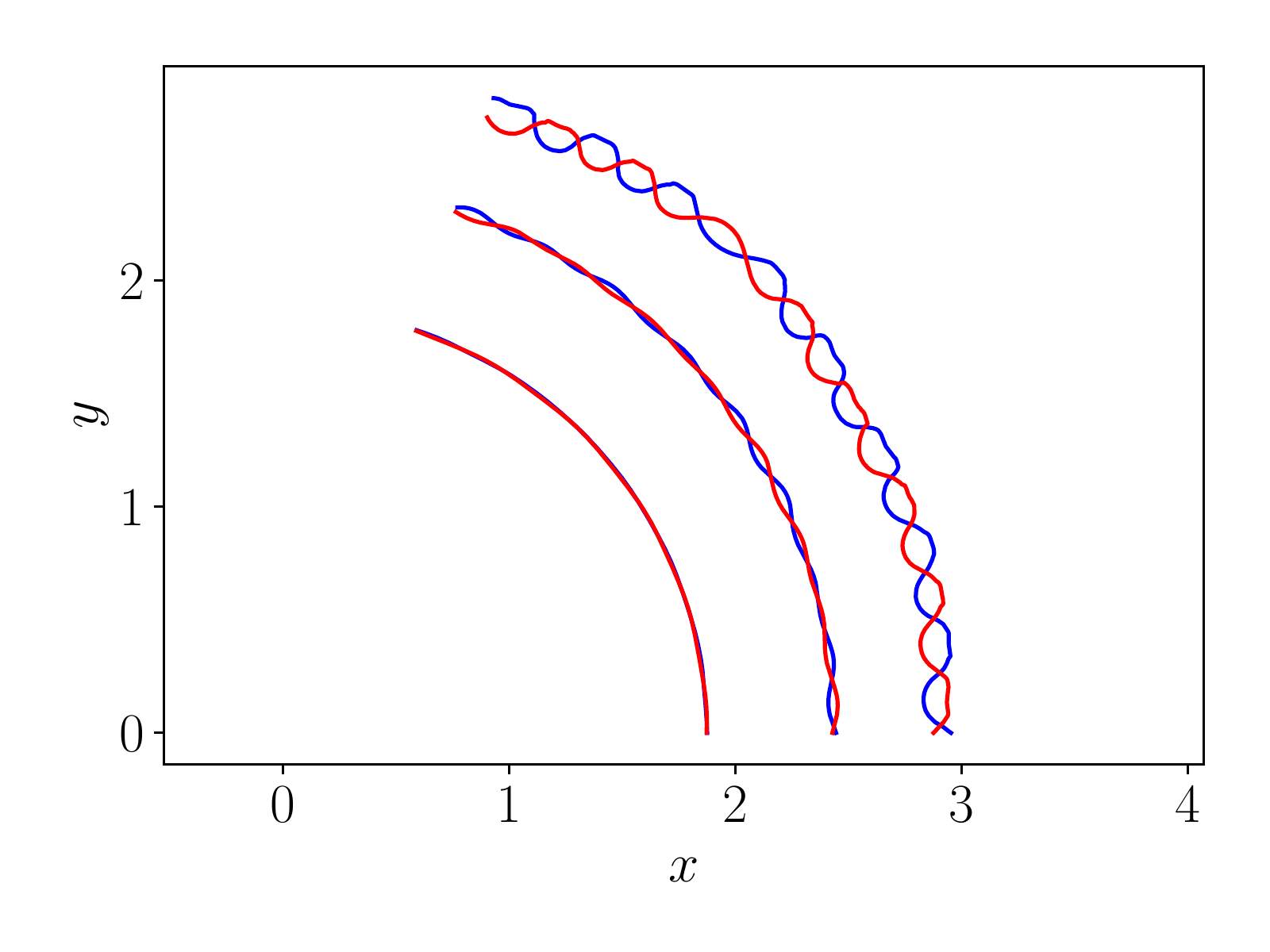}
  \includegraphics[trim={0cm 3cm 0cm 5cm},clip,width=0.49\textwidth]{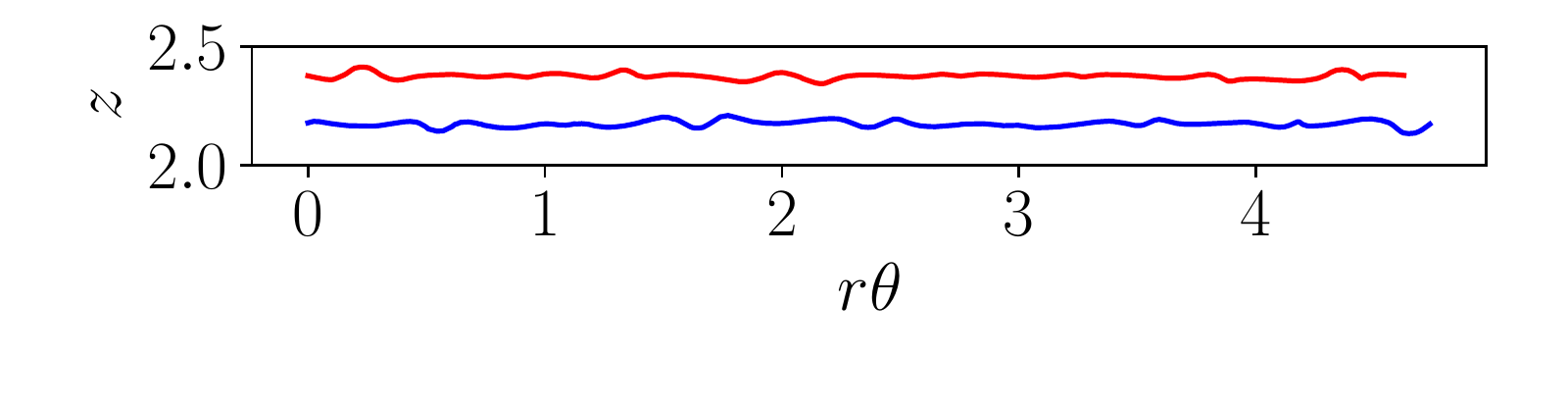}
 \includegraphics[trim={0cm 3cm 0cm 5cm},clip,width=0.49\textwidth]{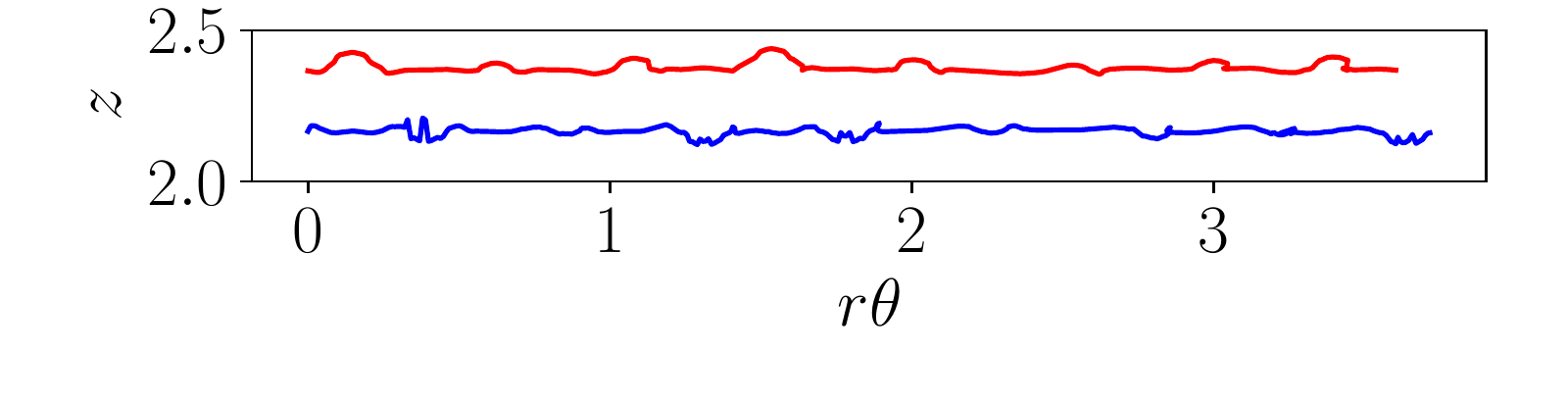}
 \caption{Extracted vortex cores at several instances of time for $\Lambda=0.1$. Left top panel: cores at $t=6$, $7$, $8$, $9$ and $10$ for $Re_\Gamma=2000$. Right top panel: cores at $t=6$, $7$, and $7.8$ for $Re_\Gamma=3500$. Bottom left panel: side view of the cores at $t=10$ for $Re_\Gamma=2000$. Bottom right panel: side view of the cores at $t=7.8$ for $Re_\Gamma=3500$.  }
 \label{fi:fils1}
\end{figure}

Having established the character of the instability, we continue by isolating the vortex cores by following Ref.~\cite{keo20} and using the pressure minimum to define the position of the core. We show the position of the core as a function of time for $Re_\Gamma=2000$ (left column) and $Re_\Gamma=3500$ (right column) in Figure \ref{fi:fils1}. As the two vortices stretch each other in the radial direction, they show the signature of a growing anti-symmetric instability. The azimuthal wavenumber $m=40$ ($k=8$) results in an instability wavelength 
of $2\pi R(t)/m \approx0.2-0.4$, depending on the instantaneous value of $R(t)$. In any case, these wavelengths 
are very close to our core vortex radius. This further confirms that is indeed the same elliptical instability arising for both cases. We also note that the rings deform predominantly in the collision plane, and not perpendicular to it, i.e.~in the $z$ direction. This is unlike the deformation seen in tubes for the elliptical instability, where the cores deform in all planes \cite{lew98,keo20}.

\begin{figure}
 \centering
  \includegraphics[width=0.32\textwidth]{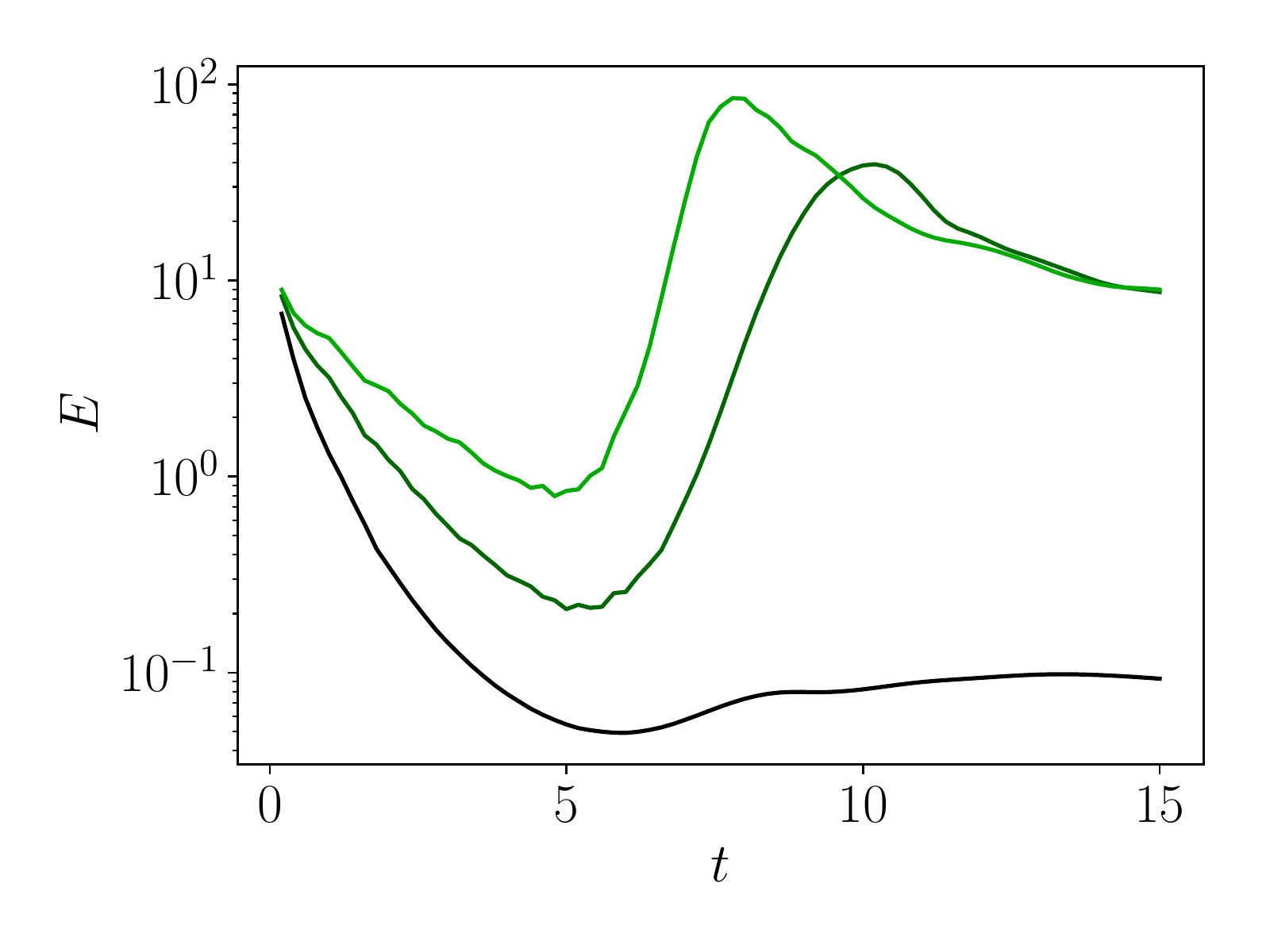} 
 \includegraphics[width=0.32\textwidth]{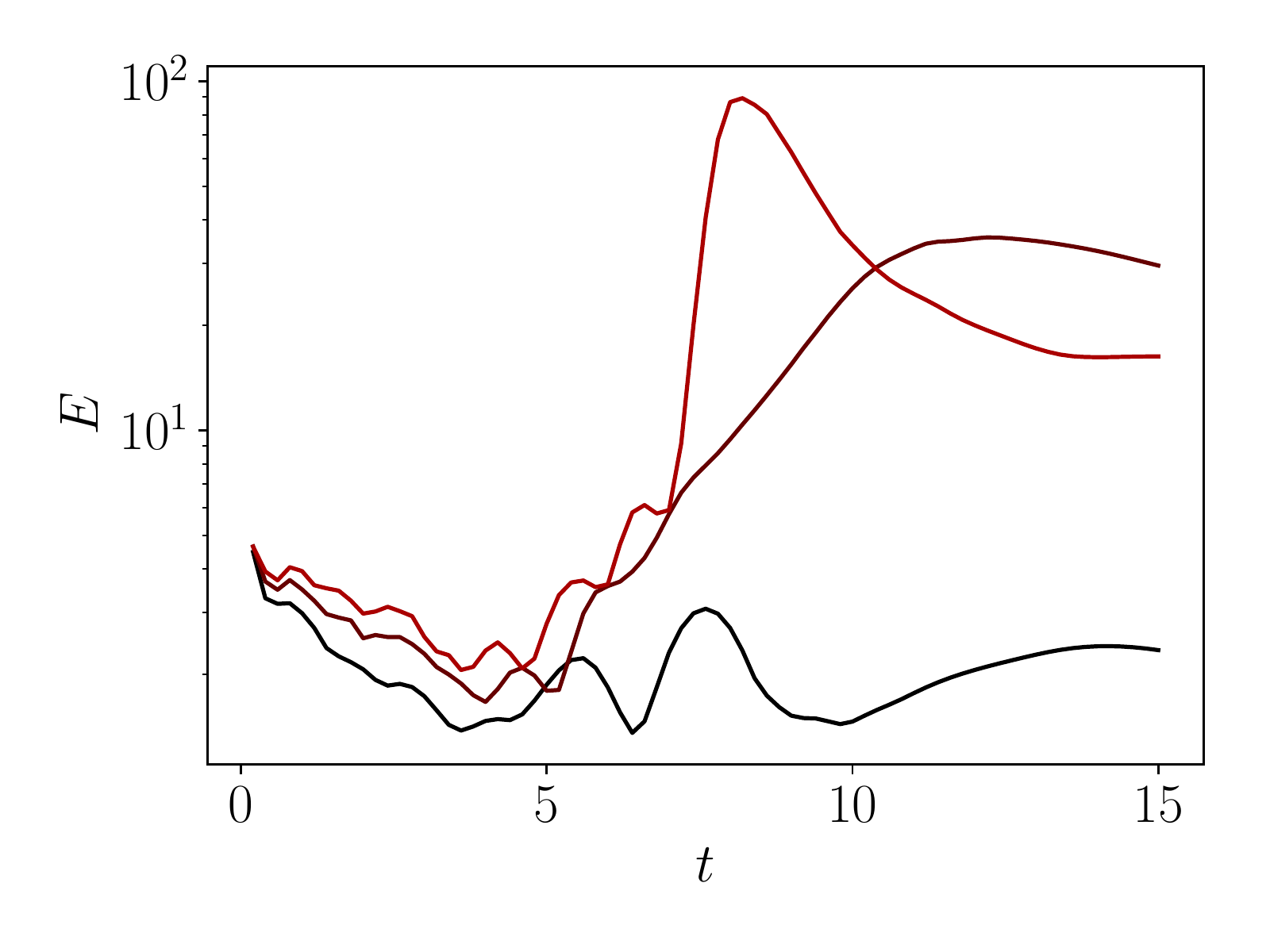} 
  \includegraphics[width=0.32\textwidth]{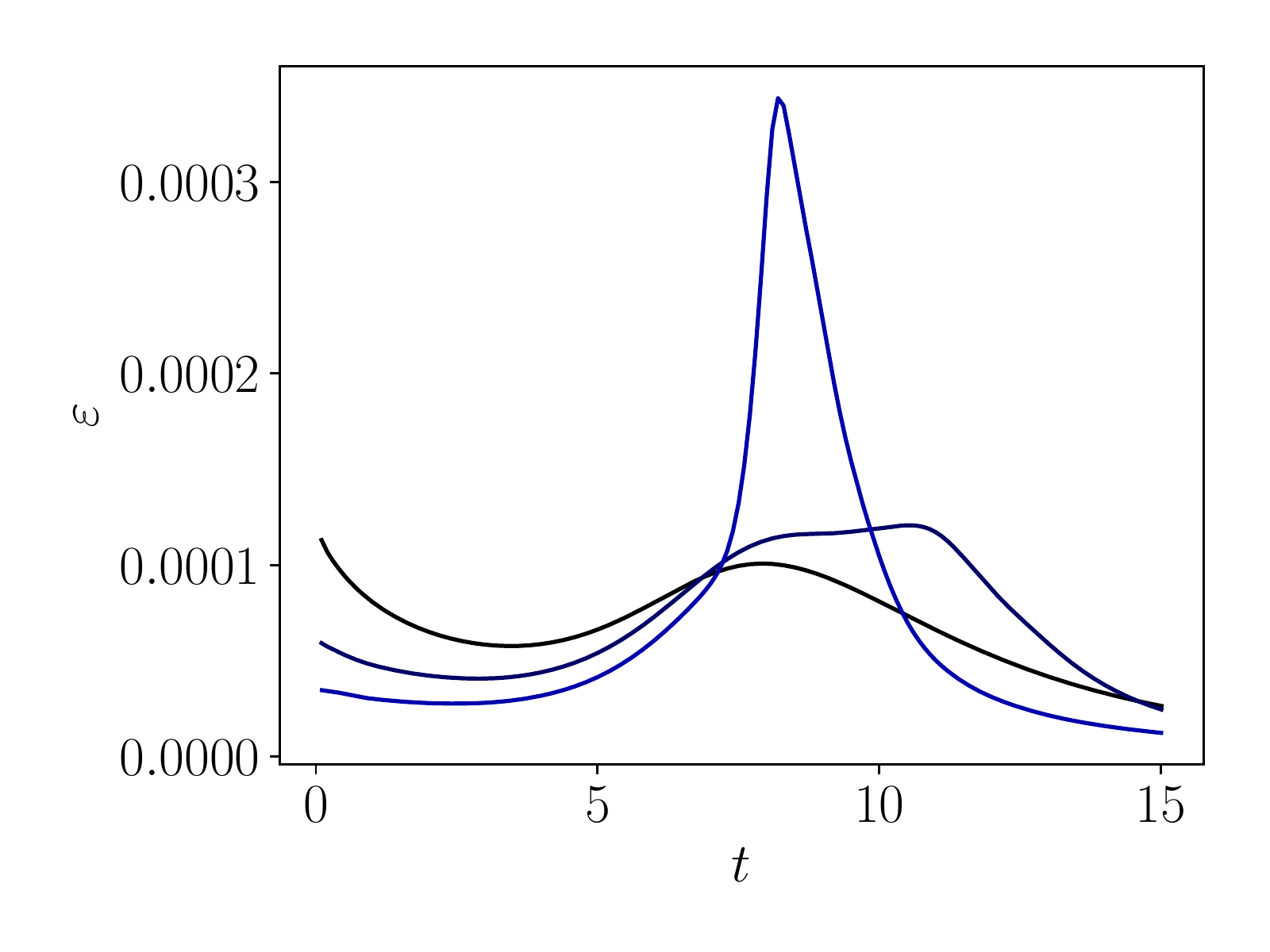}
 \caption{Left and center panels: energy of the $m=40$ (left) and $m=10$ mode (center) against time for $\Lambda=0.1$ and $Re_\Gamma=1000$ (black), $Re_\Gamma=2000$ (dark red/green) and $Re_\Gamma=3500$ (light red/green). Right: dissipation against time for the same cases with white noise. }
 \label{fi:ellipticalcrp1}
\end{figure}

Furthermore, the deformation pattern appears earlier at $Re_\Gamma=3500$, consistent with the fact that the elliptical instability has a slower growth rate at intermediate $Re_\Gamma$, and does not achieve viscosity-independent growth rates until $Re\approx 3000$ \cite{lew16}. We can analyze the energy of the $m=40$ mode across time to estimate the growth rate of the instability. This is shown in the left panel of Figure \ref{fi:ellipticalcrp1}, which corroborates the fact that the elliptical instability is absent at low Reynolds numbers, and has a $Re_\Gamma$-dependent growth rate for the values of $Re_\Gamma$ studied. By using a fit, we can estimate the exponential growth rates of the elliptical instability as $\sigma \approx 1.8$ at $Re_\Gamma=2000$ and $\sigma \approx 2.4$ at $Re_\Gamma=3500$. To compare this value to the one in the experiments of Ref.~\cite{lew98}, we estimate the inter-vortex distance as approximately $b\approx 0.2$ (see section \ref{sec:IVa}), so $\sigma$ can be expressed in the same non-dimensionalization as $0.6$ for $Re_\Gamma=3500$. This value is consistent with the theoretical asymptotic/inviscid growth rate of the instability \cite{lew98,lew16}, but lower than the one observed in the experiment in Ref.~\cite{lew98}.

In the center panel of Figure \ref{fi:ellipticalcrp1}, we show the energy of the $m=10$ ($k=2$) mode, which is associated to the longer-wavelength Crow instability. The growth rates are much smaller, and the jump in energy for $Re_\Gamma=3500$ is associated to the disintegration of the vortex and the transition to turbulence which happens at $t\approx 8$. The original rings no longer remain at this time as seen in Figure \ref{fi:vortcp1}, so this growth cannot be associated to a long-wavelength instability of the rings.

We can further corroborate that $Re_\Gamma=3500$ is in the asymptotic regime by showing the viscous dissipation $\varepsilon$ in the right panel of \ref{fi:ellipticalcrp1}. While the dissipation rate does not change significantly between $Re_\Gamma = 1000$ and $Re_\Gamma = 2000$, the ten-fold increase of $\varepsilon$ for $Re_\Gamma=3500$ coincides with the asymptotic behaviour of $\varepsilon$ seen in Ref.~\cite{keo20} for two vortex tubes decaying through the elliptical instability. We can thus anticipate that the behaviour of the collision is asymptotic at $Re_\Gamma=3500$, and that further increasing $Re_\Gamma$ will only produce finer length scales without modifying the external dynamics much, such as seen for $Re_\Gamma=4500$ in Ref.~\cite{keo20}. Energy is swiftly transferred from the large-scales to the finer scales, and this is reflected in the very fast growth of dissipation.

\subsection{Head-on collision with colored noise}
\label{sec:IIIb}

Having established this picture, we can now contrast the white noise simulations with the coloured noise simulations, i.e.~those where the noise in the two longest azimuthal wavelengths noise is started at a tenfold higher level ($\langle \epsilon_1^2 \rangle = \langle \epsilon_2^2 \rangle = 3 \times 10^{-2}$). This is done in an attempt to find the parameter space where the Crow instability becomes important, and can cause vortex ring to reconnect more locally. 

\begin{figure}
 \centering
 \includegraphics[trim={3cm 0 3cm 0},clip,width=0.32\textwidth]{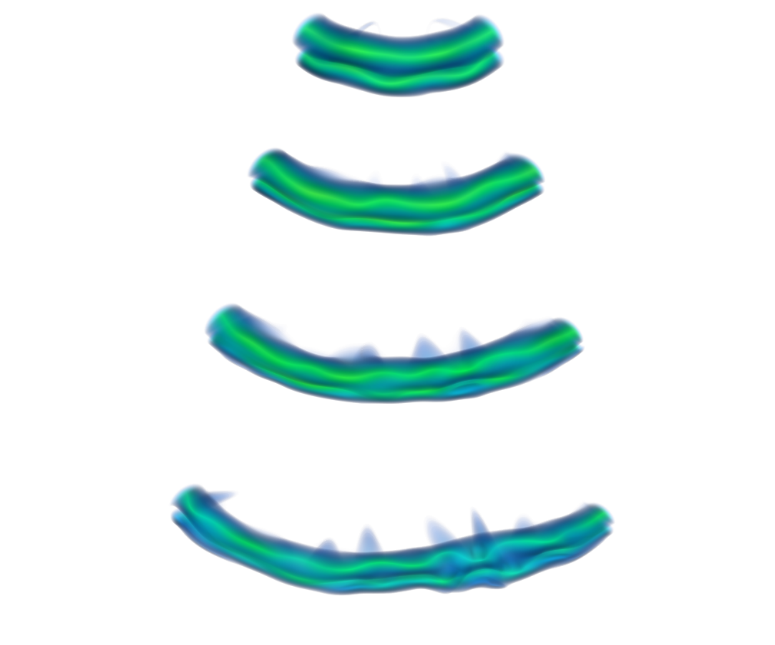}
 \includegraphics[trim={3cm 0 3cm 0},clip,width=0.32\textwidth]{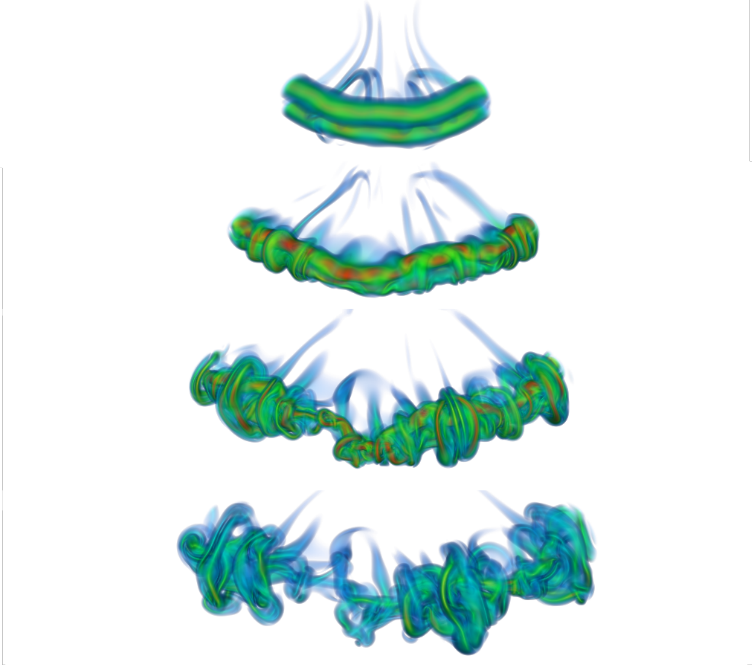}
 \includegraphics[trim={3cm 0 3cm 0},clip,width=0.32\textwidth]{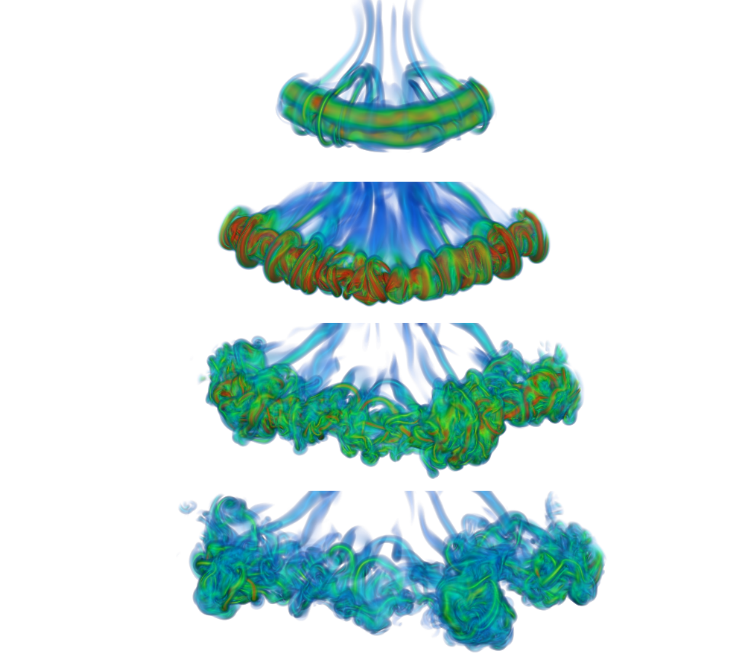}
 \caption{Vorticity volume visualization at several instances of time for $\Lambda=0.1$ and coloured noise perturbations. Reynolds number increases from left to right ($Re_\Gamma=1000$, $Re_\Gamma=2000$ and $Re_\Gamma=3500$), while time increases from top to bottom ($t=6$, $8$, $10$ and $12$). Red denotes regions of particularly high vorticity, while blue denotes regions of low vorticity. The same color map is used across all graphs.}
  \label{fi:vortcrowcp1}
\end{figure}

Figure \ref{fi:vortcrowcp1} shows a volume visualization of the vorticity modulus for all cases with $\Lambda=0.1$ and colored noise. As seen above, for $Re_\Gamma=1000$, left column, the rings stretch out radially and are slowly dissipated by viscosity, even if some more perturbations to the core can be seen due to the 
higher noise levels. For $Re_\Gamma=2000$, middle column, we see the formation of secondary vortex rings, 
with strands of perpendicular vorticity filaments attached to them. These do not appear to be as clean as in the experiments of Ref.~\cite{lim92}, which can be expected: the dye used in the experiments does not track vorticity amplification, and these perpendicular filaments contain very amplified vorticity \cite{keo20}. Finally, at $Re_\Gamma=3500$, right column, we can see the disintegration of the rings. The elliptical instability now  dominates the Crow instability, but the long-wavelength we have superimposed causes the production of fine-scales to be azimuthally inhomogeneous. The cores are first brought together at certain points, and this locally increases the growth rate of the elliptical instability. A similar flow phenomenology was also observed for two counter-rotating vortex tubes in the experiments of Ref.~\cite{lew98} and the simulations of Ref.~\cite{lap00}, where short- and large-wavelength perturbations caused decay to occur more rapidly in certain areas than others.

\begin{figure}
 \centering
 \includegraphics[width=0.49\textwidth]{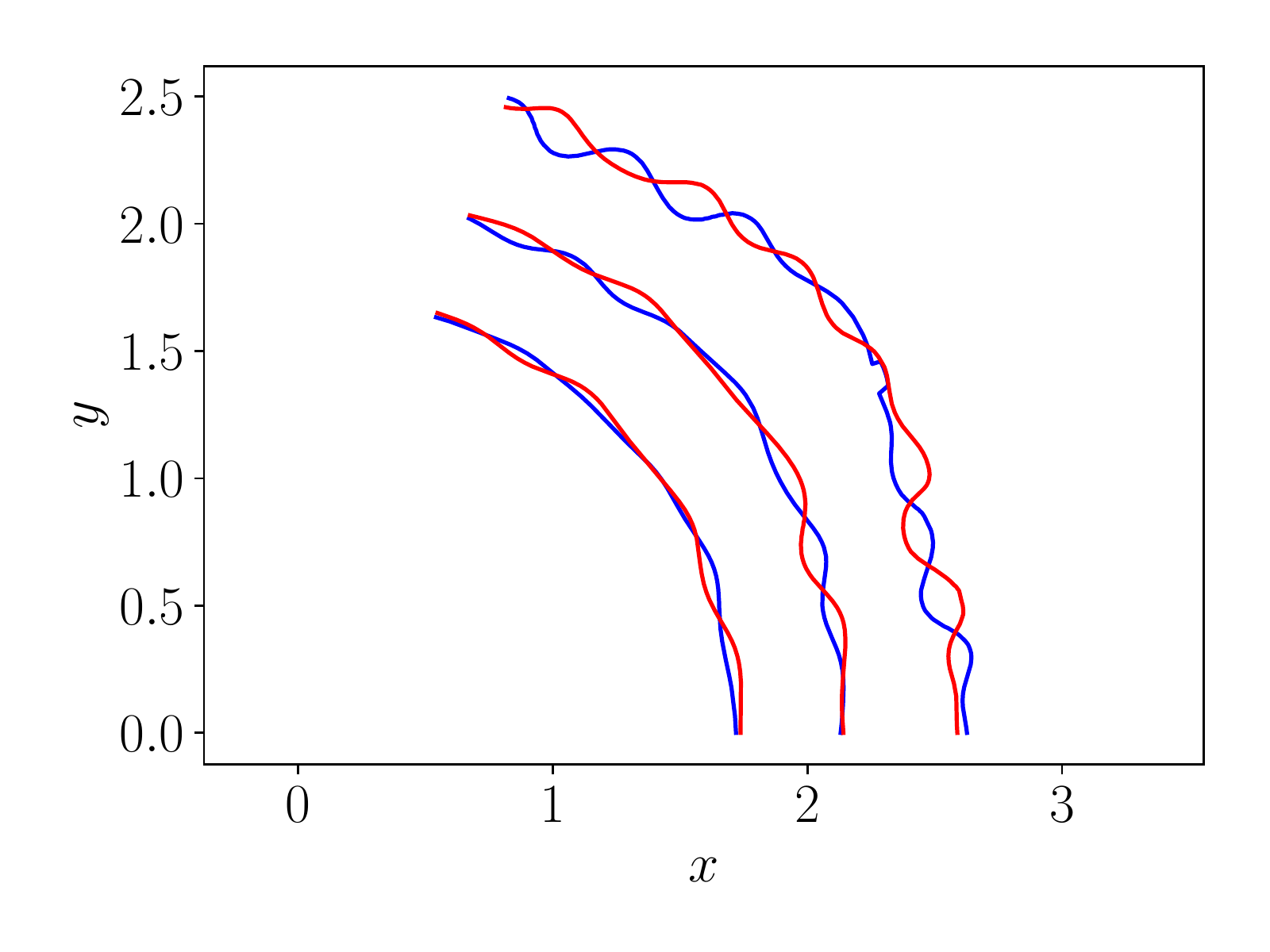}
 \includegraphics[width=0.49\textwidth]{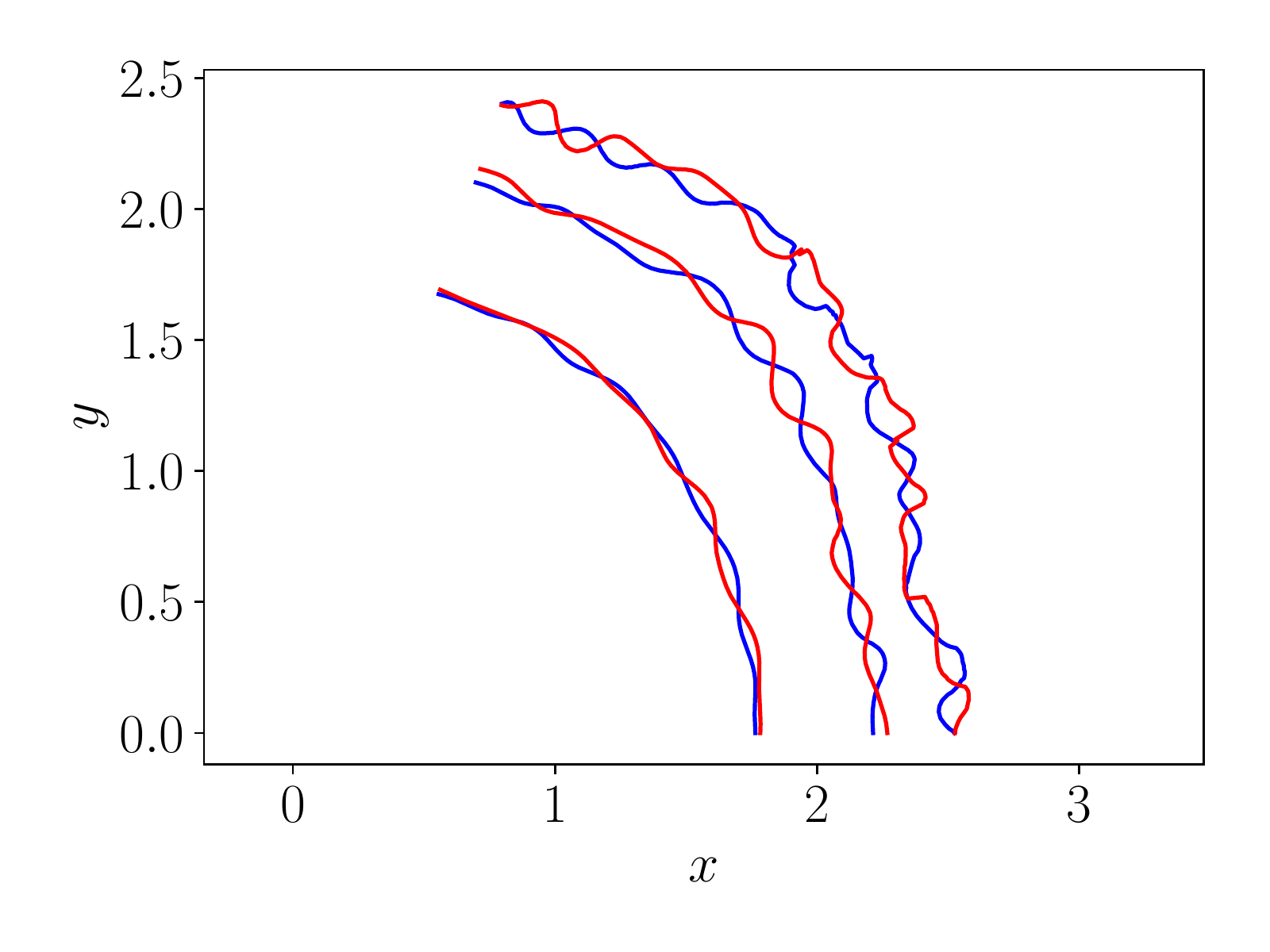}
   \includegraphics[trim={0cm 3cm 0cm 5cm},clip,width=0.49\textwidth]{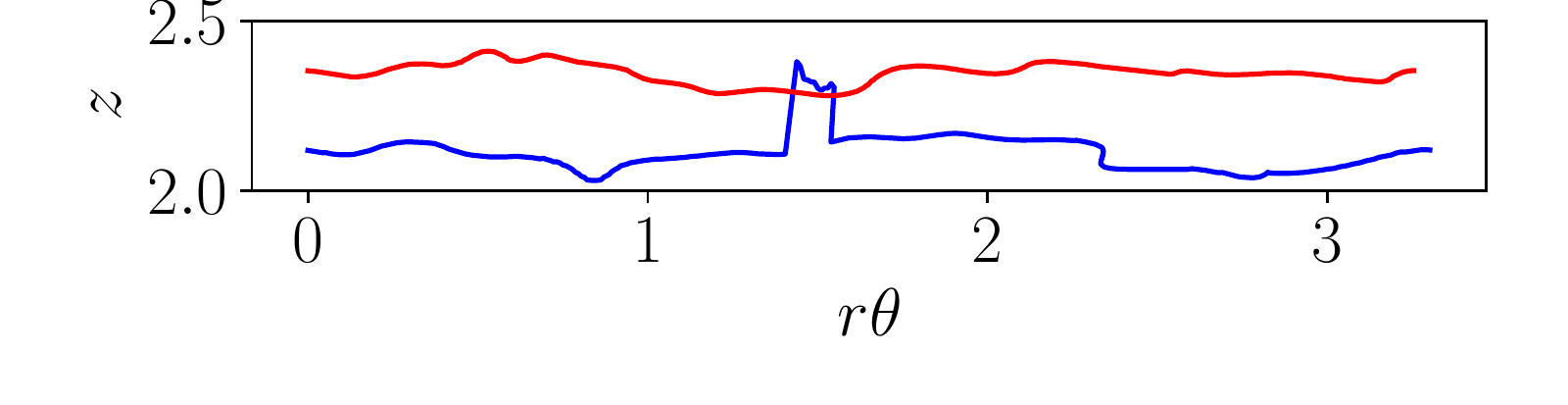}
 \includegraphics[trim={0cm 3cm 0cm 5cm},clip,width=0.49\textwidth]{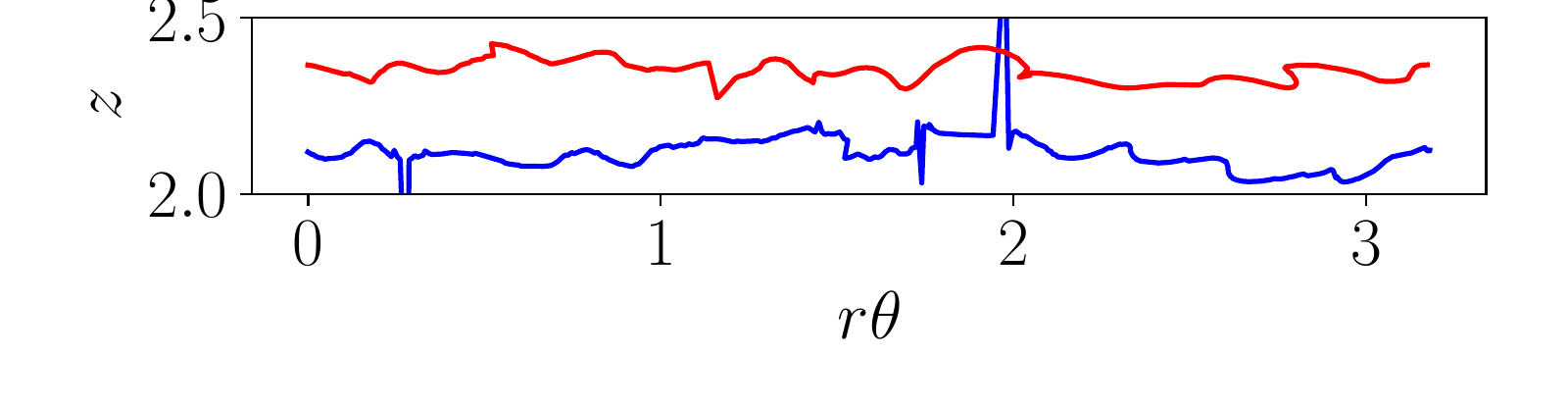}
 \caption{Extracted vortex cores at several instances of time for $\Lambda=0.1$ and coloured noise. Left top panel: cores at $t=6$, $7$ and $8$ for $Re_\Gamma=2000$. Right top panel: cores at $t=6$, $7$, and $8$ for $Re_\Gamma=3500$. Bottom left panel: side view of the cores at $t=8$ and $Re_\Gamma=2000$. Bottom right panel: side view of the cores at $t=8$ and $Re_\Gamma=3500$. }
 \label{fi:fils2}
\end{figure}

The inhomogeneous growth of the elliptical instability can be better observed by looking at the centroid of the interacting vortex tubes, shown in Figure \ref{fi:fils2}. For both $Re_\Gamma=2000$, left column and $Re_\Gamma=3500$, right column, the growth of the perturbation happens first at certain sites, which coincide with the sites where the rings are brought together by the Crow instability, before extending to the full azimuthal domain. This leads to the uneven generation of small-scales seen in the final panel. As in the case with white noise, the centroids of the rings only present significant deformation in the collision plane, and not perpendicular to it.

\begin{figure}
 \centering
  \includegraphics[width=0.32\textwidth]{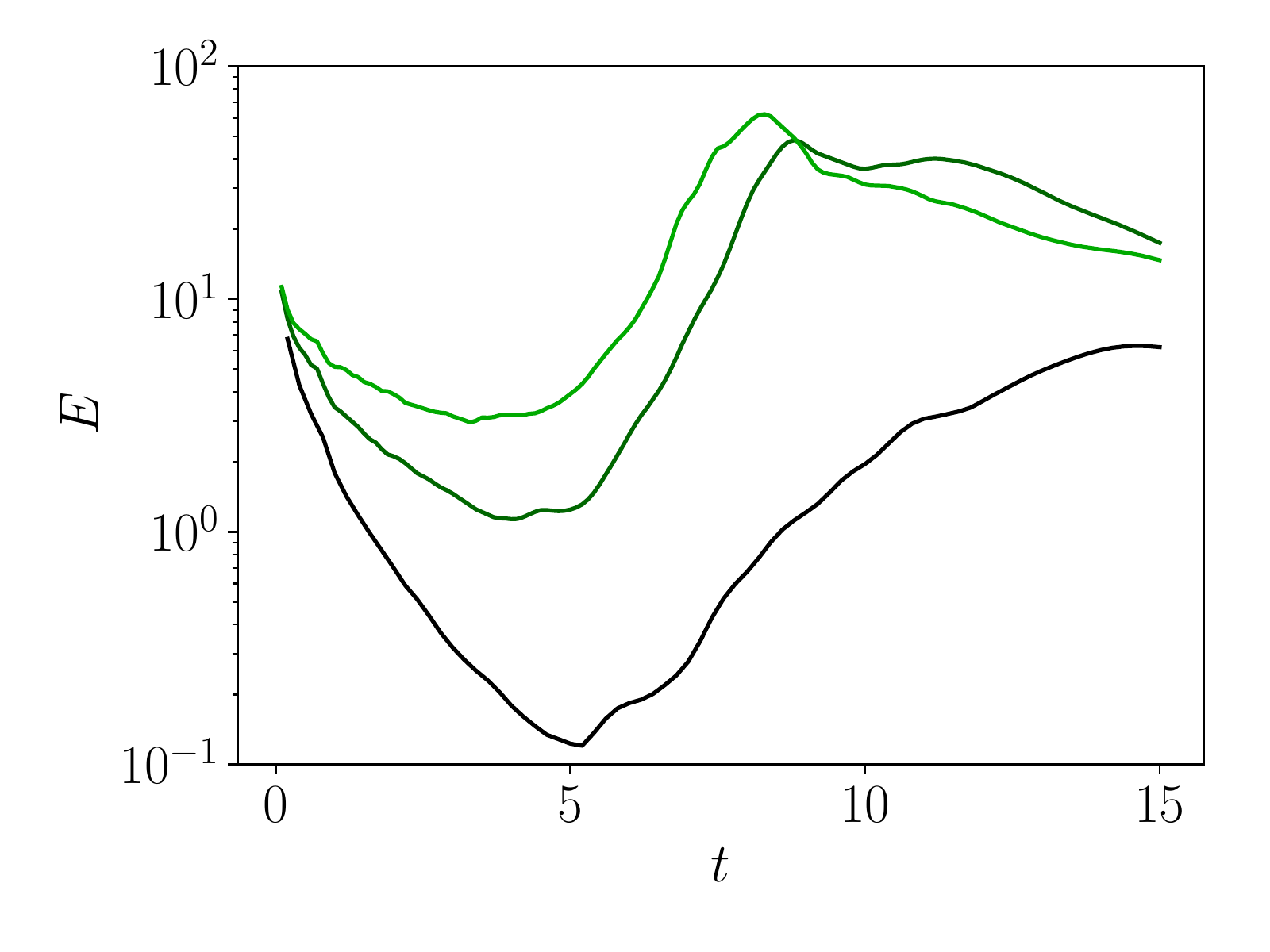}
 \includegraphics[width=0.32\textwidth]{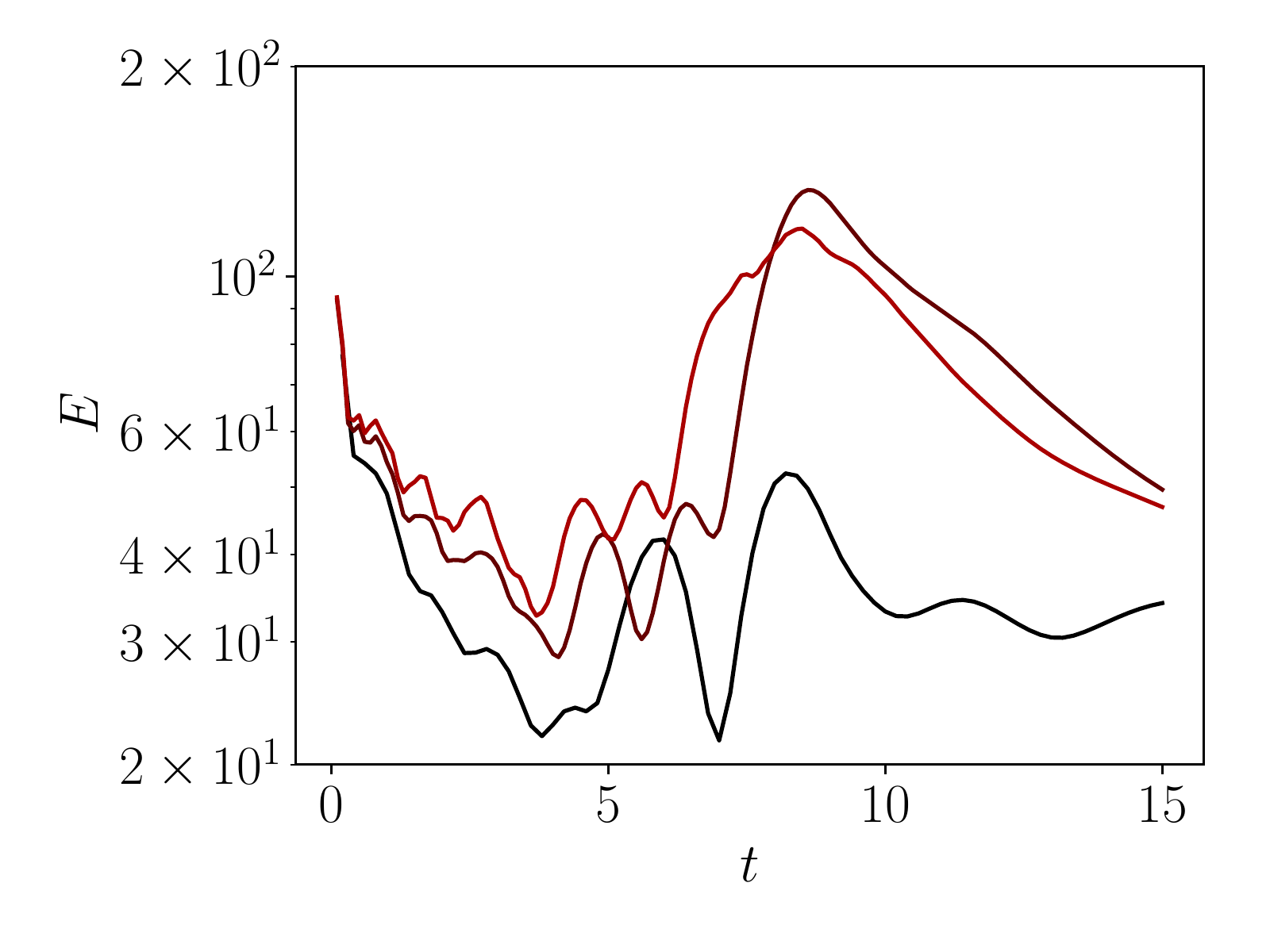}
  \includegraphics[width=0.32\textwidth]{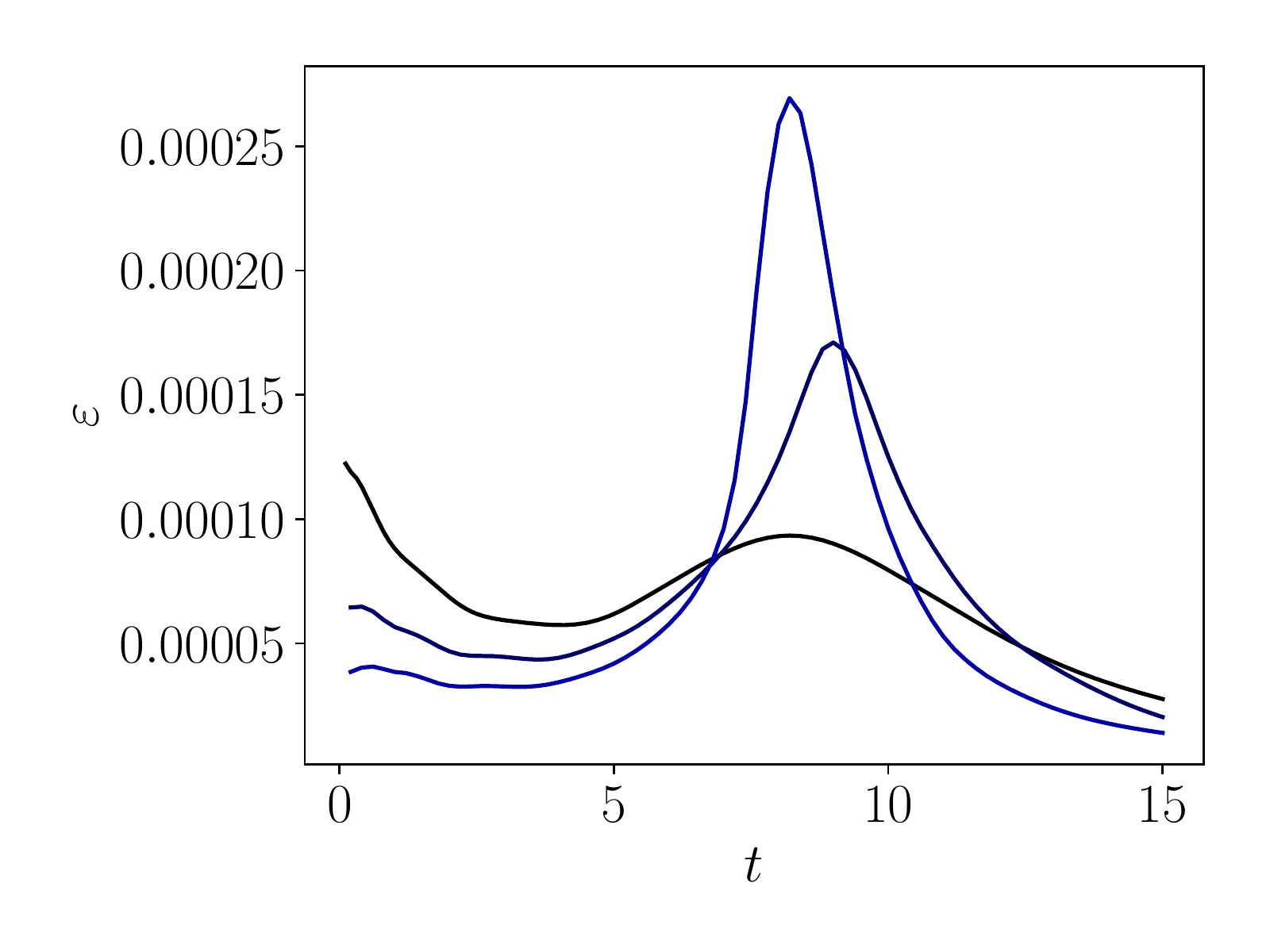}
 \caption{Left and center panels: energy of the $m=40$ (left) and $m=10$ mode against time for $\Lambda=0.1$ and $Re_\Gamma=1000$ (black), $Re_\Gamma=2000$ (dark red/green) and $Re_\Gamma=3500$ (light red/green) and colored noise. Right: dissipation against time for the same cases. }
 \label{fi:crowcrp1}
\end{figure}

It thus appears that different seeding can produce slightly different outcomes. This was already observed by Laporte \& Corjon \cite{lap00} in numerical simulations of the interaction between two vortex tubes. By controlling the ratio of large-wavelength to small-wavelength noise, they were able to show both local tube reconnection and tube disintegration in their simulations. We can quantify this effect for our case the by showing the temporal evolution of the energy in the $m=40$ and $m=10$ modes in the first two panels of \ref{fi:crowcrp1}. Because the energy of $m=10$ now starts off from a higher level (contrast the initial value of E to that of Fig.~\ref{fi:ellipticalcrp1}), the Crow instability is able to grow enough to initiate reconnection, despite its smaller growth rate. The elliptical instability can be seen to kick in around the same time ($t\approx5$) as in the white noise case, but it is not able to grow fast enough to prevent reconnection for $Re_\Gamma=2000$ and two secondary rings can be seen to form. These rings advect away from the primary ring's axis, 
as it was the case in the experiment of Ref.~\cite{lim92}. After this short time, they are damped by viscosity.

For $Re_\Gamma=3500$, see the top-left panel of Fig.\ref{fi:fils2}, we obtain a mixture of elliptical and Crow: there is an azimuthally inhomogeneous production of small scales, and the rings end in total disintegration. The end result is similar to the asymptotic regime seen for white noise, even if the disintegration is inhomogeneous. To corroborate this, we show the instantaneous dissipation in the right panel of Figure \ref{fi:crowcrp1}. The large dissipation values show that there is a large transfer of energy from the large to the small scales. It is also worth noting that the peak of dissipation happens earlier in this case than in the white noise case, indicating that the local growth of the elliptical instability induced by the Crow instability hastens the disintegration process.

From these results, we can conclude that the process which produces secondary rings which survive for a short time is prevalent in a rather small region of $Re_\Gamma$ and hinges on the fact that longer wavelengths are seeded more than others, so the Crow instability can develop fully before the elliptical one overtakes it. We note that our study was somewhat limited: we only increased the ratio between seedings to be tenfold, but preliminary simulations which increased even more the seeding of the long-wavelength modes resulted in dynamics that were very dependent on the phase difference between both instabilities and barely reproducible, so we did not pursue this further. To further study the competition between instabilities, we instead measure their growth rates. In the case of the Crow instability, this cannot be done from the curves shown in Figure \ref{fi:crowcrp1} due to the presence of the elliptical instability as a confounder. But by making a ring collide against a stress-free wall, we enforce symmetry and severely mitigate the elliptical instability, as we discuss in the next section.

\subsection{Stress-free wall}
\label{sec:IIIc}

As in the previous section, we first focus on the cases with $\Lambda=0.1$ and white noise, and study the effect of $Re_\Gamma$. As mentioned in the introduction, impact against a stress-free wall can also be understood as collision with the mirror ring, so we expect similar dynamics in the initial phases of the collision. 

\begin{figure}
 \centering
 \includegraphics[trim={3cm 0 7cm 0},clip,width=0.32\textwidth]{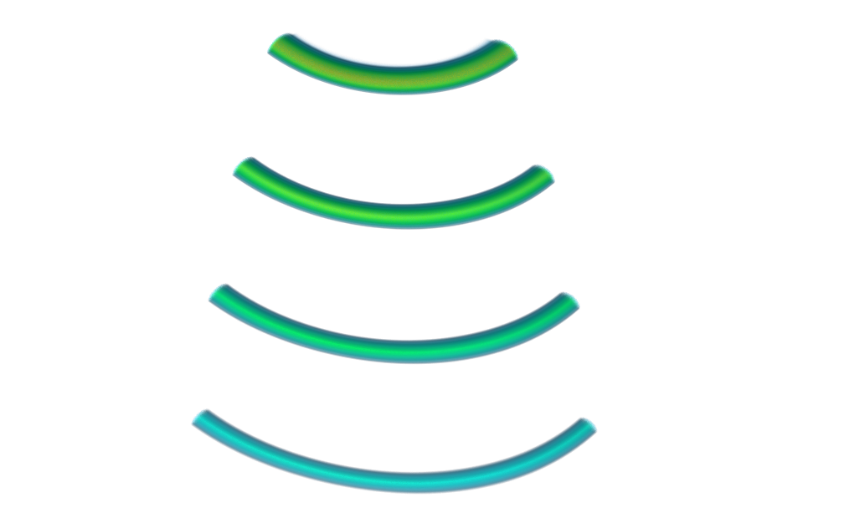}
 \includegraphics[trim={3cm 0 7cm 0},clip,width=0.32\textwidth]{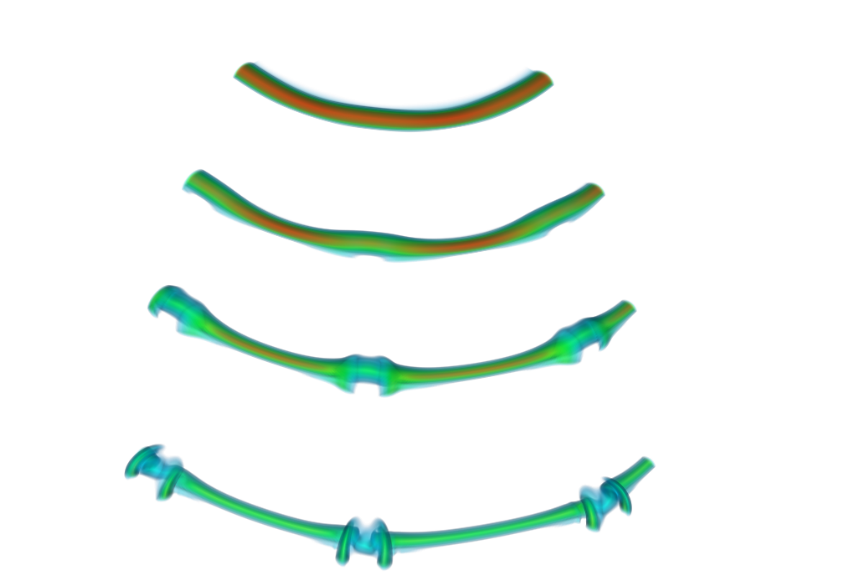}
 \includegraphics[trim={3cm 0 7cm 0},clip,width=0.32\textwidth]{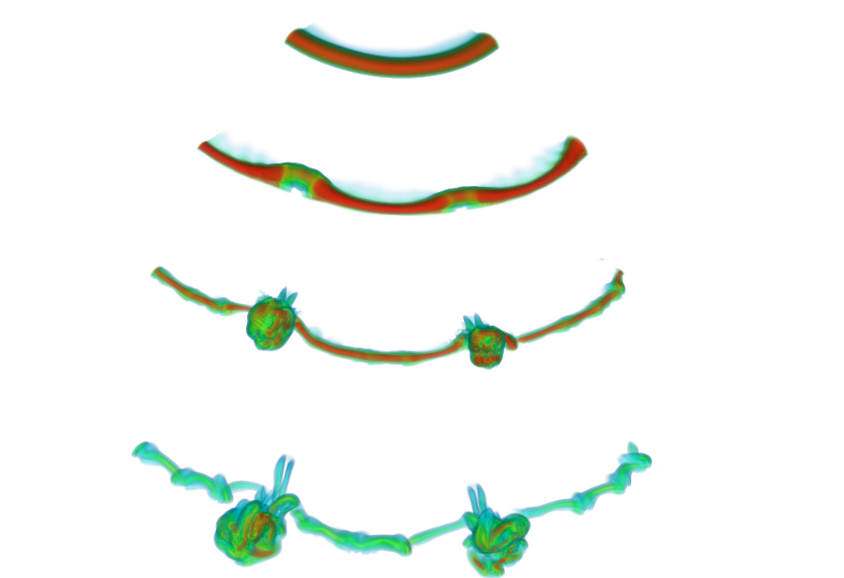}
 \caption{Vorticity volume visualization at several instances of time for ring impact against a stress-free wall at  $\Lambda=0.1$ and white noise perturbations. Reynolds number increases from left to right ($Re_\Gamma=1000$, $Re_\Gamma=2000$ and $Re_\Gamma=3500$), while time increases from top to bottom ($t=18$, $22$, $26$ and $30$ for left and center, $t=16$, $20$, $24$ and $28$ for right). Red denotes regions of particularly high vorticity, while blue denotes regions of low vorticity.  }
 \label{fi:vortcp1-fsw}
\end{figure}

Figure \ref{fi:vortcp1-fsw} shows a volume visualization of the vorticity magnitude for the three Reynolds numbers studied as the ring interaction with the stress-free wall proceeds. As was the case for the head-on collision, for $Re_\Gamma=1000$, left panel, the ring impacts the wall (mirror ring), and stretches out, remaining approximately axisymmetric. The total vorticity magnitude first increases as the vortex is stretched to conserve circulation, and then rapidly decreases as vorticity dissipates the ring. 

For $Re_\Gamma=2000$, center panel, a long-wavelength instability arises. Because the ring is stretching at the same time as the instability is growing, the base wavelength is constantly changing, and the instability is barely able to develop until the point where the ring and its mirror image come into contact, a process that leads to a local reconnection to form (half) a secondary vortex. By $t=28$, proto-structures of secondary rings, consisting of 
parallel filaments, can be observed very close to the point of contact between the ring and the wall. However, these are rapidly dissipated due to viscosity. We note that viscosity acts much later than during the head-on collisions considered earlier, so the ring has up to $7$ to $8$ times from its initial radius, compared to 4-5 ring radii earlier. 

To actually see (the top half of the) secondary rings forming, as done experimentally in Ref.~\cite{lim92}, we have to increase $Re_\Gamma$ to $Re_\Gamma=3500$. We can then observe in the figure that at the last stage, two secondary rings are created, see the right column of Fig.~\ref{fi:vortcp1-fsw}. The vorticity of the secondary rings quickly dissipate, in agreement with the head-on collision videos of Ref.~\cite{lim92} which show that the red-blue secondary rings stop travelling, and the previous case at $Re_\Gamma=2000$ and coloured noise. In the experiments, the dye remains, but the vorticity has dissipated and the ring ceases to exist. We note that the reconnection procedure here is qualitatively
similar to that seen for vortex tubes in for example Refs.~\cite{hus11,ost21}. There, the two tubes touch at certain points, forming very thin sheets. These sheets then rapidly dissipate and leave behind a changed topology. This is especially apparent in the right panel of Fig.~\ref{fi:vortcp1-fsw}, where the formation of thin sheets with high vorticity that later dissipate can be appreciated, and the two half-rings are left behind. Furthermore, the rest of the vortex filament is highly distorted, with the formation of very short-wavelength perturbations, as it was the case in Ref.~\cite{arc10}. This points to the resilience of the elliptical instability, despite the enforcement of the seemingly unfavorable symmetry induced by the stress free wall.

\begin{figure}
 \centering
  \includegraphics[width=0.32\textwidth]{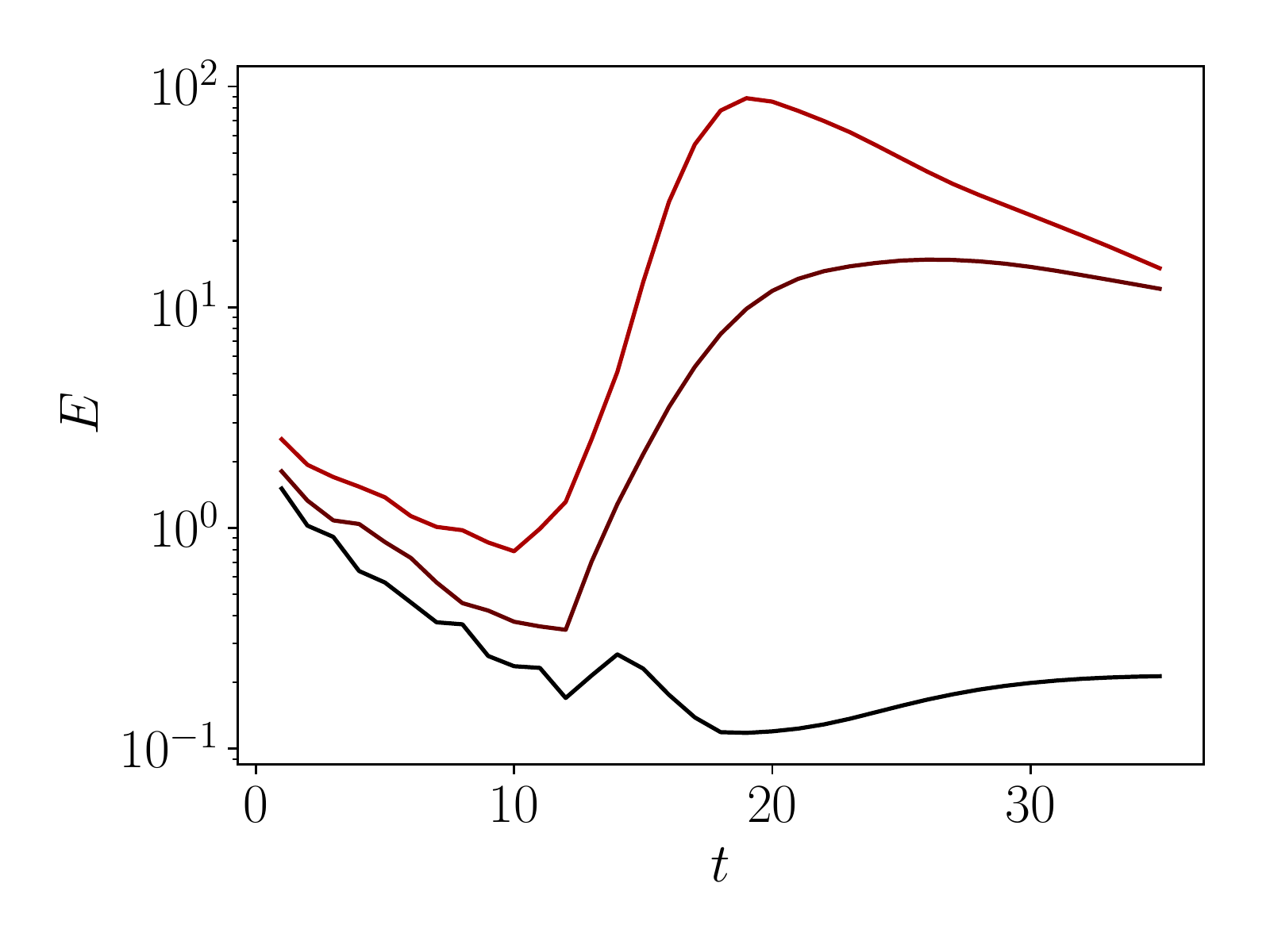}
 \includegraphics[width=0.32\textwidth]{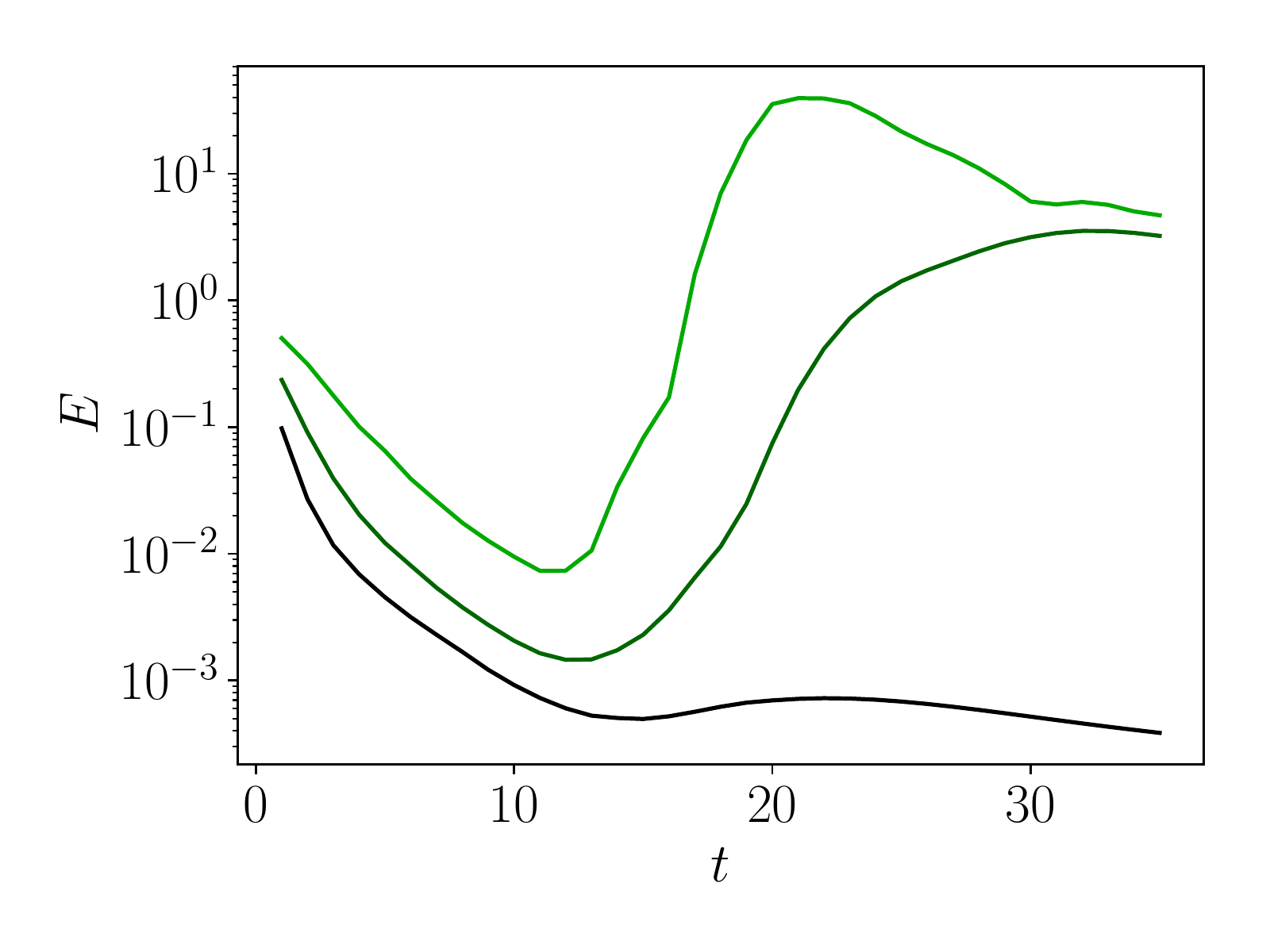}
  \includegraphics[width=0.32\textwidth]{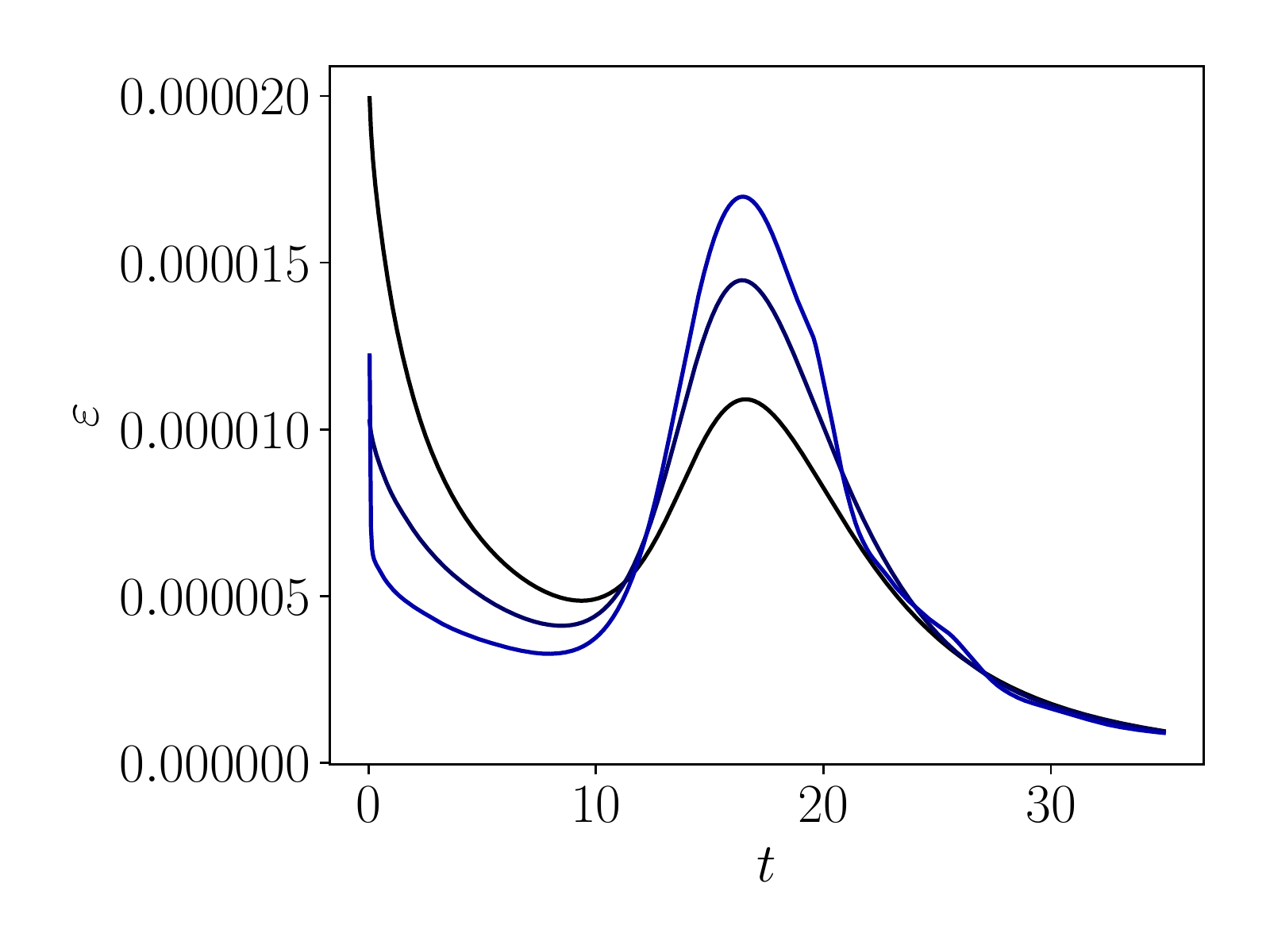}
 \caption{Left and center panels: energy of the $m=10$ (left) and $m=40$ mode against time for stress-free wall impact for $\Lambda=0.1$ and $Re_\Gamma=1000$ (black), $Re_\Gamma=2000$ (dark color) and $Re_\Gamma=3500$ (light color) and impact . Right: dissipation against time for the same cases. }
 \label{fi:fsstuffcrp1}
\end{figure}

We can now measure the growth rate of the long-wavelength perturbation associated to the Crow instability by looking at the energy of the $m=10$ mode. This is shown in the in the left panel of Figure \ref{fi:fsstuffcrp1}. The growth rates of the instability are much smaller than what we observed in the previous section. We obtain growth rates of  $\sigma \approx 0.5$ at $Re_\Gamma=2000$ and $\sigma \approx 0.8$ at $Re_\Gamma=3500$ for the long-wavelength instability. For comparison, we show in the center panel of Figure \ref{fi:fsstuffcrp1} the energy in the $m=40$ mode, from which we can obtain a growth rate of  $\sigma \approx 1.2$ at $Re_\Gamma=3500$. 
By using the stress-free wall, we have halved the growth rate of short wavelength instabilities and this has allowed us to observe the reconnection process and the formation of secondary rings. It is unclear whether the short wavelength instability is associated to the elliptical instability (as postulated in Ref.~\cite{arc10}) but it is clear that the core is deformed at the largest $Re_\Gamma$ studied. We estimate the growth rate of the instability
to be about 50\% larger than that of the Crow instability, and this points to
the relevance of short wavelength instabilities for both stress-free and head-on collision. 

As the ring stretches out as it interacts with the wall, and there is no fast disintegration, the production of small scales is heavily curtailed in this case. This can be quantified through the temporal behaviour of the dissipation, shown in the right panel of Figure \ref{fi:fsstuffcrp1}. The dissipation first decreases with time, as the ring approaches the wall. As the ring begins to interact with the wall, vorticity increases and dissipation follows, until a peak is reached. The observed peaks in Fig.~\ref{fi:fsstuffcrp1} are $\sim 10$ times smaller than those observed in Fig.~\ref{fi:ellipticalcrp1} and \ref{fi:crowcrp1}. After the peak, further increases in vorticity are overwhelmed by smaller energy available, so dissipation begins to decrease. This is the case not only at the lower $Re_\Gamma$, where the ring remains axisymmetric, but also when azimuthal instabilities, and reconnection happen. This shows that the formation of secondary rings through local reconnection is a very different process from the disintegration through the elliptical instability seen in the section above, as was analyzed for vortex tubes in Ref.~\cite{ost21}. The temporal behaviour of the dissipation is very similar to the head-on collision at $Re_\Gamma=1000$, which remained axisymmetric, and the peak values of dissipation obtained are drastically below those seen for the head-on collision, even when the noise is coloured and local reconnections are present. 

Unlike in the previous section, the cases with colored (long-wavelength) noise do not present significant differences from the white noise cases. The main difference we saw was that in some cases five half-rings (one per rotational symmetry wedge) were generated, rather than the ten (two per rotational symmetry wedge) seen above. However, this depends mainly on the level of noise of the two long wavelength modes ($m=5$ and $m=10$). The growth rates of the instabilities remained relatively constant. We expect that in a real scenario, or in the absence of any rotational symmetry, the number of secondary rings generated will be between $5$ and $10$. 

\section{The effect of ring thickness on the instabilities}
\label{sec:IV}

\subsection{Head-on collision}
\label{sec:IVa}

Up to now, we have not considered the effect of ring thickness (or equivalently, $\Lambda$) on the dynamics. As shown earlier in Figure \ref{fi:vortprof}, the larger core sizes result in smaller velocities for the same circulation. Thicker rings also self-advect slower than thin rings, with the dependence being approximately logarithmic on the radius \cite{saf79}. Finally, it is unclear how the thickness of the ring is reflected in the effective non-dimensional ratio between the tube size and distance which is an important parameter in determining the growth rate of the elliptical instability. This parameter is usually denoted as $a/b$ where $a$ is the tube radius and $b$ the distance between tube centers \cite{lew16}. 

\begin{figure}
 \centering
  \includegraphics[width=0.32\textwidth]{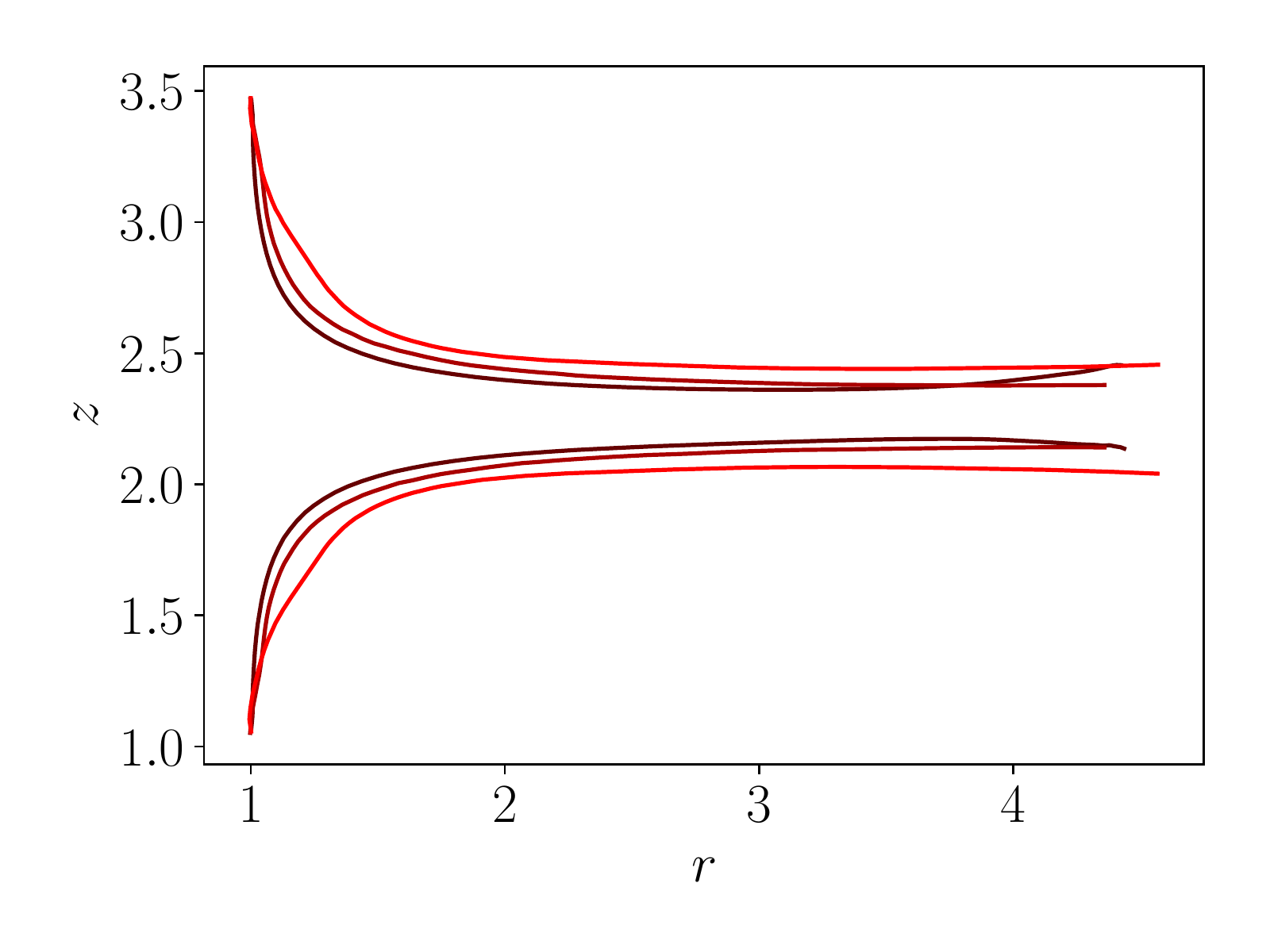}
 \includegraphics[width=0.32\textwidth]{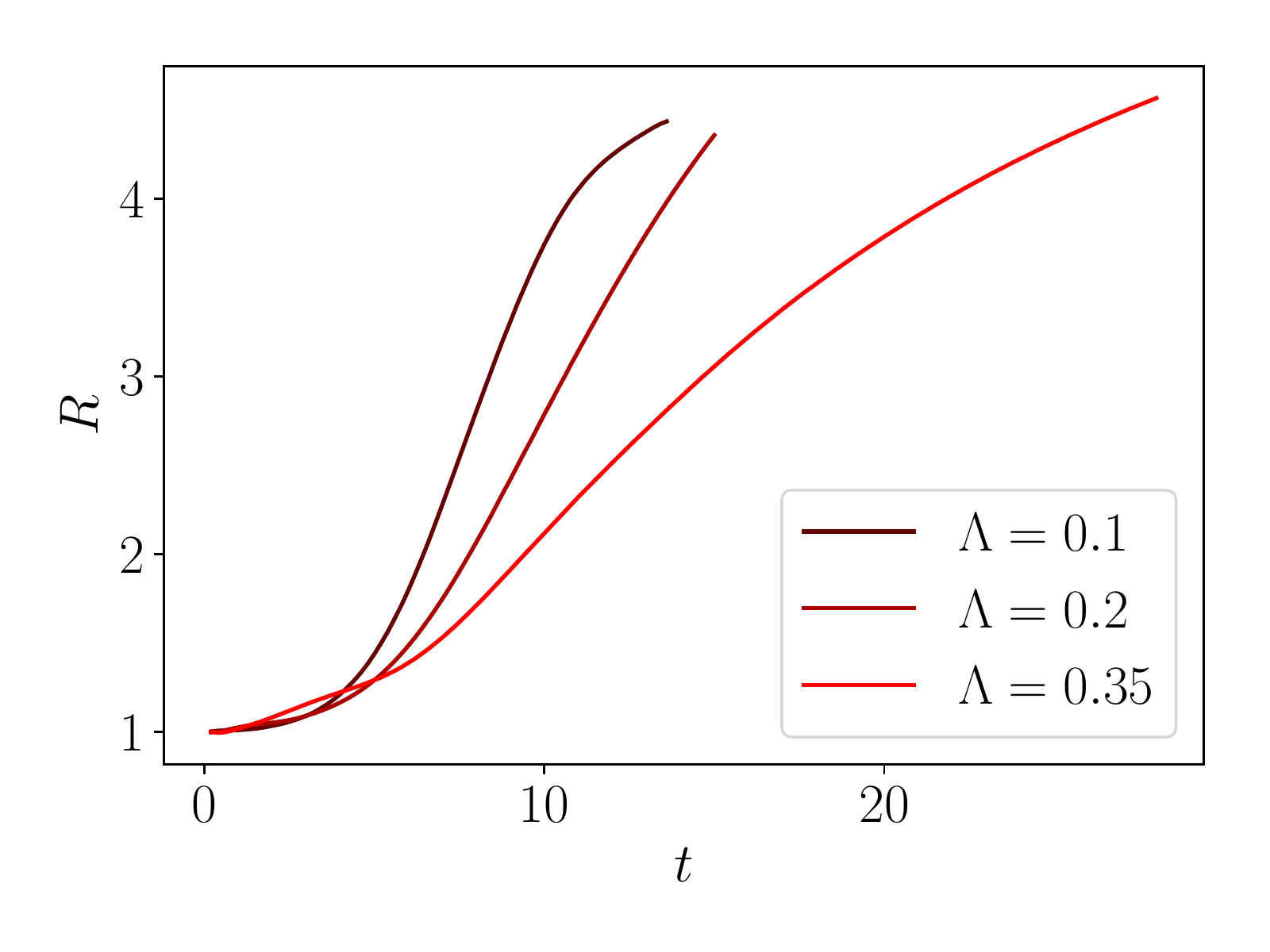}
 \includegraphics[width=0.32\textwidth]{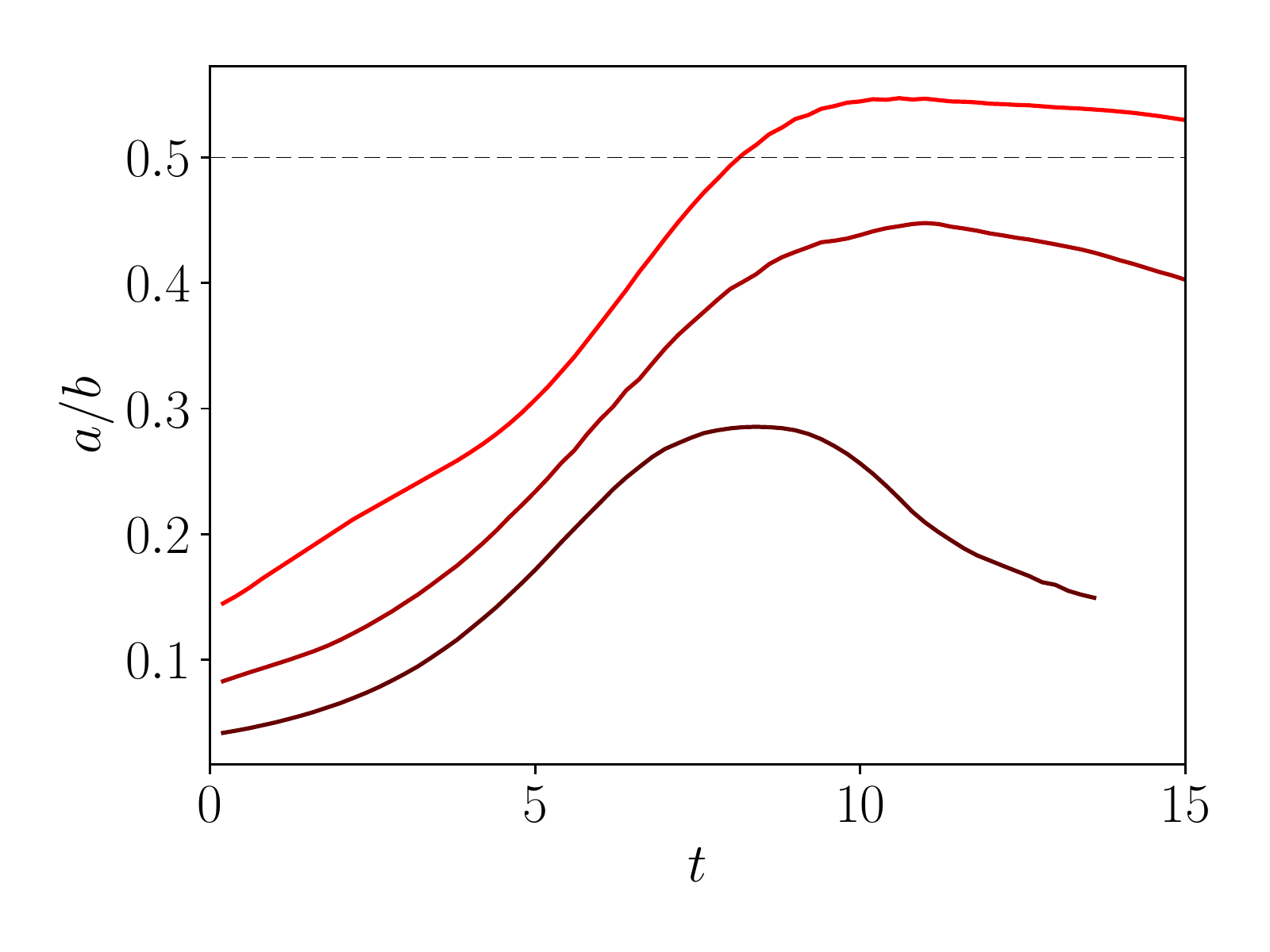}
 \caption{Left panel: Azimuthally averaged ring trajectory for $Re_\Gamma=1000$. Center panel: temporal evolution of the outer radius for several $\Lambda$ at $Re_\Gamma=1000$. Right: estimation of $a/b$ parameter for the same cases. A dashed line which marks the theoretical limit of $a/b=0.5$ is shown. }
 \label{fi:as_vs_cr_res}
\end{figure}

To study this, we now vary the ring thickness ratio $\Lambda$, and simulate cases with $\Lambda=0.2$ and $\Lambda=0.35$. We then contrast the results with those presented above with $\Lambda = 0.1$.
We first note that the $Re_\Gamma=1000$ simulations all remain relatively axisymmetric for both white and colored noise, similar to the cases with $\Lambda=0.1$. The rings come into contact forming a ring-dipole, then stretch each other before diffusing away due to the effect of viscosity. However, the resulting dynamics differ when starting off with rings of different thickness, and result in a vortex dipole of different characteristics. We can thus use these simulations to quantify the effect of ring thickness on the $a/b$ parameter mentioned above. We isolate the centroids of the rings as done before, and azimuthally average over the small deviations to obtain the radial and axial position of the two vortex cores. The resulting trajectories are shown in the left panel of Figure \ref{fi:as_vs_cr_res}. We also show the temporal evolution of the radial coordinate $R(t)$ of the top ring in the center panel of Figure \ref{fi:as_vs_cr_res}, to confirm the slower evolution of the cases with thicker rings, which take longer to come into contact and spread each other. 

To estimate the evolution of the $a/b$ parameter, we first obtain $b$, the distance between the two ring centers, directly from the axial coordinates of the vortex centroids. Obtaining the core radius $a$ is harder, we approximate it using $a/a_0=\sqrt{R_0/R}$, which comes from conservation of volume of the ring. This method is inaccurate, as it assumes that the ring retains an approximately constant shape. This is later reflected in the fact that the value of $a/b$ exceeds the theoretical limit of $\frac{1}{2}$ in one case, which would mean overlapping rings. Despite its obvious shortcoming, our method 
allows us to draw a qualitative picture of the behaviour of $a/b$. We show $a/b$ for the three cases in the right panel Figure \ref{fi:as_vs_cr_res}, where we can observe that the $a/b$ parameter is larger for thicker rings, and at first approximation, this will cause the growth rate of the elliptical instability to be higher at moderate $Re_\Gamma$ \cite{lew16}.

\begin{figure}
 \centering
 \includegraphics[trim={3cm 0 3cm 0},clip,width=0.32\textwidth]{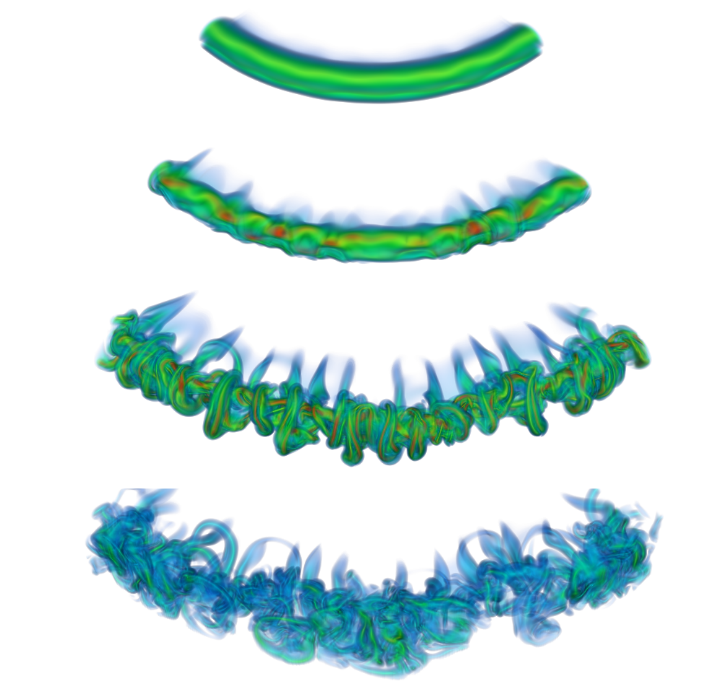}
  \includegraphics[trim={3cm 0 3cm 0},clip,width=0.32\textwidth]{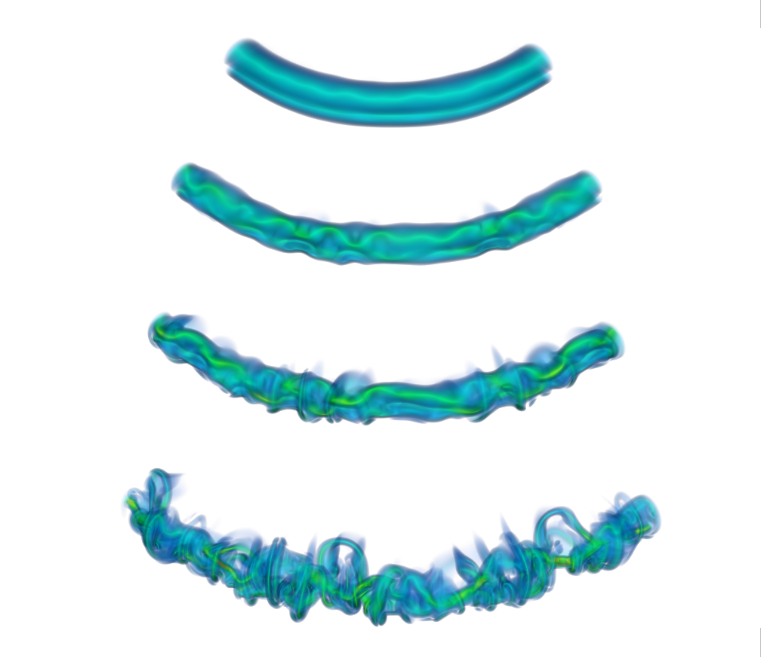}
 \caption{Vorticity volume visualization at several instances of time for $Re_\Gamma=3500$ and white noise perturbations. Left is $\Lambda=0.2$, while right is $\Lambda=0.35$. Time increases from top to bottom (left: $t=12$, 14, 16, 18, right: 14, 18, 20 and 24). Red denotes regions of particularly high vorticity, while blue denotes regions of low vorticity. }
 \label{fi:vortres3-cps}
\end{figure}

\begin{figure}
 \centering
  \includegraphics[width=0.49\textwidth]{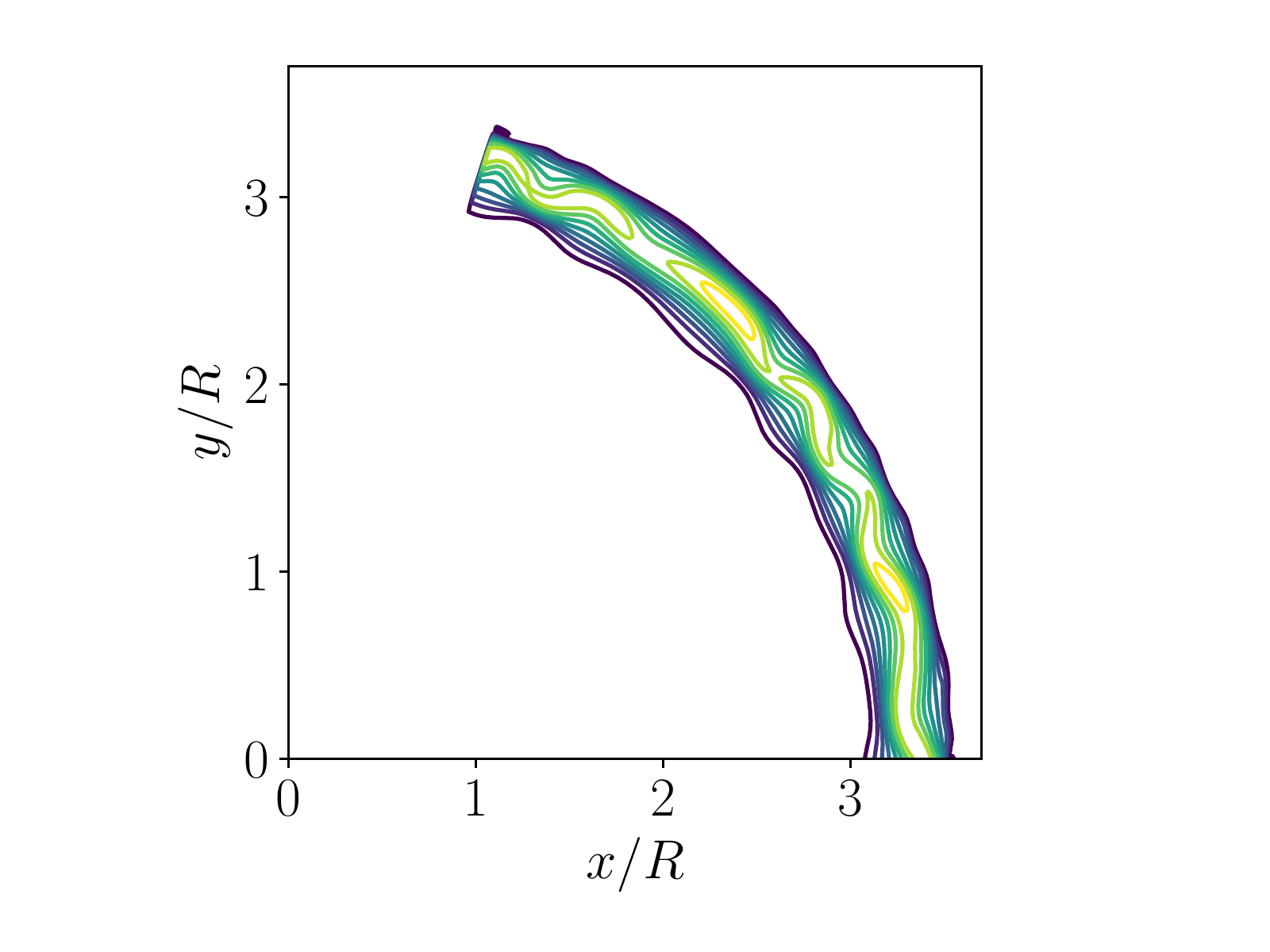}
  \includegraphics[width=0.49\textwidth]{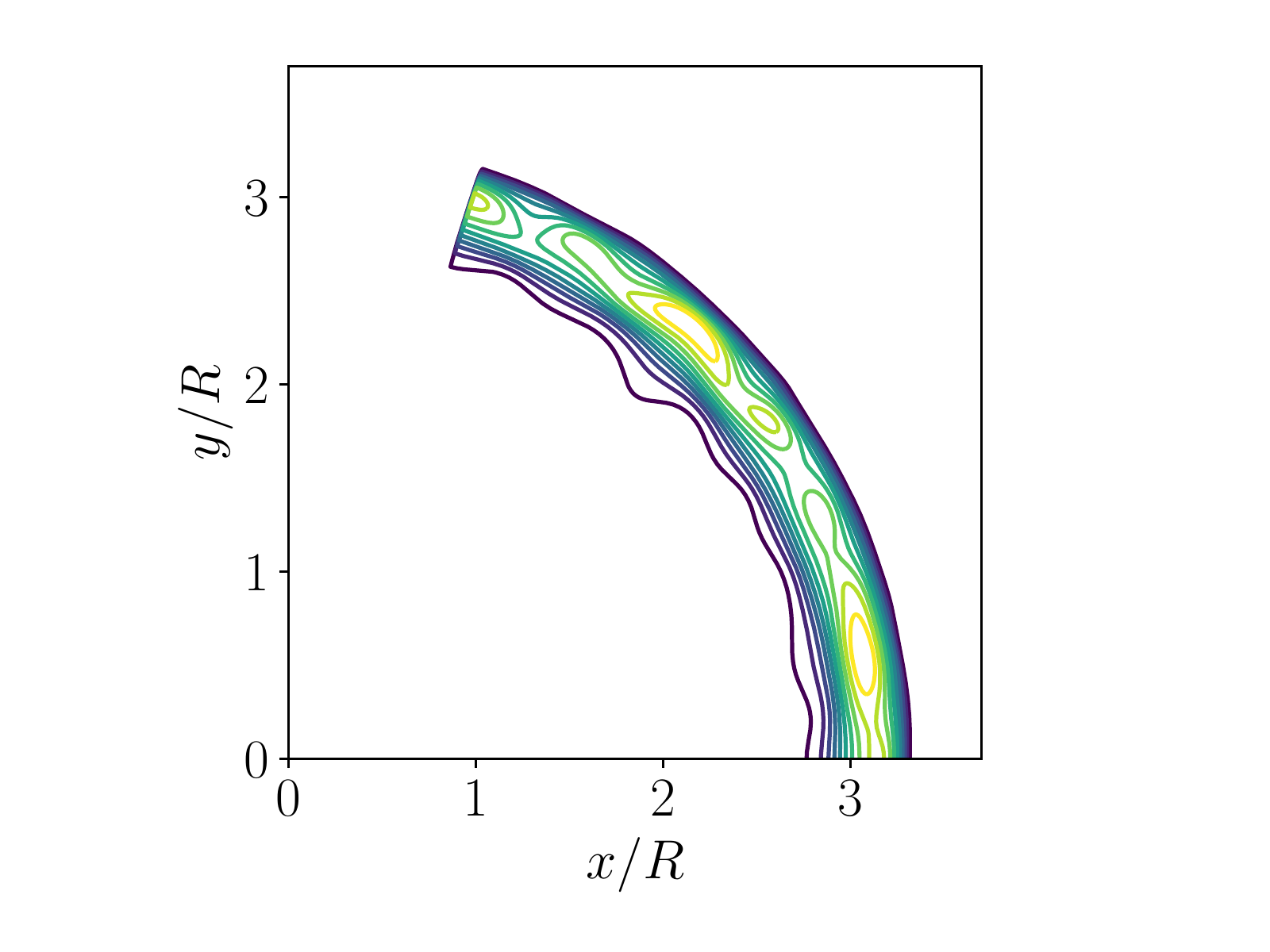}
 \caption{Contour plots of vorticity modulus at constant $z$ for the top vortex at $Re_\Gamma=3500$ and $\Lambda=0.2$ at $t=11$ (left) and $\Lambda=0.35$ at $t=16$ (right).}
 \label{fig:contwwcrps}
\end{figure}

\begin{figure}
 \centering
 \includegraphics[width=0.32\textwidth]{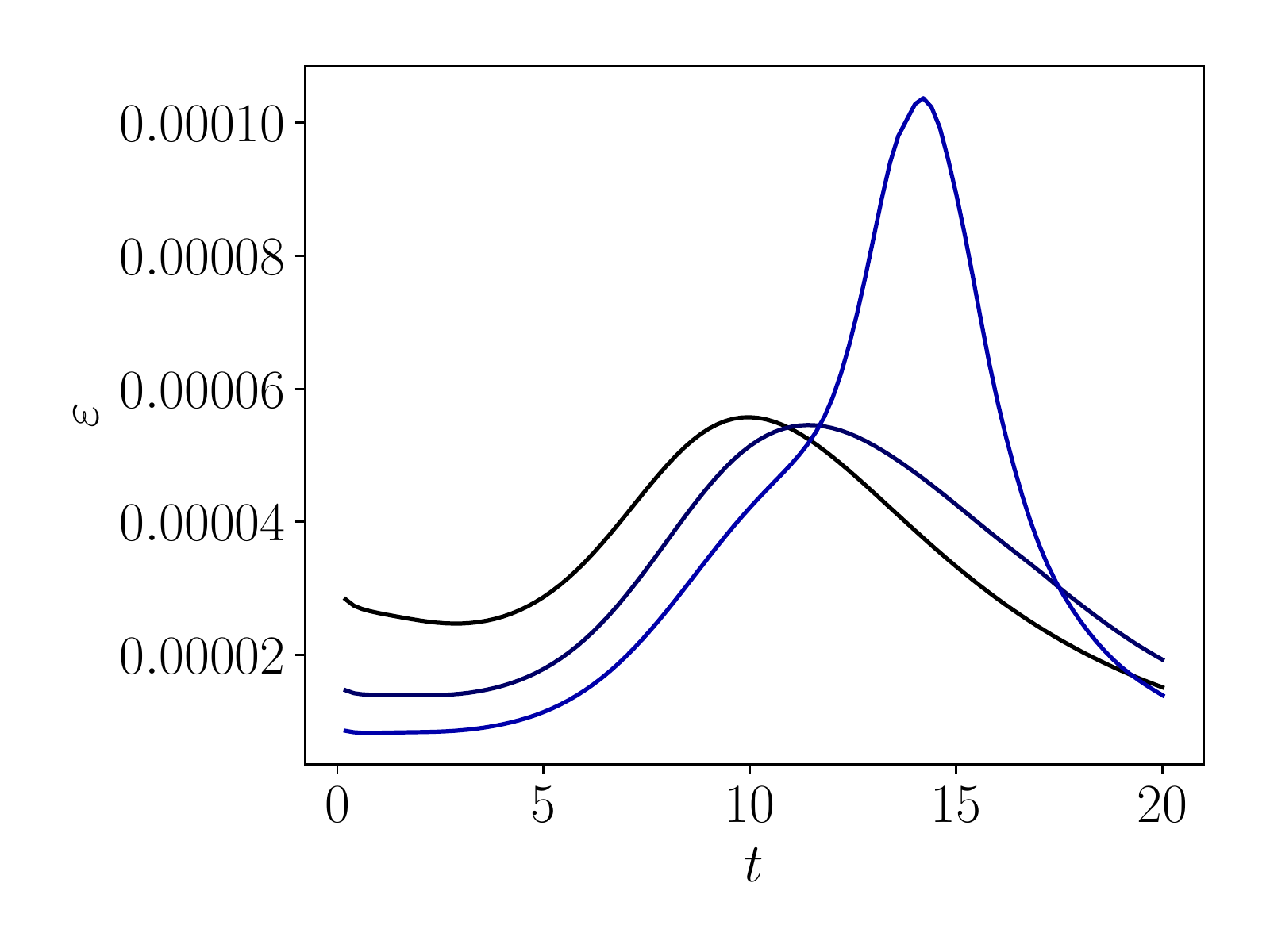}
 \includegraphics[width=0.32\textwidth]{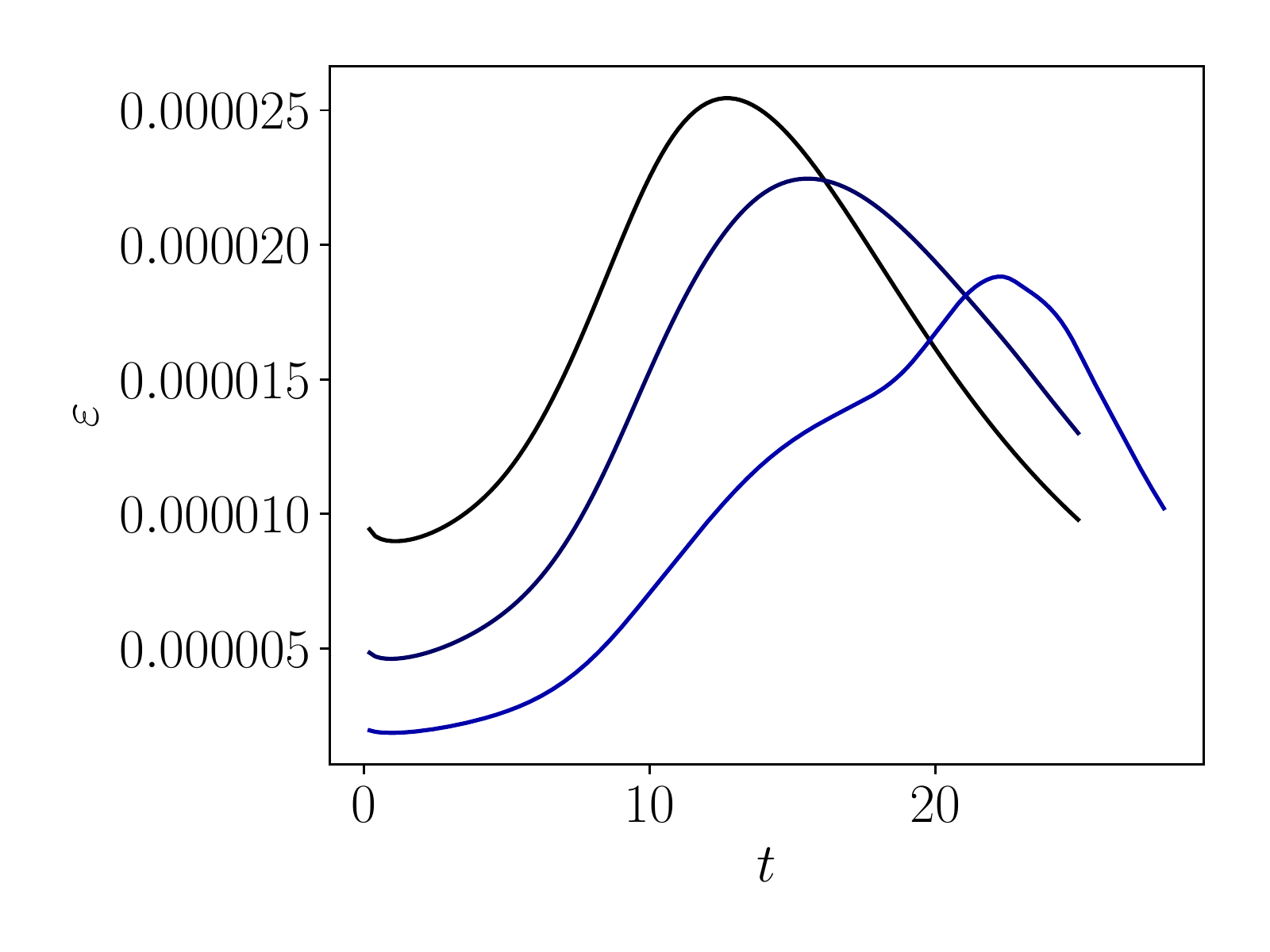}
 \includegraphics[width=0.32\textwidth]{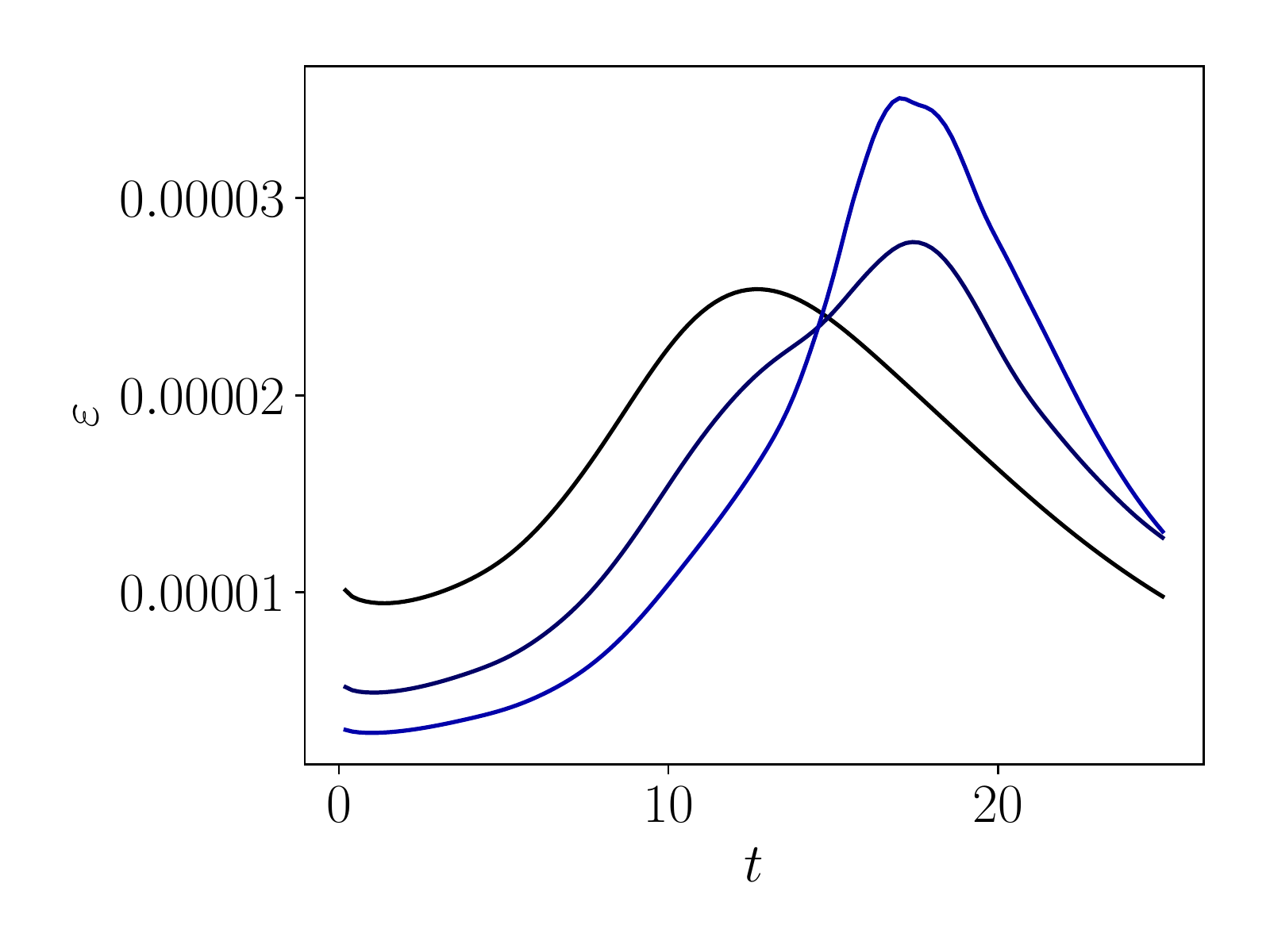}
 \caption{Temporal evolution of the dissipation and $Re_\Gamma=1000$ (black), $Re_\Gamma=2000$ (dark color) and $Re_\Gamma=3500$ (light color) for $\Lambda=0.2$ and white noise perturbations  (left) and $\Lambda=0.35$ and white noise (center) and colored noise (right) perturbations. }
 \label{fi:diss-crps}
\end{figure}

However, this initial intuition is not borne out in the simulations. For white noise, the cases with $Re_\Gamma=2000$ remain relatively axisymmetric for $\Lambda=0.2$ and $\Lambda=0.35$. Unlike the $\Lambda=0.1$ case, significant distortions of the ring or the formation of parallel vortices cannot be observed. Despite their lower value of $a/b$, thinner rings are more unstable and undergo disintegration at lower $Re_\Gamma$. To see the azimuthal instabilities set in and change the evolution of the system we have to increase the Reynolds number further to $Re_\Gamma=3500$, as 
shown by the visualization of the instantaneous vorticity in Figure \ref{fi:vortres3-cps}, the vorticity contours through the top ring in the left panel of Figure \ref{fig:contwwcrps}, and the temporal evolution of the dissipation in the left and center panels of Figure \ref{fi:diss-crps}. The $\Lambda=0.2$ case at $Re_\Gamma=3500$, illustrated by the left column of Figure~\ref{fi:vortres3-cps}, shows rapid disintegration through the elliptical instability, and we can again observe deformations of the vortex consistent with the shape that the the invariant streamtube takes during an elliptical instability in Fig.~\ref{fig:contwwcrps} \cite{lew98}. The core can be first
seen to distort with a short-wavelength pattern in the cooperative stage of the instability. This is followed by the emergence of perpendicular filaments, and finally the ring disintegrates. The dissipation profiles corroborate the mechanism, as we can observe a large increase in the instantaneous dissipation consistent with what was seen for $\Lambda=0.1$. 

On the other hand, the $\Lambda=0.35$ case at $Re_\Gamma=3500$, shown in the right column of Figure~\ref{fi:vortres3-cps}, presents a more complicated picture. While short wavelength patterns arise on the ring, the cores significantly distort around the collision plane and at times come closer to each other, something not observed at lower values of $\Lambda$.  We can also compare the contours of vorticity shown in the right panel of Fig.\ref{fig:contwwcrps} to the characteristic deformations of for the theoretical invariant streamtube in Ref.~\cite{lew98}. 
Whereas the deformation pattern at $\Lambda = 0.1$ (Fig.~\ref{fig:contwwcrp1}) was indicating that the core and periphery regions deform in opposite directions, this trend is largely absent at $\Lambda = 0.35$ (right panel of Fig.~\ref{fig:contwwcrps}).
We note that even if these contours are shown for a single time instant, this corresponds to the instant where the deformation can be most clearly appreciated. At no other point in time does the shape of the vortex resemble the characteristic shape seen earlier.

Furthermore, at the later stages ($t=18$), a clear long-wavelength pattern can be observed, even when this is not seeded at a higher level than the short wavelength. While some phenomena, from visual inspection, can be interpreted as effects of the late-stage elliptical instability, such as significant in-plane core distortion, the rings do not generate intense perpendicular filaments throughout the entire azimuthal extent as the other two cases did. Furthermore, the instantaneous dissipation does not show the a large instantaneous increase contrary to what was seen for other $\Lambda$ once the transition to the elliptical instability sets in (c.f.~Fig.~\ref{fi:diss-crps}). The interaction between tubes happens through diverse mechanisms, consonant with what was studied in Ref.~\cite{keo18}, which showed the generation of small scales through the forming of vortex sheets. The dynamics for thick rings are more complicated than simple Crow-like and elliptical-like instabilities, none of the two can grow fast enough, and there appear to be more complicated phenomena happening as the curvature of the vortex line becomes important.

\begin{figure}
 \centering
 \includegraphics[width=0.32\textwidth]{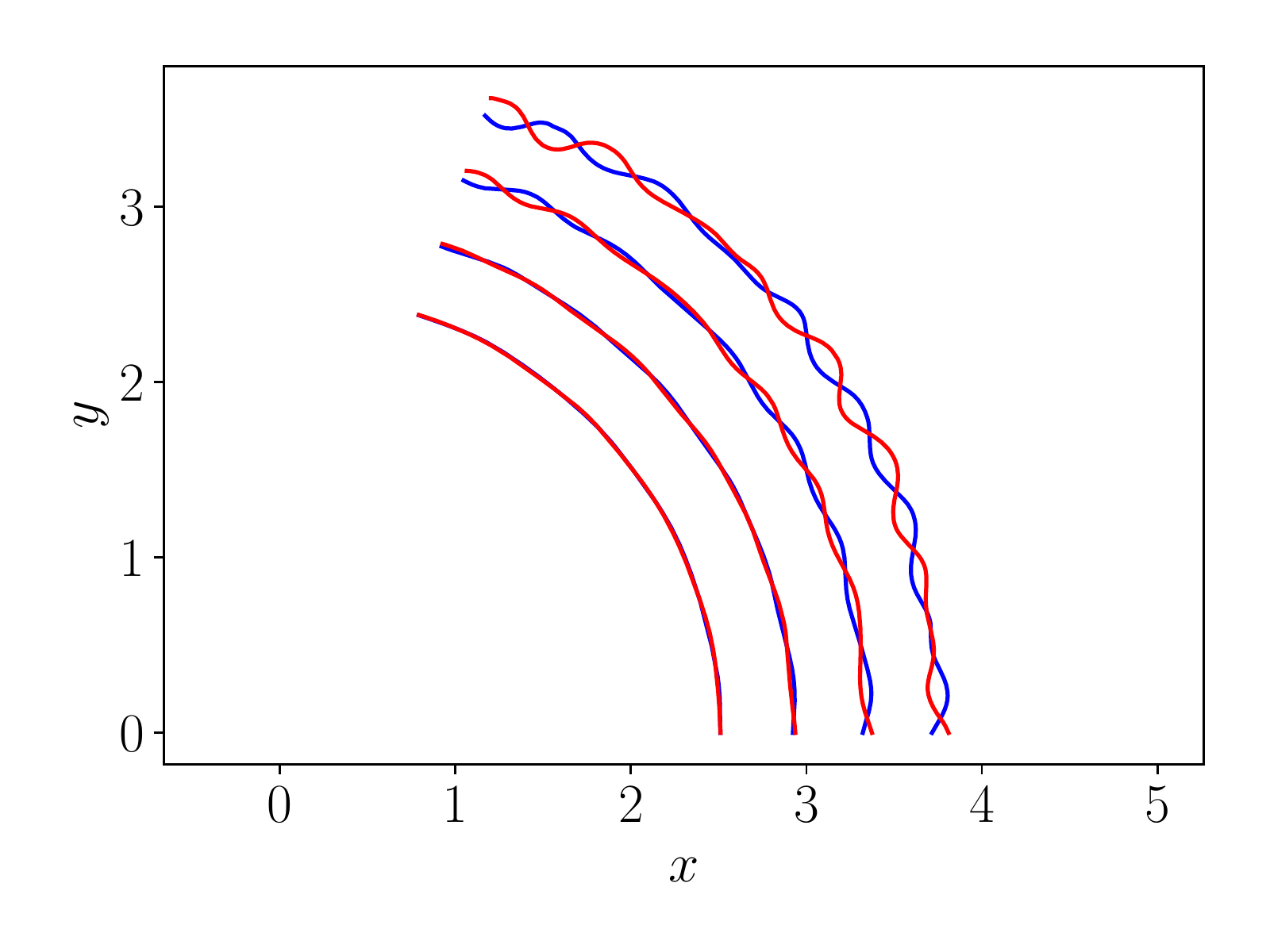}
  \includegraphics[width=0.32\textwidth]{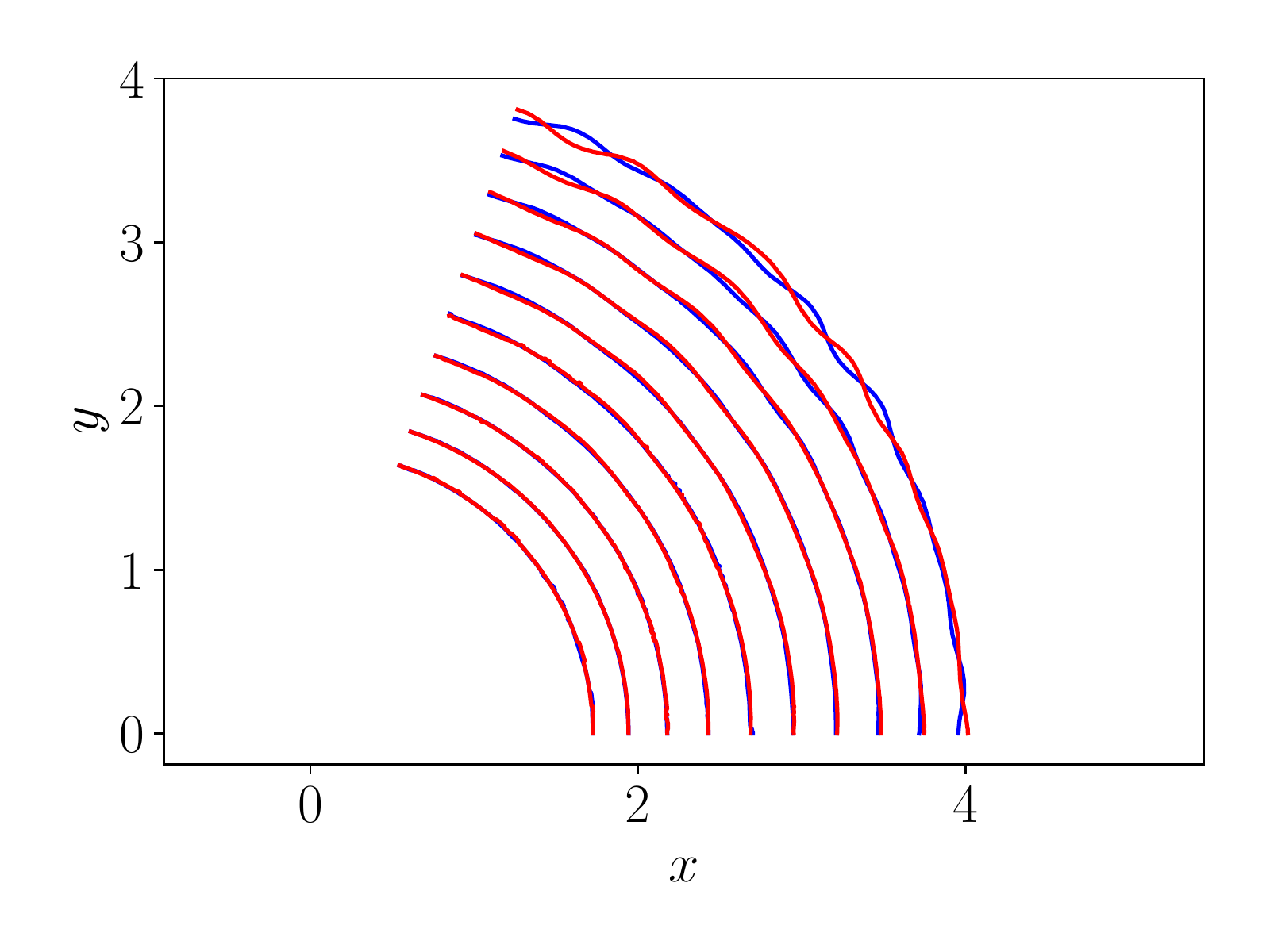}
   \includegraphics[width=0.32\textwidth]{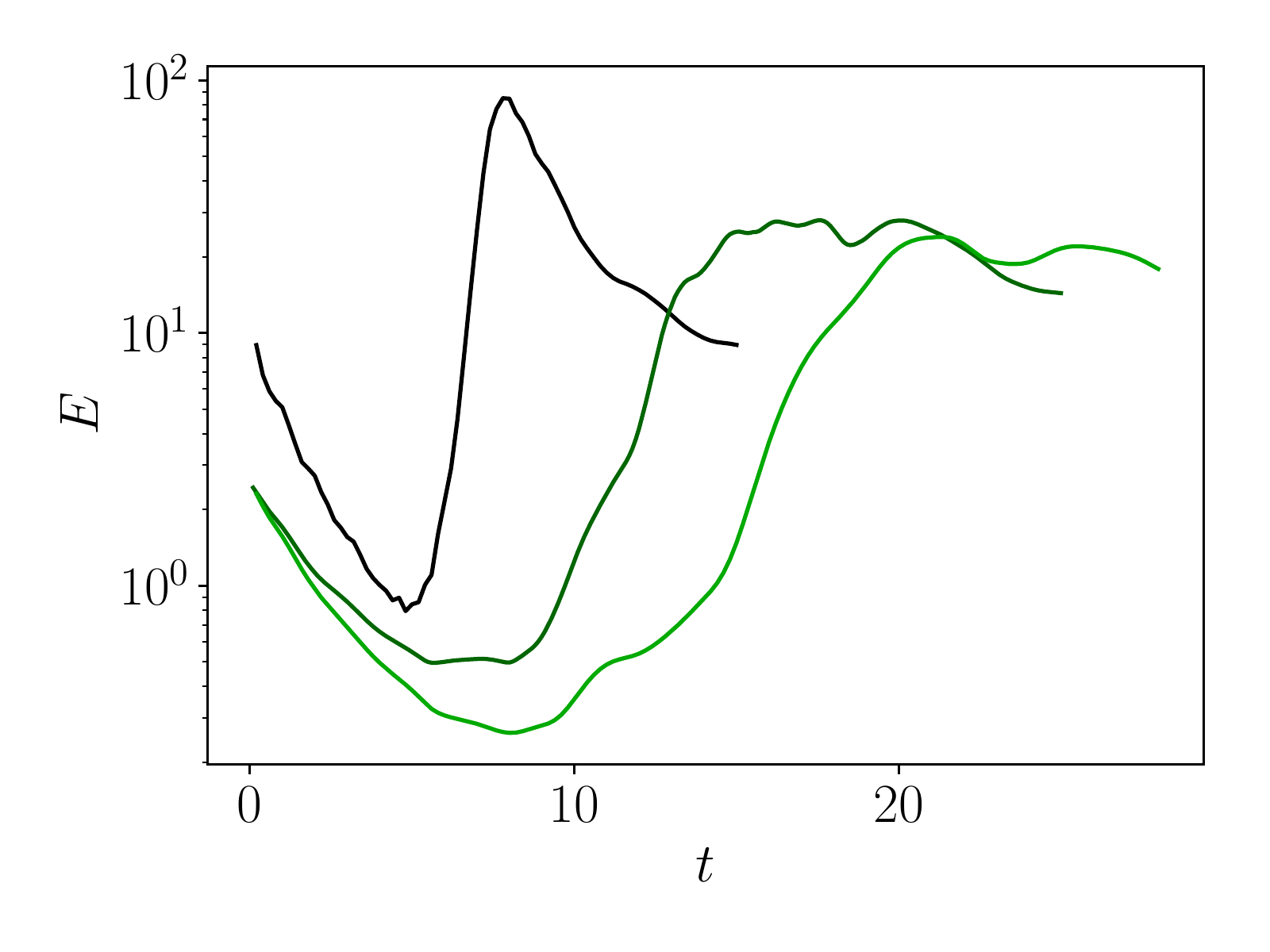} 
 \caption{Left and center: Evolution of the ring filaments for $Re_\Gamma=3500$ and several time instants. Left panel: a $\Lambda=0.2$ at times $t=9$, 10, 11, 12. Center: $\Lambda=0.35$ at times from $t=8$ to $t=17$ increasing in unit intervals. Right: Energy of the $m=40$ mode for $Re_\Gamma=3500$ and $\Lambda=0.1$ (black), $\Lambda=0.2$ (dark green) and $\Lambda=0.35$ (light green).}
 \label{fi:fils-crps}
\end{figure}

We can analyze the short wavelength instabilities by looking at the ring filaments shown in Figure \ref{fi:fils-crps}. The suspicion that the wavenumber mode of the elliptical instability had to decrease with increasing ring thickness is invalidated by this figure. Both cases show a $m=35$ or a $m=40$ pattern, which are very close to those observed earlier for $\Lambda=0.1$. What changes with increasing $\Lambda$ is that the elliptical instability sets in later once the ring has expanded beyond a certain radius, and that this setting radius becomes larger with increasing $\Lambda$. As a larger ring (outer) radius means that the core (inner) radius is smaller, and the core radius sets the length-scale for the elliptical instability, this results in the observed instability wavenumber that is essentially constant. This is further corroborated in the right panel of Figure \ref{fi:fils-crps}, which shows the energy of the $m=40$ mode for all the cases at $Re_\Gamma=3500$. The exponential growth phase of the instability, present in all cases, comes in later for the thicker rings, and with a reduced growth rate ($\sigma\approx0.8$ for $\Lambda=0.2$ and $\sigma\approx0.6$ for $\Lambda=0.35$, as compared to $\sigma\approx 2.4$ for $\Lambda=0.1$). 
These results urge for caution when extending the intuition gained from studying the elliptical instability for tubes to vortex rings.

\begin{figure}
 \centering
 \includegraphics[width=0.32\textwidth]{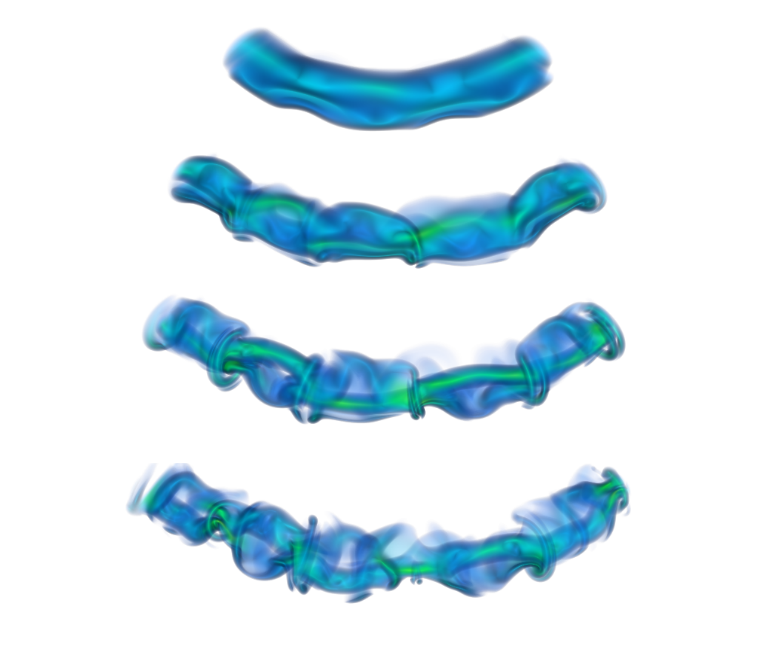}
 \includegraphics[width=0.32\textwidth]{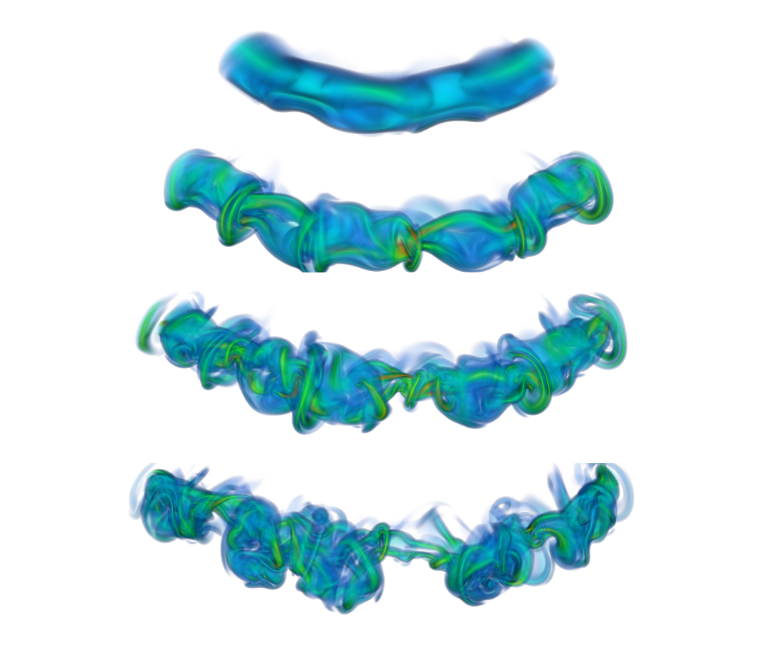}
 \caption{Vorticity volume visualization at several instances of time for $\Lambda=0.35$ and colored noise perturbations. Left is $Re_\Gamma=2000$, while right is $Re_\Gamma=3500$. Time increases from top to bottom ($t=12$, 16, 18, 20). Red denotes regions of particularly high vorticity, while blue denotes regions of low vorticity. }
 \label{fi:vortcp3-crow}
\end{figure}

We turn to the colored noise cases to analyze the effect of seeding the long-wavelengths at a higher level. The $\Lambda=0.2$ cases do not differ substantially from what was seen for $\Lambda=0.1$ in Figure \ref{fi:vortcrowcp1}: the formation of secondary rings for $Re_\Gamma=2000$, and an azimuthally inhomogeneous disintegration for $Re_\Gamma=3500$. The $\Lambda=0.35$ cases do present a distinct behaviour, which can be seen in the volume visualizations of vorticity of figure \ref{fi:vortcp3-crow} for $Re_\Gamma=2000$ and $Re_\Gamma=3500$. The vortices can be seen to come together at certain places, and to flatten as sheets. However, these sheets are not the precursor of local reconnection, as was seen for $\Lambda=0.1$ and $\Lambda=0.2$. Instead, the sheets tear apart, and roll back to form one, or two vortices as was seen in the simulations and experiments of Ref.~\cite{keo18}. 
This mechanism of tubes evolving into sheets evolving into tubes, suggested in Ref.~\cite{bre16} as a possible mechanism for the turbulent cascade, is very elusive in parameter space: it 
clearly appears only for relatively thick rings, and only once the levels of long-wavelength noise are large enough to trigger the process. This mechanism results in larger values for the instantaneous dissipation, as shown in the right panel of figure \ref{fi:diss-crps} for the coloured noise case than for the white noise case. However, even if these are larger than the corresponding values for white noise, they are still much smaller than the large increase seen when the elliptical instability dominates. Overall it is hard to conclude much about the universality of the ring collision process for $\Lambda=0.35$. It no longer resembles the relatively simpler picture of vortex tube interaction, where the elliptical instability takes over everything, and instead physical phenomena only found with two rings interacting can be seen. We expect this complicated picture to still hold for even larger values of $\Lambda$.

\subsection{Free slip wall}
\label{sec:IVb}

\begin{figure}
 \centering
 \includegraphics[width=0.32\textwidth]{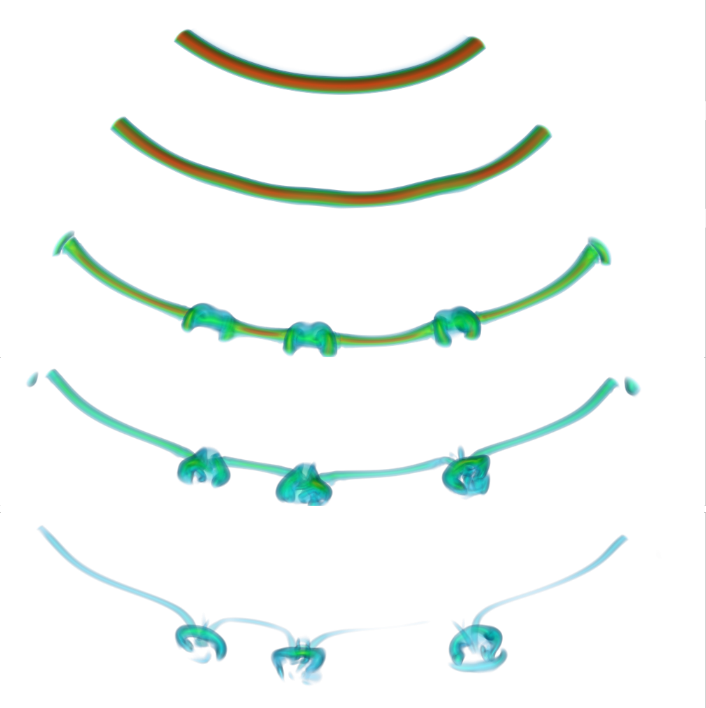}
  \includegraphics[width=0.37\textwidth]{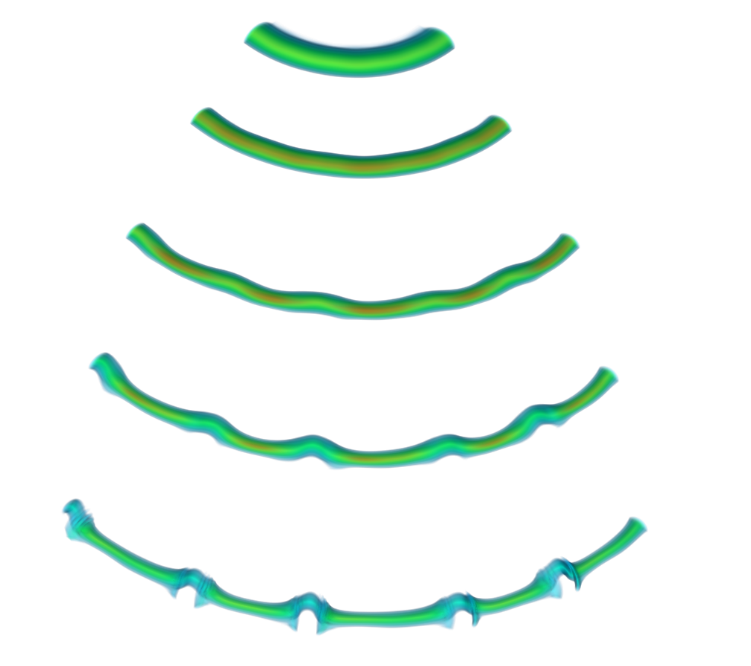}
 \caption{Vorticity volume visualization at several instances of time for $Re_\Gamma=3500$ and impact against a free-slip wall. Left is $\Lambda=0.2$, while right is $\Lambda=0.35$. Time increases from top to bottom (left: $t=22$, 28, 36, 42 and 48). Red denotes regions of particularly high vorticity, while blue denotes regions of low vorticity. }
 \label{fi:vortcp3-fs}
\end{figure}

As in the previous section, we also ran simulations where the vortex ring would impact a flat free-slip wall to best analyze the long-wavelength instabilities. We find that ring thickness also has an important effect on the behavior. For low $Re_\Gamma$, the rings remain relatively axisymmetric as was seen for $\Lambda=0.1$. A long wavelength instability, which appears as $Re_\Gamma$ increases, 
leads to the formation of secondary half-ring vortices. Figure \ref{fi:vortcp3-fs} provides a visualization of the vorticity field at high $Re_\Gamma$ for the two cases $\Lambda = 0.2$ (left column) and $\Lambda = 0.35$ (right column) at $Re_\Gamma=3500$. The first visible difference 
is that unlike the $\Lambda=0.1$ case, there is no short-wavelength instability leading to the disintegration of the remants of the primary ring that connect the secondary vortices (compare with the right column of Figure \ref{fi:vortcp1-fsw}). 
Thus, the remnants remain in place even at later times, until viscosity damps them.

\begin{figure}
 \centering
  \includegraphics[width=0.32\textwidth]{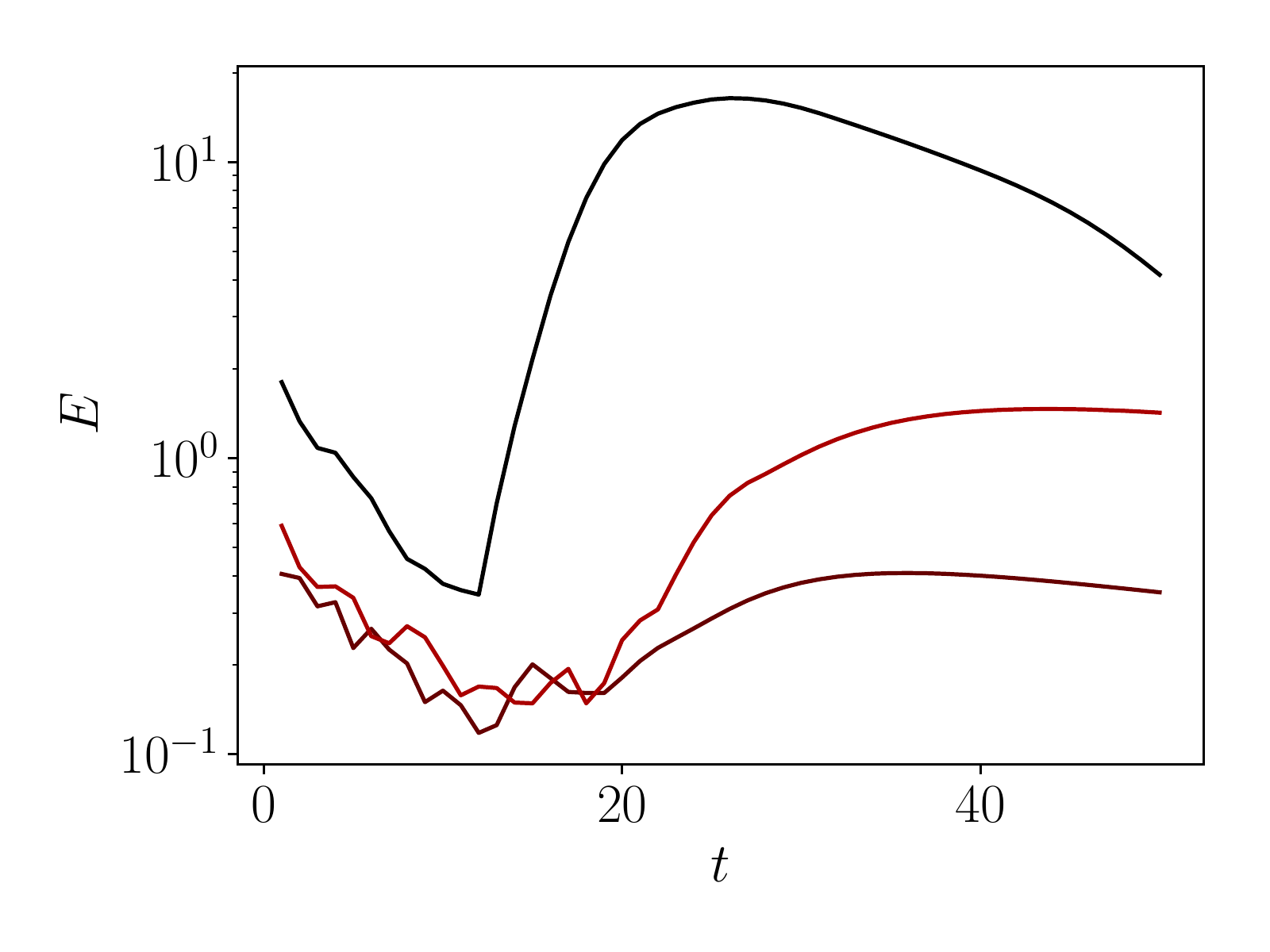}
    \includegraphics[width=0.32\textwidth]{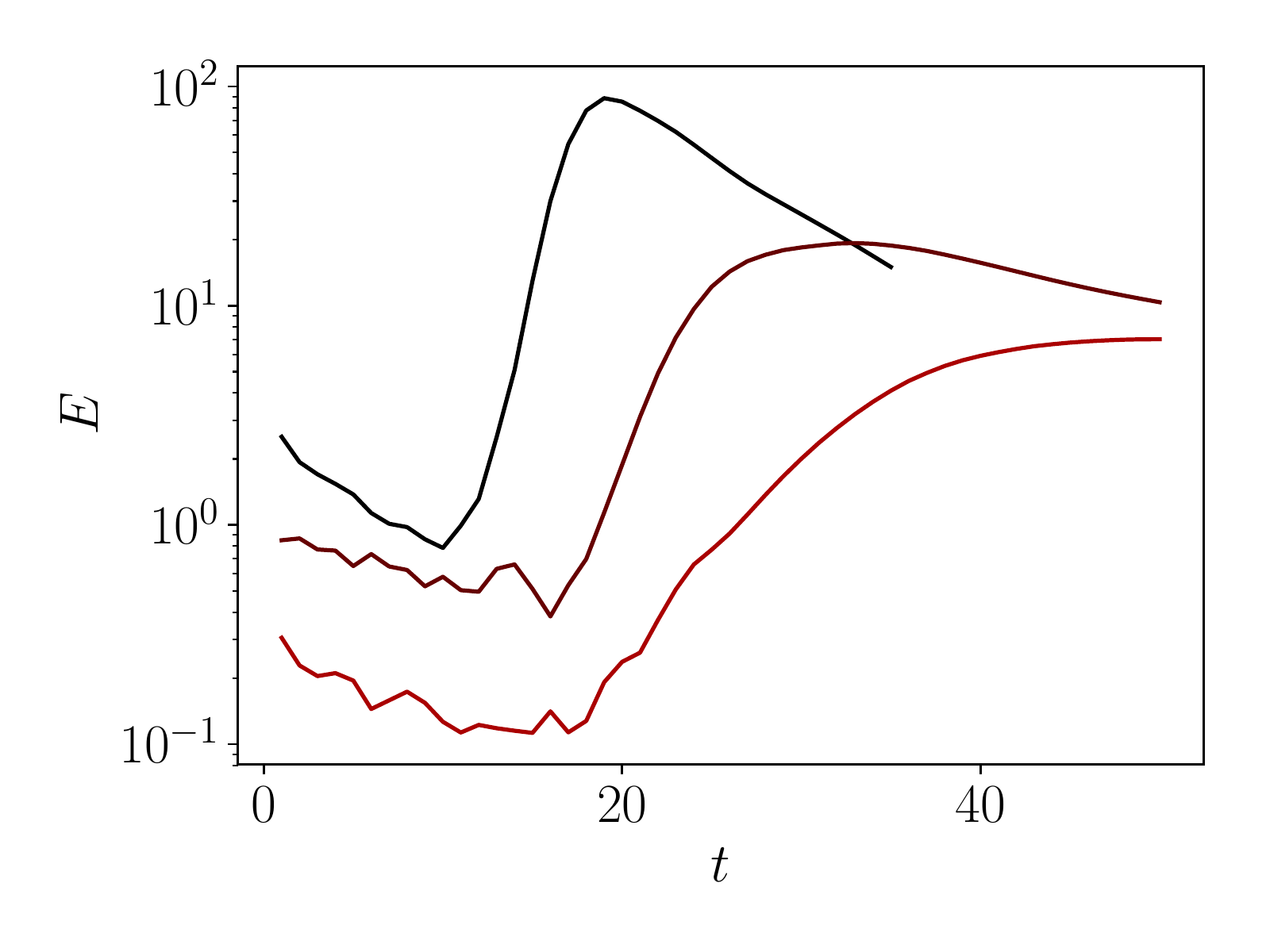}
    \includegraphics[width=0.32\textwidth]{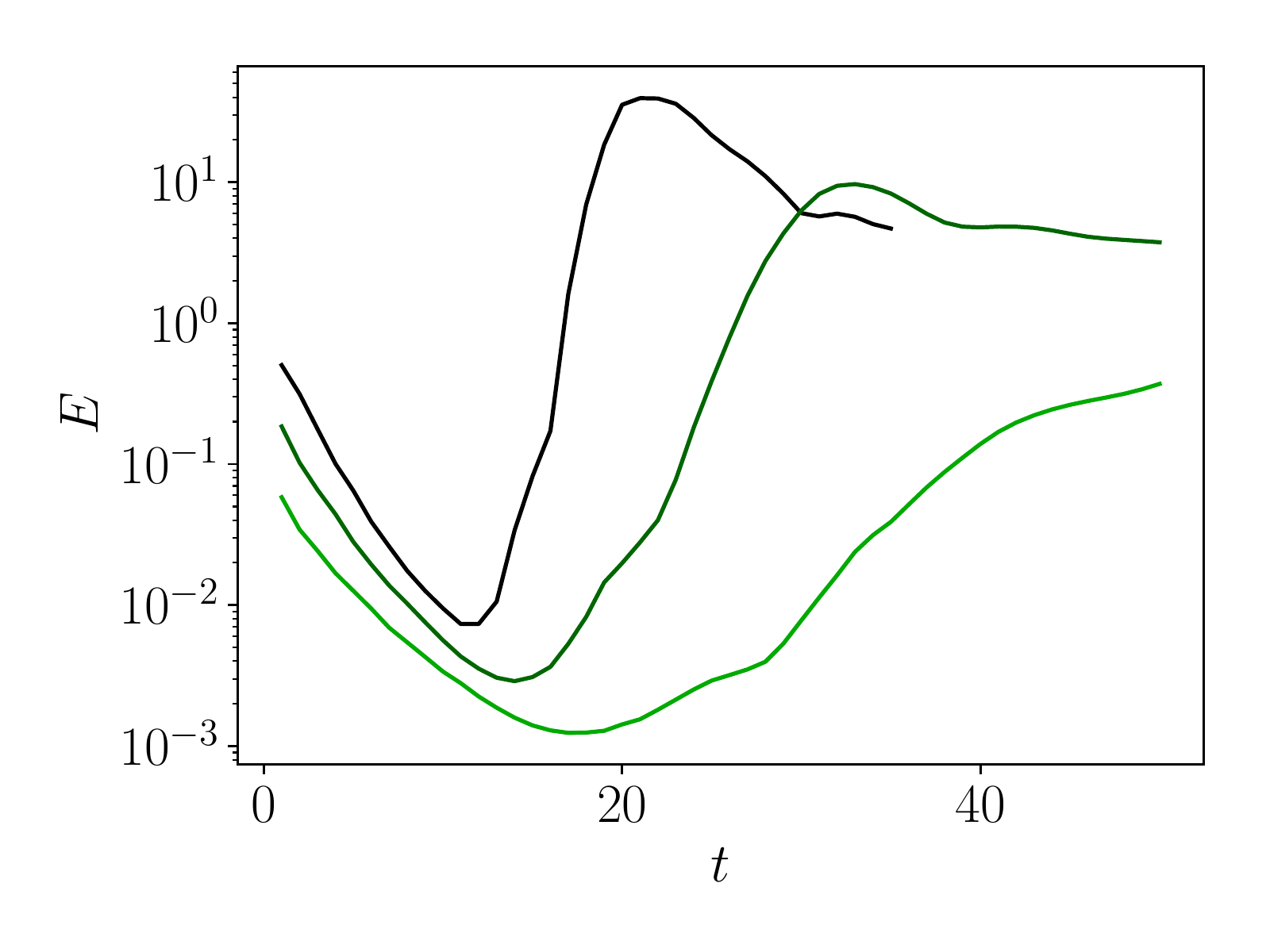}
 \caption{Left and center: Energy of the $m=10$ mode against time for $Re_\Gamma=2000$ (left) and $Re_\Gamma=3500$ (center) for $\Lambda=0.1$ (black), $\Lambda=0.2$ (dark color) and $\Lambda=0.35$ (light color) Right: Energy of the $m=40$ mode for $Re_\Gamma=3500$ and the same cases. All simulations are for impact against a stress-free wall. }
 \label{fi:fsstuffcrps}
\end{figure}

To further quantify this, we analyze how the energy in the $m=10$ mode and the $m=40$ modes are affected by the increased ring thickness. Figure \ref{fi:fsstuffcrps} shows the temporal evolution of these energies for a set of parameter values. The instability growth rate for the $m=10$ mode becomes significant for $\Lambda = 0.35$ only for $Re_\Gamma=3500$, and it is halved from the $\Lambda=0.1$ case: we can estimate $\sigma\approx 0.4$ for $\Lambda = 0.2$ and $\sigma\approx 0.2$ for $\Lambda=0.35$ at $Re_\Gamma=3500$, which are smaller than the previously obtained value of $\sigma\approx0.8$ for $\Lambda=0.1$. Second, the short wavelength modes are practically inexistent for $\Lambda=0.35$, and present a very mild growth for $\Lambda=0.2$ even at the highest $Re_\Gamma$. This is consistent with what is observed from the vorticity visualizations and what was observed for the head-on collision where the elliptical instability grew slower for thicker rings.

To conclude, we can state that ring thickness adversely affects the growth rate of both Crow-like long-wavelength instabilities that lead to local reconnection, and of elliptical-like short-wavelength instabilities that lead to disintegration. However, the former are dampened less than the latter, so long-wavelength instabilities predominate, and as these are the precursors to local reconnection, we can expect to observe more instances of local reconnection for thicker rings.

\section{No-slip wall}
\label{sec:V}

\subsection{Thin rings $\Lambda = 0.1$}
\label{sec:Va}

In this final section, we analyze the collision of a ring against a no-slip wall. This case appears to have attracted more numerical and experimental attention~\cite{wal87,orl93,swearingen1995dynamics,che10,har12} than those studied in Sec.~\ref{sec:III} and \ref{sec:IV}. 
The process through which a secondary vortex is generated from a lifted boundary layer is well understood. As long as axisymmetry is preserved, which can either be imposed in the equations or naturally happens naturally at low $Re_\Gamma$, this liftoff process can be repeated, producing tertiary vortices \cite{orl93}. This process leads to an interaction between the lifted secondary vortex and the primary vortex that ends up producing a turbulent cloud at high $Re_\Gamma$ \cite{wal87,har12}. The primary ring and the secondary ring 
lifted off from the wall have circulations which differ not only in magnitude, but also, more crucially, in sign. As such, we can expect instabilities governed by large core deformation to appear at large $Re_\Gamma$. 
Indeed, such instabilities have already been noticed and analysed in simulations \cite{swearingen1995dynamics,lut97,che10} and in experiments \cite{har12}. However, distinguishing the precise azimuthal instability responsible for the disintegration has been a matter of debate. Whereas Ref.~\cite{swearingen1995dynamics} attributes the disintegration of the vortex rings to the elliptical instability, Ref.~\cite{har12} argues that due to the mean rotation of the system, the elliptical instability is superseded by a displacement bending type of instability, consistent with the Crow mechanism.

\begin{figure}
 \centering
 \includegraphics[trim={0cm 0 7cm 0},clip,width=0.32\textwidth]{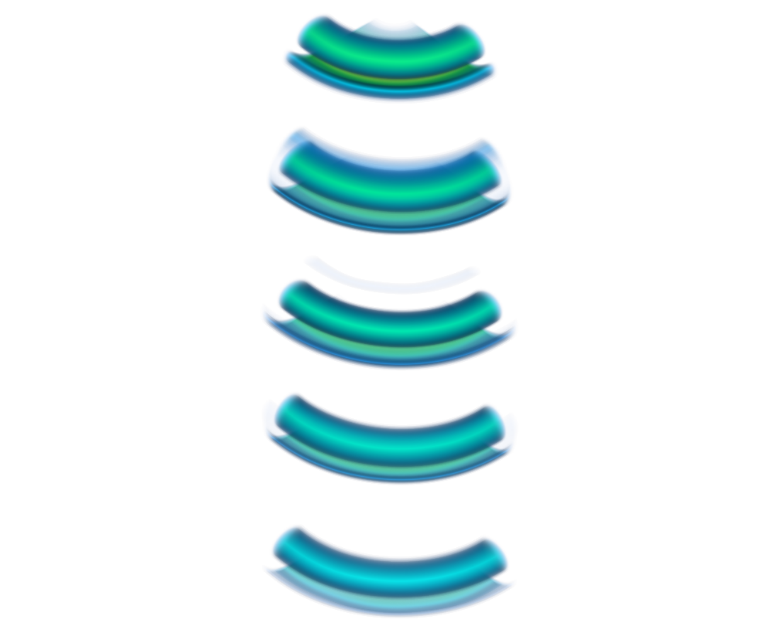}
 \includegraphics[trim={0cm 0 7cm 0},clip,width=0.32\textwidth]{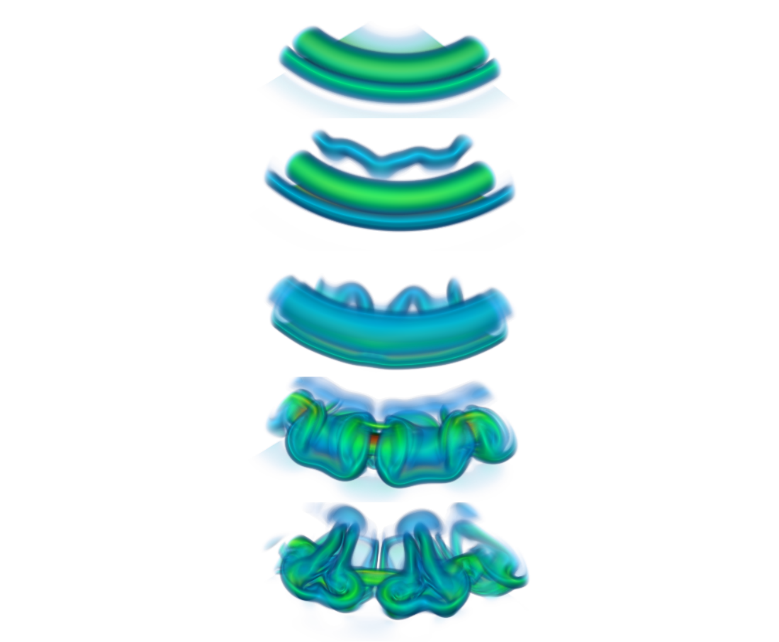}
 \includegraphics[trim={0cm 0 7cm 0},clip,width=0.32\textwidth]{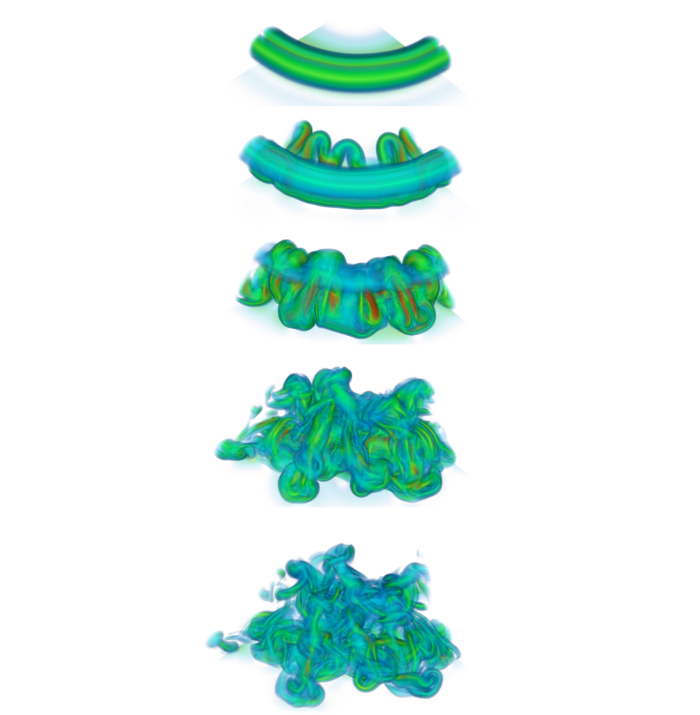}
 \caption{Vorticity volume visualization at several instances of time for ring impact against a no-slip wall at  $\Lambda=0.1$ and white noise perturbations. Reynolds number increases from left to right ($Re_\Gamma=1000$, $Re_\Gamma=2000$ and $Re_\Gamma=3500$), while time increases from top to bottom ($t=12$, $18$, $21$, $23$ and 28 for left, 12, 16, 18, 21, 24 for center and 12, 15, 18, 21, 23 for right). Red denotes regions of particularly high vorticity, while blue denotes regions of low vorticity.  }
 \label{fi:vortcp1-nsw}
\end{figure}

In this study, we focus on how $Re_\Gamma$ and $\Lambda$ affect the total disintegration of the system, while bracketing out the precise identification of the instability.
Figure \ref{fi:vortcp1-nsw} shows the vorticity magnitude for the three Reynolds numbers studied as the ring interaction with the no-slip wall proceeds. As the ring approaches the wall, it generates a boundary layer with opposite-signed vorticity. With increasing Reynolds number, the boundary layer becomes thinner and has stronger vorticity. As time evolves, the primary ring stretches out against the wall, and the boundary layer eventually lifts up. If $Re_\Gamma$ is sufficiently large, the boundary layer curls up as a secondary vortex with a much smaller circulation and radius than the primary ring. Two vortex tubes of unequal circulation rotate around a center which is located closer to the vortex with a higher circulation. In practice, this means that the secondary vortex will be rapidly pushed around the primary vortex, while the primary vortex will be slightly separated from the wall, which is usually interpreted as a rebound \cite{wal87}. We also note that during this interaction, the primary ring also stops expanding, having reached only 2-3 times its initial radius, a much smaller stretching than in the other cases \cite{wal87,che10}.

\begin{figure}
 \centering
 \includegraphics[trim={0 2cm 0cm 2cm},clip,width=0.32\textwidth]{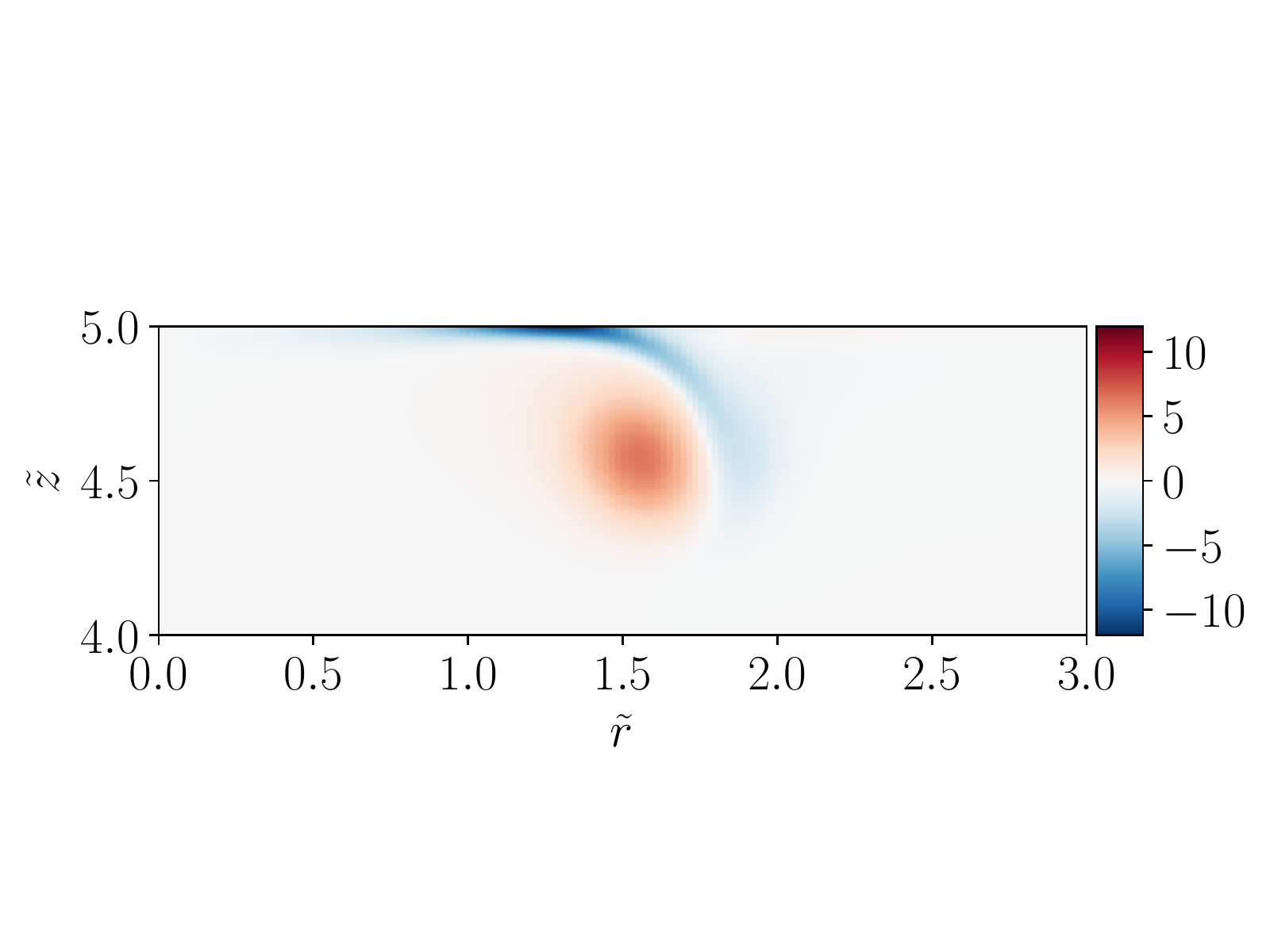}
 \includegraphics[trim={0 2cm 0cm 2cm},clip,width=0.32\textwidth]{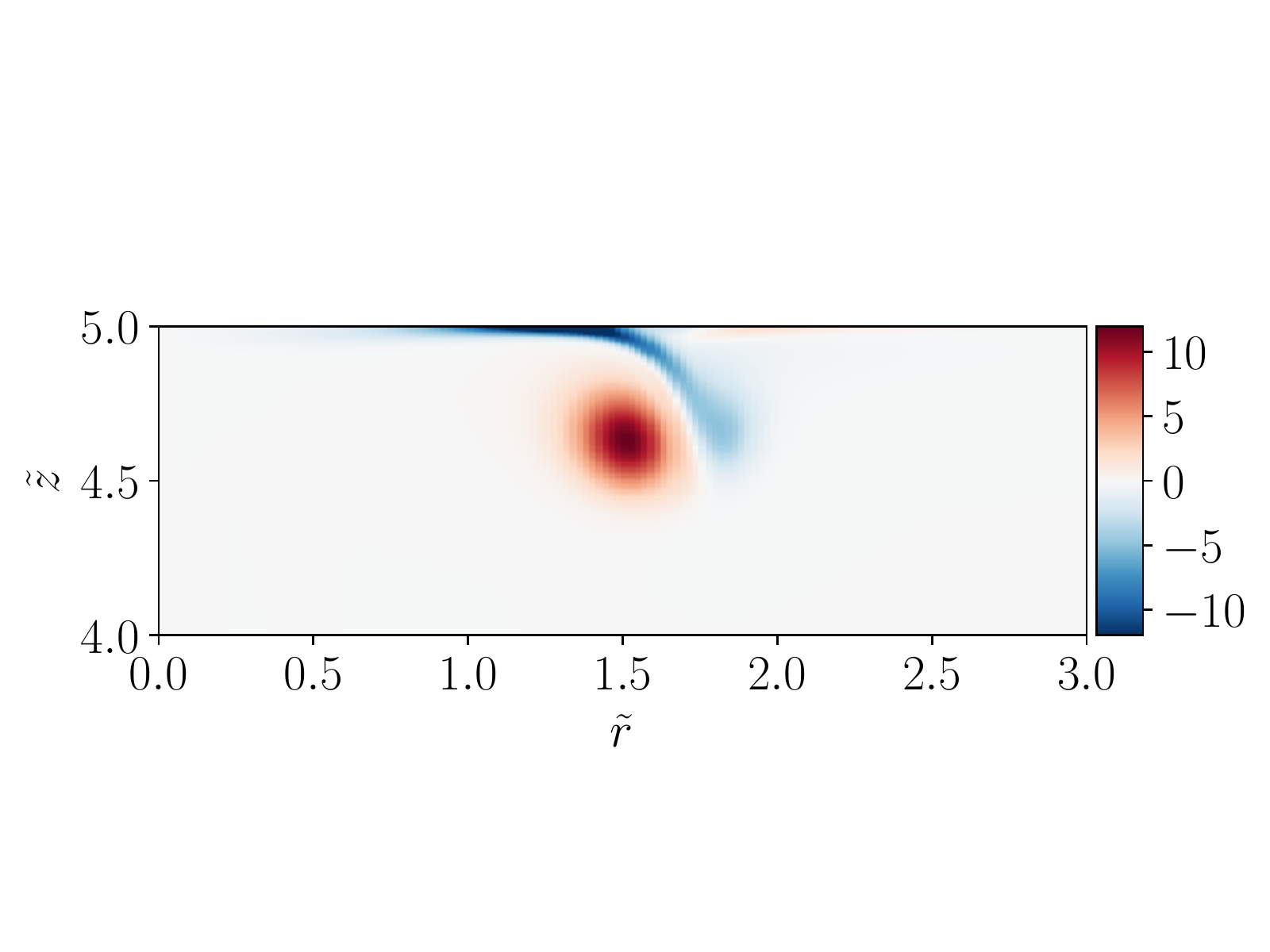}
 \includegraphics[trim={0 2cm 0cm 2cm},clip,width=0.32\textwidth]{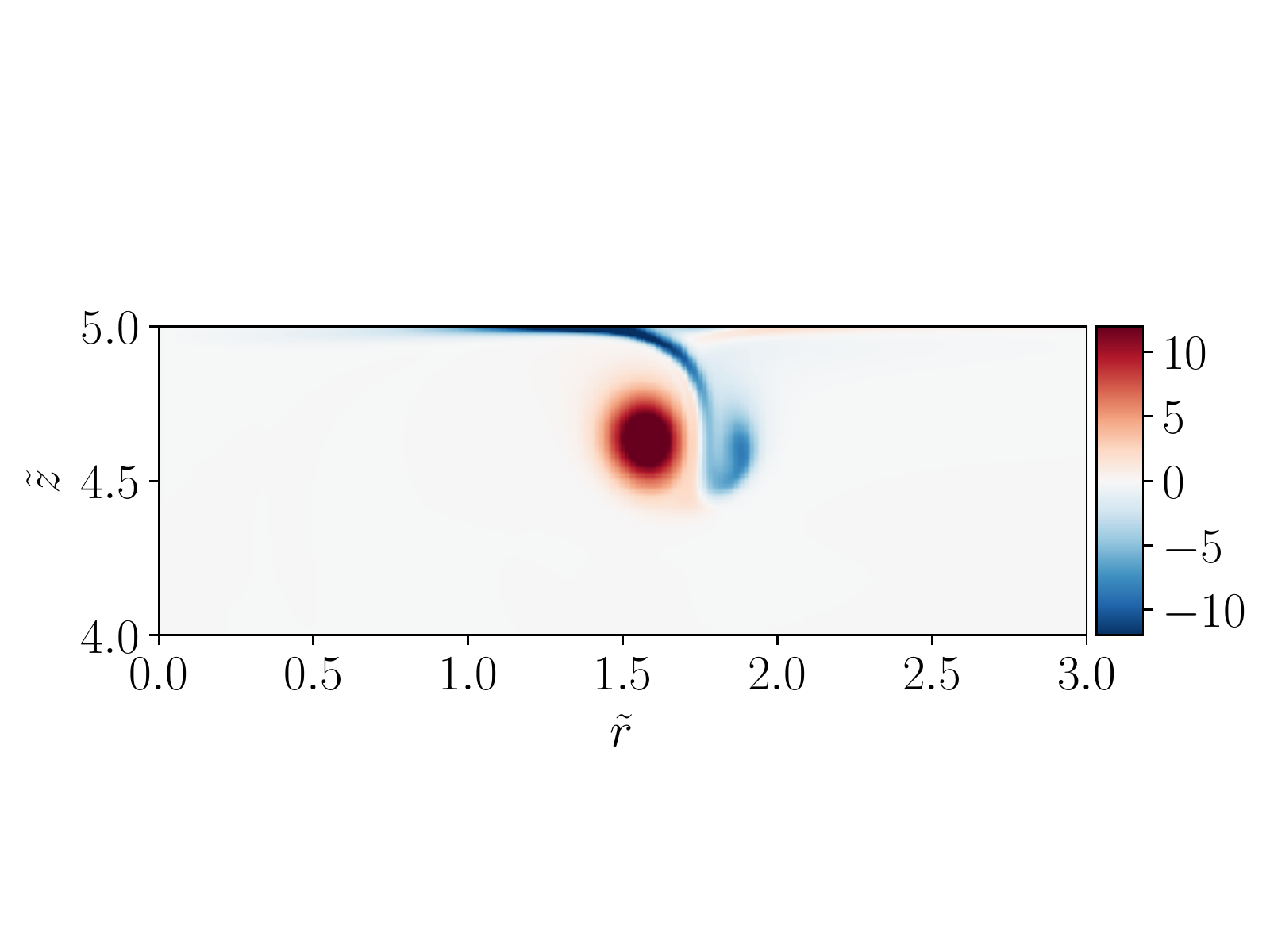}
 \caption{Azimuthal vorticity in an axisymmetric simulation of impact against a no-slip wall (at $z=5$) for $\Lambda=0.1$ at the moment of liftoff. Reynolds number increases from left to right ($Re_\Gamma=1000$, $Re_\Gamma=2000$ and $Re_\Gamma=3500$). Time is $t=13$ in the left panel and $t=12$ for the center and right panels. The circulation of the secondary vortex increases as $Re_\Gamma$ increases, up to approximately $25-30\%$ of the primary vortex ring's circulation for the right panel. }
 \label{fi:vortaxisymns-crp1}
\end{figure}

We first start by characterizing the size and strength of the secondary vortex as a function of $Re_\Gamma$. For this we conduct a series of simulations of the impact with an imposed axisymmetry. In Figure \ref{fi:vortaxisymns-crp1} we show the vorticity at the time where the boundary layer has lifted off and, for larger $Re_\Gamma$, curled up into a secondary vortex. The difference in circulation between both vortices can be really appreciated in this picture. For the highest $Re_\Gamma$, can estimate the circulation of the secondary vortex to be only between 25-30\% that of the primary vortex. This number is in line with what was reported in Ref.~\cite{har12} for finite $Re_\Gamma$, who also postulate that the circulation of the secondary vortex asymptotically tends to 40\% of the initial circulation as $Re_\Gamma\to\infty$.

Having established this, we return to the fully three-dimensional cases, which are shown in Figure \ref{fi:vortcp1-nsw}. As was revealed in the axisymmetric simulations, for $Re_\Gamma=1000$, see the left column, the lifted boundary layer is very weak, and does not form a secondary vortex. It is rapidly damped by viscosity. No centrifugal instabilities of this uncurled boundary layer are present along the lines seen in Ref.~\cite{thompson2007sphere}. The ring keeps on producing a boundary layer, which again lifts off at later times. The vorticity in the system rapidly decreases and the flow approaches equilibrium, while remaining axisymmetric. For $Re_\Gamma=2000$, see middle column, the secondary vortex is more intense, as reflected in the figure. It also undergoes a short wavelength azimuthal deformation, similar to the patterns seen before for the case of the elliptical instability, in a similar manner to what was observed in Ref.~\cite{lut97} for a tube and a no-slip wall. However, the contamination due to the presence of the no-slip wall, as well as the mean rotation due to the unequal vortex strengths makes it hard to attribute the dynamics to this instability. Furthermore, the experimental results from Ref.~\cite{har12} show that some characteristic features of the elliptical instability are missing, and we note that our results much more closely resemble those in Ref.~\cite{har12}, attributed to a Crow-like displacement instability, than those in Ref.~\cite{swearingen1995dynamics}, attributed to an elliptical instability. For lack of a definite characterization, we will refer to the corresponding mechanism as short-wavelength instability. 

The secondary ring then wraps around the primary ring, tearing it apart and generating small scales, while a tertiary vortex is being generated. This whole process becomes predominantly turbulent, even if large-scale structures can still be distinguished. For $Re_\Gamma=3500$, right column, the secondary vortex is further destabilized, and the thin, perpendicular filaments characteristic of the late-stage elliptical instability can be seen at $t=18$. The system rapidly evolves 
to turbulence as fine structure predominates at later times. The flow phenomenology seen here matches very well that observed experimentally in Ref.~\cite{wal87}, where the distortion of the secondary vortex is seen to increase with increasing $Re_\Gamma$ until the secondary vortex cannot really be discerned as it rapidly destabilizes and the flow transitions to a turbulent cloud.

For the case illustrated in Fig.~\ref{fi:vortcp1-nsw}, the fastest growing mode
can be identified visibly to be the $m=10$ mode for the thinner secondary vortex. It is unclear whether we can identify this mode with the elliptical instability as the wavelength appears to be larger than other cases, about $\approx 1.2-1.4R_0$, coincident with experiments \cite{har12} and much larger than that observed in the simulations of \cite{swearingen1995dynamics}. However, the presence of strong perpendicular filaments at the highest $Re_\Gamma$ shows that there must be some role played here by either this instability, or by the strains which are generated by two anti-parallel vortex tubes and is regularly associated to it.

In any case, the presence of this relatively weak vortex is enough to feed-back on the primary vortex and aid in its disintegration. This process shows the importance the late-stage dynamics of elliptical-like instabilities take for thin vortex rings in the three kinds of collisions considered here. In the three cases, the evolution largely follows the same pattern: an initial stretching, which remains largely axisymmetric at low Reynolds numbers. As the Reynolds number increases, azimuthal instabilities kick in which at the highest Reynolds numbers are dominated by elliptical-like instabilities that lead to rapid disintegration of the rings. 

\subsection{Effect of ring thickness}
\label{subsec:ring_thick}

\begin{figure}
 \centering
 \includegraphics[trim={0 2cm 0cm 2cm},clip,width=0.32\textwidth]{mov_01200.pdf}
 \includegraphics[trim={0 2cm 0cm 2cm},clip,width=0.32\textwidth]{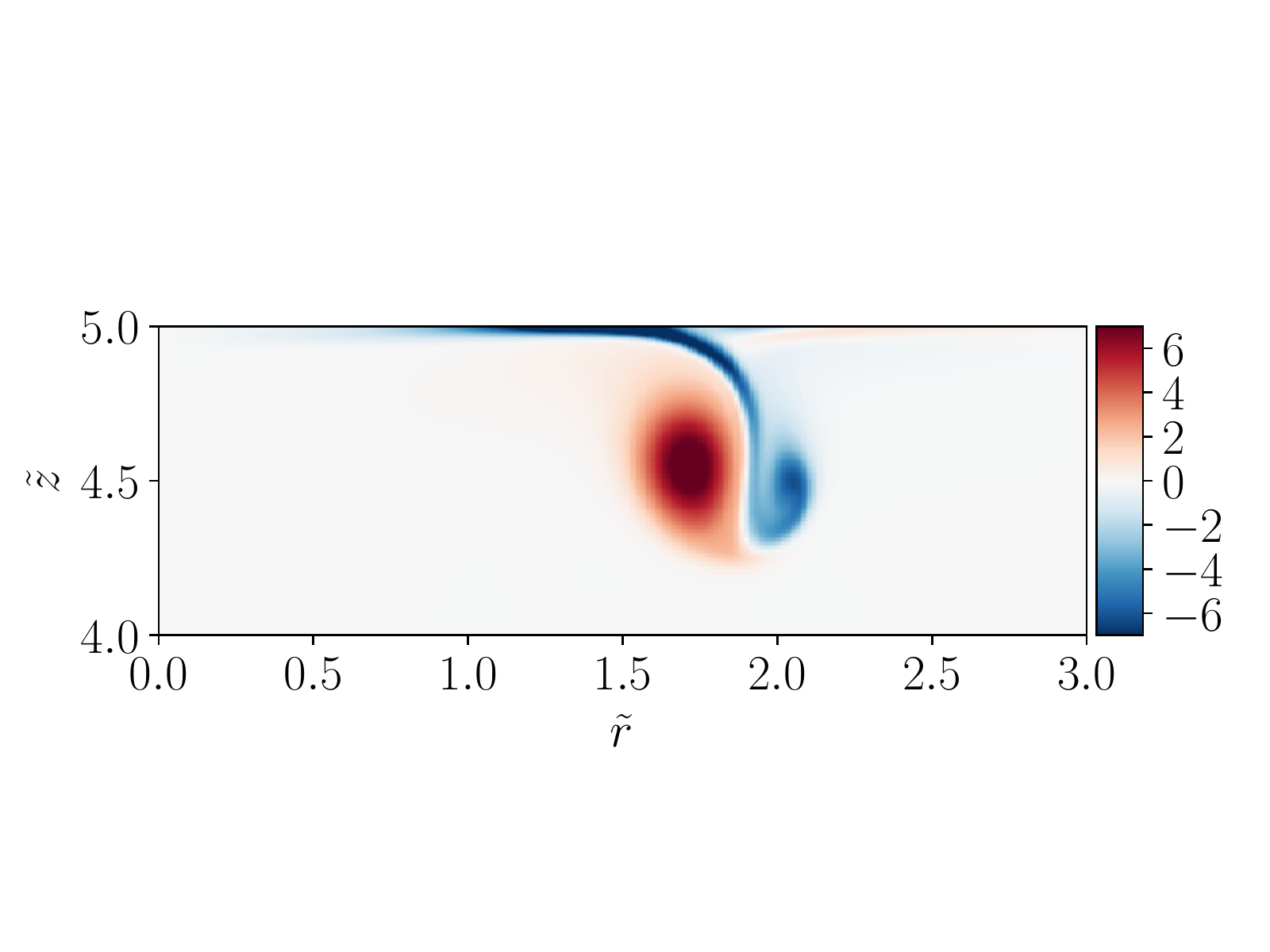}
 \includegraphics[trim={0 2cm 0cm 2cm},clip,width=0.32\textwidth]{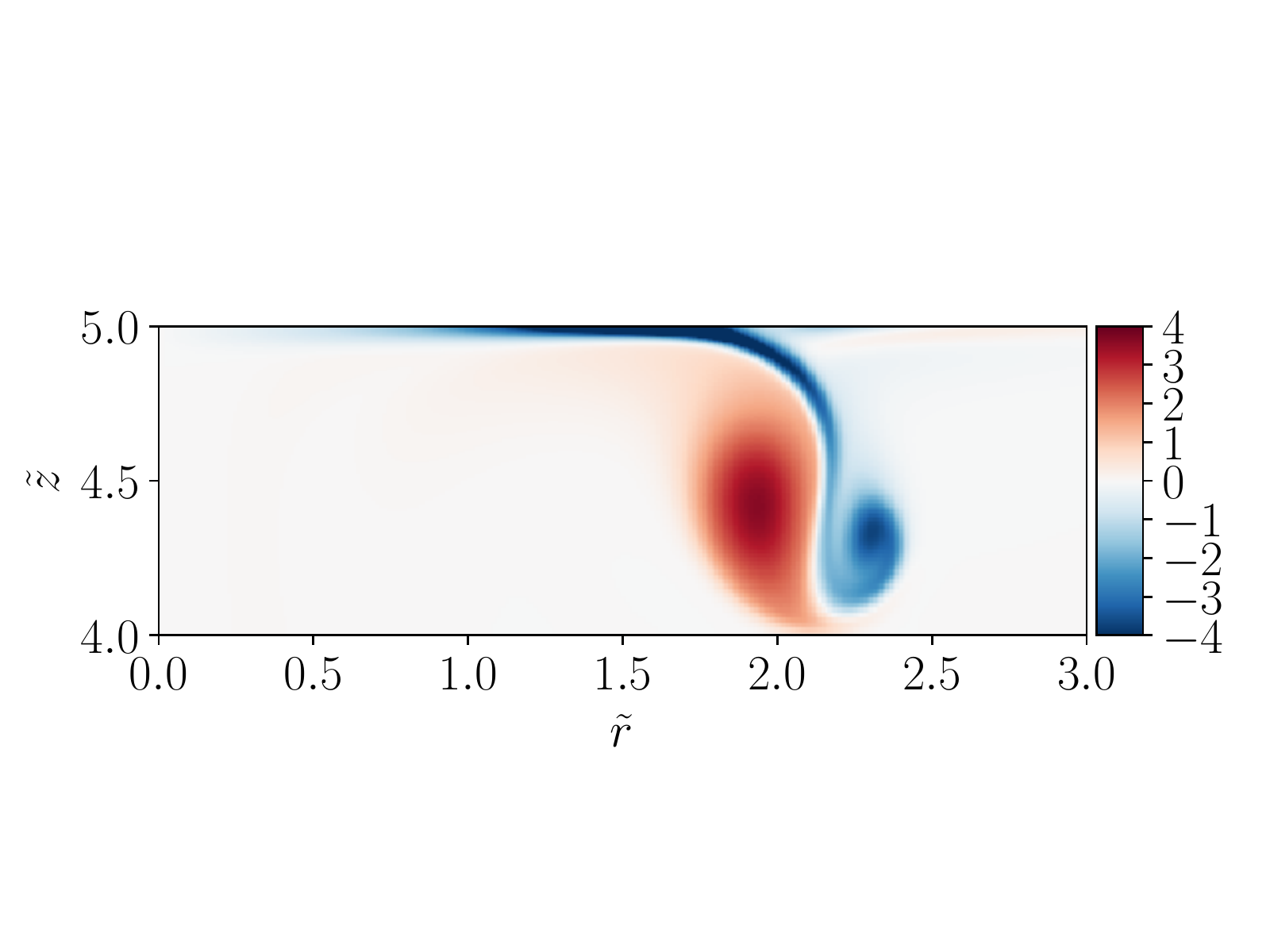}
 \caption{Azimuthal vorticity in an axisymmetric simulation of impact against a no-slip wall for $Re_\Gamma=3500$ at the moment of liftoff. Slenderness ratio increases from left to right ($\Lambda=0.1$, $\Lambda=0.2$ and $\Lambda=0.35$). Time is $t=12$ in the left panel, $t=15$ for the center and $t=20$ for the right panel.  The circulation of the secondary vortex is approximately the same in all panels, $25-30\%$ of the primary vortex rings's circulation.}
 \label{fi:vortaxisymns-crps}
\end{figure}

For the final set of simulations, we vary $\Lambda$ for the cases of no-slip impact considered above. To start, we repeat the axisymmetric simulations to characterize how the size and strength of the secondary vortex changes as $\Lambda$ varies. In Figure \ref{fi:vortaxisymns-crps} we show the azimuthal vorticity for $Re_\Gamma=3500$, at the time where the boundary layer has lifted off and curled into a secondary vortex which is to the side of the primary vortex. This time increases with $\Lambda$,
as the ring travels slower. We notice the increasing thickness of both the primary and the secondary vortex, as well as the lower values of vorticity. Interestingly, the relative circulation of the secondary vortex remains approximately constant, at 25-30\% of the value of the primary circulation independently of $\Lambda$. The size ratio between primary and secondary vortex also seems independent of $\Lambda$ from visual inspection, even if adequately measuring this is a challenge due to the sensitivity of this statistic on the choice of where the cut is made between secondary vortex and boundary layer. 

\begin{figure}
 \centering
 \includegraphics[trim={0cm 0 7cm 0},clip,width=0.32\textwidth]{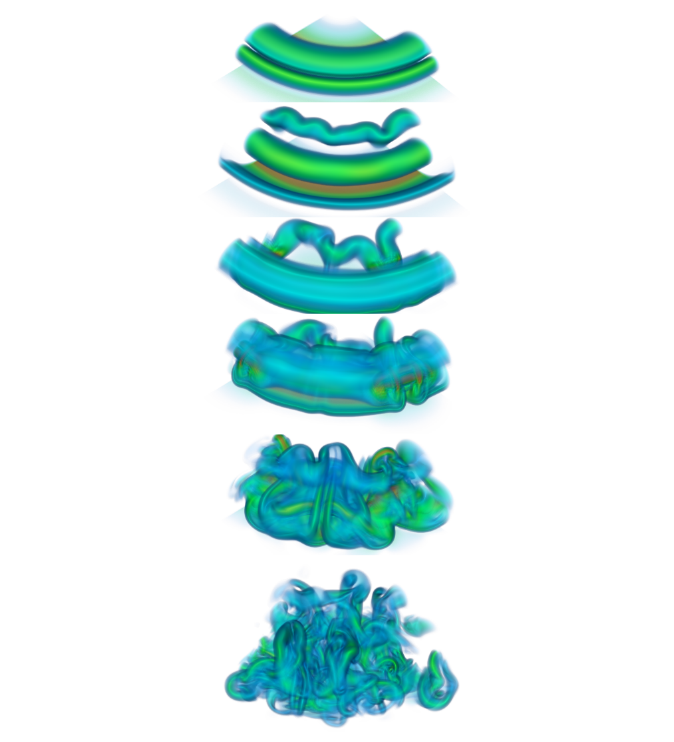}
  \includegraphics[trim={0cm 0 7cm 0},clip,width=0.32\textwidth]{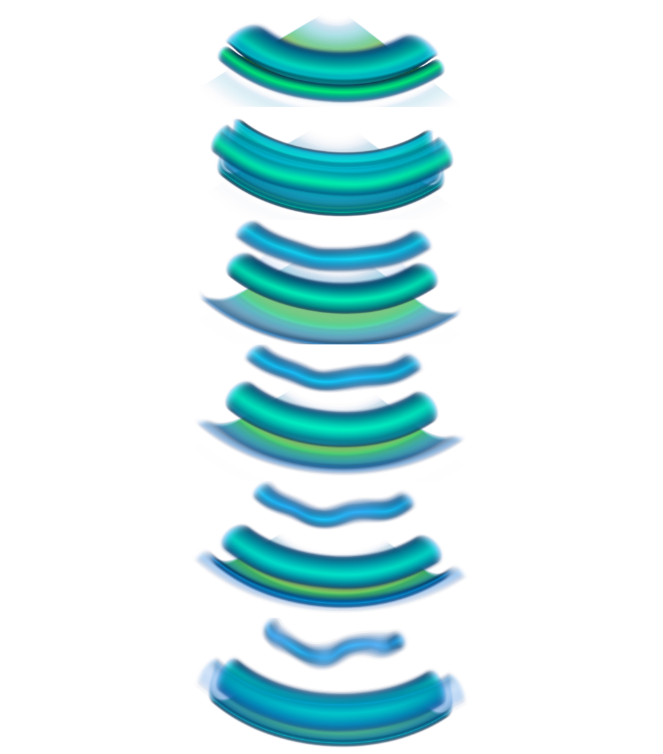}
 \caption{Vorticity volume visualization at several instances of time for ring impact against a no-slip wall at  $Re_\Gamma=3500$  and white noise perturbations. Left panel: $\Lambda=0.2$, Right panel: $\Lambda=0.35$. Time increases from top to bottom (Left: $t=14$, 19, 22, 24, 28, and 36. Right: $t=12$, 18, 21, 28, 32 and 36). Red denotes regions of particularly high vorticity, while blue denotes regions of low vorticity.  }
 \label{fi:vort-nsw}
\end{figure}

Having established this, we return to the three-dimensional simulations. For $\Lambda=0.2$, the simulation at $Re_\Gamma=1000$ remains relatively axisymmetric, while the cases at $Re_\Gamma=2000$ and $Re_\Gamma=3500$ end up with the disintegration of the vortices after a fast interaction between primary and secondary vortices. In the left panel of Figure \ref{fi:vort-nsw}, we visualize the vorticity magnitude for the largest Reynolds number simulated, $Re_\Gamma=3500$, and $\Lambda=0.2$ as the ring interaction with the no-slip wall proceeds. Indeed, it can be seen how for the secondary vortex is strongly deformed by the primary vortex, wrapping around it, until they both disintegrate into a turbulent cloud together. 

For $\Lambda=0.35$, the $Re_\Gamma=1000$ case again remains axisymmetric. For $Re_\Gamma=3500$, we do not observe a disintegration of the vortices. The only thing we can observe is how the secondary vortex deforms, as shown in the visualization in the right panel of Figure \ref{fi:vort-nsw} at $Re_\Gamma=3500$. The increased thickness of the ring prevents destabilization in a similar manner to what was observed in the previous sections. In the absence of any other mechanisms such as the centrifugal instability observed in Ref.\cite{thompson2007sphere}, the vortex ring does not disintegrate. From the axisymmetric studies, we know that the changes cannot be attributed to just the circulation, as the circulation of the secondary vortex does not depend on $\Lambda$. Instead, this shows the combined effect of both $\Lambda$ and $Re_\Gamma$ in determining the outcome of all simulations. We conducted one additional simulation for $\Lambda=0.35$ at $Re_\Gamma=5000$ to determine whether this behaviour was asymptotic, or if disintegration would happen for further increases in $Re_\Gamma$. We found that for this case, the ring disintegrated, suggesting that at sufficiently high Reynolds numbers all configurations result in disintegration.

We note that while these results are in accordance to the experiments of Refs.~\cite{wal87,har12}, they do not match the simulations of Ref.~\cite{che10}. In particular, Ref.~\cite{che10} shows vortex disintegration at Reynolds numbers as low as $Re_\Gamma\approx 925$ ($Re=500$ using their definition), for $\Lambda=0.21$, while our simulations remain axisymmetric at $Re_\Gamma=1000$. Following Ref.~\cite{ver96} we hypothesize that the main reason for this is their use of rectangular coordinates, versus our use of cylindrical coordinates. We can clearly observe that the main mode of instability in their simulations is $m=4$, associated to the use of a rectangular coordinate frame which artificially increases the instability for that mode \cite{ver96} (as a comparison, the experiments in Ref.~\cite{wal87,har12} show instabilities at higher values of $m$, or smaller wavelenghts). Meanwhile our use of cylindrical coordinates avoids this artificial forcing, and the ring is able to preserve its axisymmetry. We also note that when $Re_\Gamma$ is insufficiently large to curl up the boundary layer into a secondary vortex, we do not see any indication of the centrifugal instabilities observed in Ref.~\cite{thompson2007sphere}. As the wavenumber observed there is rather large, this absence cannot be attributed to the enforced symmetry.

\section{Conclusions and Outlook}

We have conducted an exploration of the parameter space of two vortex rings colliding head-on, and of a vortex ring impacting a no-slip and a free-slip wall. 
For thin rings, we have consistently found that the elliptical instability dominates the interactions at sufficiently large Reynolds numbers, following the general evolution pattern in Ref.~\cite{lew98}, and resulting in the formation of very fine scales consistent with Ref.~\cite{keo20}. Modifying the relative sizes of the initial noise levels triggers inhomogeneous structure formation, but at sufficiently large Reynolds numbers the ring still disintegrates. At intermediate Reynolds number, we can capture the formation of secondary rings through a long-wavelength instability which can be assimilated to the Crow instability following Ref.~\cite{lim92}. As the rings become thicker, the interactions become more complex, and for $\Lambda=0.35$, we can see vortex behaviour that cannot be simply reduced to either Crow-like or elliptical-like instabilities, and instead relates to the formation and tearing up of sheets, and the formation of vortices from rolled up sheets, similar to the mechanisms postulated in Ref.~\cite{bre16} and observed in Ref.~\cite{keo18}. 

For the cases with head-on impact against a free-slip wall, we were able to obtain local reconnection through a Crow-like instability quite robustly as long as the Reynolds number was large enough. This allowed us to estimate the Crow instability's growth rate, which was found to be approximately three times smaller than that of the elliptical instability at large Reynolds number, providing an indication of why this instability always dominates the asymptotic behaviour. Making the rings thicker weakened the Crow instability, but the flow behavior did not change much as local reconnection always prevailed.

Finally, we studied the impact of a ring against a no-slip wall. In agreement with earlier studies, we find that as a secondary vortex appears due to the boundary layer lift-off and curling at high Reynolds number, the new vortex pair undergoes a short-wavelength instability \cite{lut97,har12}. Due to the large asymmetry between primary and secondary vortex, we cannot attribute this instability to a pure elliptical instability, instead suggesting other mechanisms such as a displacement-bending Crow-like instability \cite{har12}. We also found that for the thickest rings, the secondary vortex shows some indications of instability, but the significant deformations do not occur. Only once $Re_\Gamma$ was increased to $Re_\Gamma=5000$ we observed a turbulent cloud.

This manuscript provides a large parameter space exploration which complements the works in \cite{keo18,keo20,ost21}, further solidifying the claim that the most important instability at high Reynolds number is related to the elliptical instability and to its associated strains in some shape or form. It also further corroborates the fact that high Reynolds number vortex ring collisions are a good framework for studying the generation of small scales and large dissipation occurring in turbulence. 

\noindent ~

\noindent {\it Acknowledgments:} We would like to thank R. Verzicco for many valuable discussions. We acknowledge the Research Computing Data Core, RCDC, at the University of Houston for providing us with computational resources and technical support.

\end{document}